\documentclass[twocolumn,showpacs,preprintnumbers,amsmath,amssymb]{revtex4}

\usepackage{graphicx}
\usepackage{dcolumn}
\usepackage{bm}
\usepackage{rotating}

\newcommand{\kslash}{k\kern-1ex /}
\newcommand{\pslash}{p\kern-1ex /}
\newcommand{\qslash}{q\kern-1ex /}
\newcommand{\lslash}{l\kern-1ex /}
\newcommand{\sslash}{s\kern-1ex /}
\newcommand{\Dslash}{D\kern-1.2ex /}

\newcommand{\tr}{{\rm tr}}

\newcommand{\beqa}{\begin{eqnarray}}
\newcommand{\eeqa}{\end{eqnarray}}
\newcommand{\Tr}{{\rm Tr}}

\newcommand{\bd}{\begin{description}}
\newcommand{\ed}{\end{description}}
\newcommand{\la}{\langle}
\newcommand{\ra}{\rangle}
\newcommand{\ben}{\begin{eqnarray}}
\newcommand{\een}{\end{eqnarray}}
\newcommand{\nn}{\nonumber}
\def\lsim{\raise0.3ex\hbox{$<$\kern-0.75em\raise-1.1ex\hbox{$\sim$}}}
\def\gsim{\raise0.3ex\hbox{$>$\kern-0.75em\raise-1.1ex\hbox{$\sim$}}}
\def\simgt{\rlap{\lower 6.0 pt\hbox{$\mathchar \sim$}}\raise 2.5pt \hbox {$>$}}
\def\simlt{\rlap{\lower 6.0 pt\hbox{$\mathchar \sim$}}\raise 2.5pt \hbox {$<$}}

\newcommand{\msbar}{{\overline {\rm MS}}}

\newcommand{\csw}{{c_{\rm SW}}}

\begin{document}

\preprint{UTCCS-P-44, UTHEP-567, HUPD-0801, KANAZAWA-08-06}

\title{2+1 Flavor Lattice QCD toward the Physical Point}

\author{
 S.~Aoki${}^{a,b}$,
 K.-I.~Ishikawa${}^{d}$,
 N.~Ishizuka${}^{a,c}$,
 T.~Izubuchi${}^{b,e}$,
 D.~Kadoh${}^{c}$,
 K.~Kanaya${}^{a}$,
 Y.~Kuramashi${}^{a,c}$,
 Y.~Namekawa${}^{c}$,
 M.~Okawa${}^{d}$,
 Y.~Taniguchi${}^{a,c}$,
 A.~Ukawa${}^{a,c}$,
 N.~Ukita${}^{c}$,
 T.~Yoshi\'e${}^{a,c}$\\
(PACS-CS Collaboration)
}
\affiliation{
 ${}^a$Graduate School of Pure and Applied Sciences, University of Tsukuba, Tsukuba, Ibaraki 305-8571, Japan\\
 ${}^b$Riken BNL Research Center, Brookhaven National Laboratory, Upton,
New York 11973, USA\\
 ${}^c$Center for Computational Sciences, University of Tsukuba, Tsukuba, Ibaraki 305-8577, Japan\\
 ${}^d$Graduate School of Science, Hiroshima University, Higashi-Hiroshima, Hiroshima 739-8526, Japan\\
 ${}^e$Institute for Theoretical Physics, Kanazawa University, Kanazawa,
       Ishikawa 920-1192, Japan
}



\date{\today}

\begin{abstract}
We present the first results of the
PACS-CS project which aims to simulate 2+1
flavor lattice QCD on the physical point 
with the nonperturbatively $O(a)$-improved
Wilson quark action and the Iwasaki gauge action.
Numerical simulations are carried out at $\beta=1.9$, corresponding to
the lattice spacing of $a=0.0907(13)$~fm, 
on a $32^3\times 64$ lattice with the use of the domain-decomposed 
HMC algorithm to reduce the up-down quark mass. Further 
algorithmic improvements make possible the simulation whose up-down
quark mass is as light as the physical value. 
The resulting pseudoscalar meson masses range from 702~MeV down to
 156~MeV, which clearly exhibit the presence of chiral logarithms.
An analysis of the pseudoscalar meson sector 
with SU(3) chiral perturbation theory reveals that
the next-to-leading order corrections are large at the physical strange quark mass.
In order to estimate the physical up-down quark mass, we employ the
SU(2) chiral analysis expanding the strange quark contributions 
analytically around the physical strange quark mass. 
The SU(2) low energy constants ${\bar l}_3$ and ${\bar l}_4$ are comparable with
the recent estimates by other lattice QCD calculations. 
We determine the physical point together with the lattice spacing   
employing $m_\pi$, $m_K$ and $m_\Omega$ as input.
The hadron spectrum extrapolated to the physical point shows
an agreement with the experimental values 
at a few \% level of statistical errors,
albeit there remain possible cutoff effects.
We also find that our results of 
$f_\pi=134.0(4.2)$~MeV, $f_K=159.4(3.1)$~MeV and $f_K/f_\pi=1.189(20)$
where renormalization is carries out perturbatively at one loop and 
the errors are statistical only, are compatible with the experimental values.
For the physical quark masses we obtain $m_{\rm ud}^\msbar=2.527(47)$~MeV and
$m_{\rm s}^\msbar=72.72(78)$~MeV extracted from the axial-vector
Ward-Takahashi identity with the perturbative renormalization factors.
We also briefly discuss the results for the static quark potential.
\end{abstract}

\pacs{11.15.Ha, 12.38.-t, 12.38.Gc}
\maketitle

\section{Introduction}
\label{sec:intro}

Lattice QCD is expected to be an ideal tool to understand 
the nonperturbative dynamics of strong interactions from first principles.
In order to fulfill this promise, the first
step should be to establish QCD as the fundamental theory of the
strong interaction by reproducing basic physical quantities, {\it e.g.,} the
hadron spectrum, with the systematic errors under control.
This is about to be attained thanks to
the recent progress of simulation algorithms and the 
availability of increasingly more powerful computational resources.
 
Among various systematic errors, the two most troublesome 
are quenching effects and chiral extrapolation uncertainties.
After the systematic studies on the hadron spectrum in
quenched and two-flavor QCD\cite{cppacs_nf0,cppacs_nf2,jlqcd_nf2}, 
the CP-PACS and JLQCD collaborations performed 
a 2+1 flavor full QCD simulation employing the nonperturbatively 
$O(a)$-improved Wilson quark action\cite{csw} and the Iwasaki gauge 
action\cite{iwasaki} on a (2 fm$)^3$ lattice 
at three lattice spacings\cite{cppacs/jlqcd1,cppacs/jlqcd2}.
While the quenching effects were successfully removed, 
we were left with a long chiral extrapolation:
the lightest up-down quark mass reached with the plain HMC algorithm
was about 67 MeV corresponding to $m_\pi/m_\rho\approx 0.6$.

The PACS-CS project, which is based on the PACS-CS (Parallel Array
Computer System for Computational Sciences) computer with a
peak speed of 14.3 Tflops developed at University of 
Tsukuba\cite{ukawa1,ukawa2,boku},
aims at calculations on the physical point to remove the ambiguity of 
chiral extrapolations. It employs the same quark and gauge actions as the
previous CP-PACS/JLQCD work, but uses a different simulation
algorithm: the up-down quark mass is reduced by using the
domain-decomposed HMC (DDHMC) algorithm 
with the replay trick\cite{luscher,kennedy}. At the
lightest up-down quark mass, which is about 3 MeV, several algorithmic
improvements are incorporated, including 
the mass-preconditioning\cite{massprec1,massprec2}, 
the chronological inverter\cite{chronological}, 
and the deflation technique\cite{deflation}.
For the strange quark part we improve the PHMC 
algorithm\cite{fastMC,Frezzotti:1997ym,phmc}
with the UV-filtering procedure\cite{Alexandrou:1999ii,ishikawa_lat06}.

So far our simulation points cover from 702 MeV to 156 MeV
for the pion mass.
While we still have to reduce the pion mass by 21 MeV to reach the 
real physical point, we consider that 
the findings so far already merits a detailed report.   
In this paper we focus on the following points: (i) several algorithmic 
improvements make possible a simulation with the up-down quark mass
as light as the physical value. (ii) The range of pion mass we have 
simulated is sufficiently light to deserve chiral
analyses with the chiral perturbation theory (ChPT), which
reveals that the strange quark mass is not small enough to be treated
by the SU(3) ChPT up to the next-to-leading order (NLO).
(iii) The SU(2) chiral analysis on the pion sector and the linear chiral
extrapolation for other hadron masses yield the hadron spectrum 
at the physical point which is compatible with the experimental values
at a few \% level of statistical errors.

This paper is organized as follows. In Sec.~\ref{sec:detail} 
we present the simulation details.  
Measurements of hadron masses, pseudoscalar meson decay constants 
and quark masses are described in Sec.~\ref{sec:measurement}.
In Sec.~\ref{sec:chiral} we make chiral analyses on the pseudoscalar 
meson sector using the SU(3) and SU(2) ChPTs. We present 
the values of low energy constants and discuss convergences of 
the SU(3) and SU(2) chiral expansions.
The results of hadron spectrum at the physical point are given
in Sec.~\ref{sec:physicalpt} together with 
the pseudoscalar meson decay constants and the quark masses.
In Sec.~\ref{sec:potential} we show
the results for the static quark potential. 
Our conclusions are summarized in Sec.~\ref{sec:conclusion}.
Appendices are devoted to describe the algorithmic details.
Preliminary results have been reported 
in Refs.~\cite{kura_lat07,ukita_lat07,kadoh_lat07}.

\section{Simulation details}
\label{sec:detail}
\subsection{Actions}
\label{subsec:action}

We employ the Iwasaki gauge action\cite{iwasaki} and 
the nonperturbatively $O(a)$-improved Wilson quark action
as in the previous CP-PACS/JLQCD work.
The former is composed of a plaquette and a $1\times 2$ rectangle loop:
\ben
S_{\rm g}=\frac{1}{g^2}\left\{ c_0\sum_{\rm plaquette}\tr U_{pl}
+c_1\sum_{\rm rectangle}\tr U_{rtg} \right\}
\label{eq:action_g}
\een
with $c_1=-0.331$ and $c_0=1-8c_1=3.648$.
The latter is expressed as
\begin{widetext}
\ben
S_{\rm quark}&=&\sum_{q={\rm  u,d,s}}\left[
\sum_n {\bar q}_n q_n 
    -\kappa_q \csw \sum_n \sum_{\mu,\nu}\frac{i}{2}
              {\bar q}_n\sigma_{\mu\nu}F_{\mu\nu}(n)q_n
\right.\nn\\
&&\left.
-\kappa_q\sum_n
\sum_\mu \left\{{\bar q}_n(1-\gamma_\mu)U_{n,\mu} q_{n+{\hat \mu}}
               +{\bar q}_n(1+\gamma_\mu)U^{\dag}_{n-{\hat \mu},\mu} q_{n-{\hat \mu}}\right\}
 \right],
\label{eq:action_q}
\een
\end{widetext}
where we consider the case of a degenerate up and down quark mass
$\kappa_{\rm u}=\kappa_{\rm d}$.
The Euclidean gamma matrices are defined in terms of
the Minkowski matrices in the Bjorken-Drell convention:
$\gamma_j=-i\gamma_{BD}^j$ $(j=1,2,3)$, 
$\gamma_4=\gamma_{BD}^0$,
$\gamma_5=\gamma_{BD}^5$ and 
$\sigma_{\mu\nu}=\frac{1}{2}[\gamma_\mu,\gamma_\nu]$.
The field strength $F_{\mu\nu}$ in the clover term
is given by
\ben 
F_{\mu\nu}(n)&=&\frac{1}{4}\sum_{i=1}^{4}\frac{1}{2i}
\left(U_i(n)-U_i^\dagger(n)\right), \\
U_1(n)&=&U_{n,\mu}U_{n+{\hat \mu},\nu}
         U^\dagger_{n+{\hat \nu},\mu}U^\dagger_{n,\nu}, \\
U_2(n)&=&U_{n,\nu}U^\dagger_{n-{\hat \mu}+{\hat \nu},\mu}
         U^\dagger_{n-{\hat \mu},\nu}U_{n-{\hat \mu},\mu}, \\
U_3(n)&=&U^\dagger_{n-{\hat \mu},\mu}U^\dagger_{n-{\hat \mu}-{\hat \nu},\nu}
         U_{n-{\hat \mu}-{\hat \nu},\mu}U_{n-{\hat \nu},\nu}, \\
U_4(n)&=&U^\dagger_{n-{\hat \nu},\nu}U_{n-{\hat \nu},\mu}
         U_{n+{\hat \mu}-{\hat \nu},\nu}U^\dagger_{n,\mu}.
\een
The improvement coefficient $\csw$ for $O(a)$ improvement was 
determined nonperturbatively in Ref.~\cite{csw}.

\subsection{Simulation parameters}
\label{subsec:params}

Our simulations are carried out at $\beta=1.90$ on a $32^3\times 64$
lattice for which we use $\csw=1.715$~\cite{csw}. This $\beta$ value is one of the three in the previous
CP-PACS/JLQCD work, whereas the lattice size 
is enlarged from $20^3\times 40$  to investigate the baryon masses.
The lattice spacing is found to be 0.0907(14) fm 
whose determination is explained later.
Table~\ref{tab:param} lists the run parameters of our simulations. 
The six combinations of the hopping parameters 
$(\kappa_{\rm ud},\kappa_{\rm s})$ are chosen 
based on the previous CP-PACS/JLQCD results. 
The heaviest combination 
$(\kappa_{\rm ud},\kappa_{\rm s})=(0.13700,0.13640)$ in this work 
corresponds to the lightest one in the previous CP-PACS/JLQCD simulations, 
which enable us to make a direct comparison of the two results with
different lattice sizes.
The physical point of the strange quark at $\beta=1.90$ was
estimated as $\kappa_{\rm s}=0.136412(50)$ 
in the CP-PACS/JLQCD work\cite{cppacs/jlqcd1, cppacs/jlqcd2}.
This is the reason why all our simulations 
are carried out with $\kappa_{\rm s}=0.13640$, 
the one exception being the run at 
$(\kappa_{\rm ud},\kappa_{\rm s})=(0.13754,0.13660)$ 
to investigate the strange quark mass dependence. 
After more than 1000 MD time for thermalization
we calculate hadronic observables solving quark propagators at every
10 trajectories for $\kappa_{\rm ud}\ge 0.13770$
and 20 trajectories for $\kappa_{\rm ud}=0.13781$,
while we measure the plaquette expectation value at every trajectory.

\subsection{Algorithm}
\label{subsec:algorithm}

Our base algorithm for penetrating into the small mass region for a 
degenerate pair of up and down quarks is the DDHMC algorithm\cite{luscher}.
The effectiveness of this algorithm for reducing the quark mass 
was already shown in the $N_f=2$ case\cite{luscher,del06,del07}. 
We found that it works down to $\kappa_{\rm ud} = 0.13770$ 
(or $m_\pi\approx 300$~MeV) on our $32^3\times 64$ lattice.  
Moving closer to the physical point, however, we found it necessary to 
add further enhancements including mass preconditioning, which we call 
mass-preconditioned DDHMC (MPDDHMC).  This is the algorithm we applied 
at our lightest point at $\kappa_{\rm ud}=0.13781$. 

The characteristic feature of the DDHMC algorithm is a geometric separation 
of the up-down quark determinant into the UV and the IR parts, which 
is implemented by domain-decomposing the full lattice into small blocks.
We choose $8^4$ for the block size, being less than (1~fm)$^4$ in
physical units and small enough to reside 
within a computing node of the PACS-CS computer.
The latter feature is computationally advantageous since the calculation 
of the UV part requires no communication between blocks so that 
the inter-node communications are sizably reduced.   

The UV/IR separation enables the application of multiple time scale 
integration schemes\cite{sexton}, which reduces the simulation cost 
substantially. In our simulation points we find that 
the relative magnitudes of the force terms are 
\ben   
||F_{\rm g}||:||F_{\rm UV}||:||F_{\rm IR}|| \approx 16:4:1,
\label{eq:force_ddhmc}
\een
where we adopt the convention $||M||^2=-2{\rm tr}(M^2)$
for the norm of an element $M$ of the SU(3) Lie algebra, and 
$F_{\rm g}$ denotes the gauge part and $F_{\rm UV, IR}$ are for the UV
and the IR parts of the up-down quarks.
The associated step sizes for the forces are controlled by
three integers $N_{0,1,2}$ introduced by  
$\delta\tau_{\rm g}=\tau/N_0 N_1 N_2,\ \  \delta\tau_{\rm UV}=\tau/N_1
N_2,\ \  \delta\tau_{\rm IR}=\tau/N_2$ with $\tau$ the trajectory
length.
The integers $N_{0,1,2}$ should be chosen such that 
\begin{eqnarray}
 \delta\tau_{\rm g} ||F_{\rm g}|| \approx \delta\tau_{\rm UV} ||F_{\rm UV}|| \approx \delta\tau_{\rm IR} ||F_{\rm IR}||.
\end{eqnarray} 
The relative magnitudes between the forces in Eq.~(\ref{eq:force_ddhmc})
tell us that $\delta\tau_{\rm IR}$ may be chosen roughly 16 times as 
large as $\delta\tau_{\rm g}$ and 4 times that of $\delta\tau_{\rm UV}$,
which means that we need to calculate $F_{\rm IR}$ an order of magnitude 
less frequently in the molecular dynamics trajectories.
Since the calculation of  $F_{\rm IR}$ contains the quark matrix
inversion on the full lattice, which is the most computer 
time consuming part, 
this integration scheme saves the simulation cost remarkably.

The values for $N_{0,1,2}$ are listed in Table~\ref{tab:param}, where
$N_0$ and $N_1$ are fixed at 4 for all the hopping parameters, 
while the value of $N_2$ is adjusted taking account of acceptance rate 
and simulation stability. The threshold for the replay 
trick\cite{luscher,kennedy} 
for dealing with instabilities of molecular dynamics 
trajectories leading to large values of $dH$ is set to be $\Delta  H>2$.

For the strange quark, we employ the UV-filtered PHMC (UVPHMC) 
algorithm\cite{ishikawa_lat06}.
The UVPHMC action for the strange quark is obtained through
the UV-filtering\cite{Alexandrou:1999ii} applied after 
the even-odd site preconditioning for the quark matrix.
The domain-decomposition is not used.
The polynomial approximation is corrected by the global Metropolis test
\cite{NoisyMetropolisMB}.
Since we find  $||F_{\rm s}||\approx ||F_{\rm IR}||$, 
the step size is chosen as $\delta\tau_{\rm s}=\delta\tau_{\rm IR}$.
The polynomial order for UVPHMC, which is denoted 
by $N_{\rm poly}$ in Table~\ref{tab:param},
is adjusted to yield high acceptance rate for the global Metropolis
test at the end of each trajectory.

The inversion of the Wilson-Dirac operator $D$ on the full lattice 
is carried out by the SAP (Schwarz alternating procedure) 
preconditioned GCR solver.  
The preconditioning is accelerated with the single-precision
arithmetic\cite{sap+gcr}. 
We employ the stopping condition $|Dx-b|/|b|<10^{-9}$ for the force
calculation and $10^{-14}$ for the Hamiltonian, which guarantees 
the reversibility of the molecular dynamics trajectories to a high
precision: $|\Delta U|<10^{-12}$ for the link variables and 
$|\Delta H|<10^{-8}$ for the Hamiltonian at 
$(\kappa_{\rm ud}, \kappa_{\rm s})=(0.13781,0.13640)$.
We describe the details of the DDHMC algorithm and the solver 
implementation used for $\kappa_{\mathrm{ud}}\leq 0.13770$ 
in Appendix~\ref{app:DDHMC}.

As we reduce the up-down quark mass, we observe a tendency that
the fluctuation of $||F_{\rm IR}||$ during the molecular dynamics 
trajectory increases, which results in 
a higher replay rate due to the appearance of trajectories with 
large $\Delta H$.
Since $\Delta  H$ is controlled by the product of
$\delta\tau_{\rm IR}$ and $||F_{\rm IR}||$, 
a possible solution to suppress the replay rate is to reduce 
$\delta\tau_{\rm IR}$. In this case, however, 
we find the acceptance becoming unnecessarily close to unity.
Another solution would be to tame the fluctuation of $||F_{\rm IR}||$, 
and we employ for this purpose the 
mass-preconditioner\cite{massprec1,massprec2} to the IR part 
of the pseudofermion action.  
The quark mass in the preconditioner is controlled by an additional
hopping parameter $\kappa_{\rm ud}^\prime=\rho\kappa_{\rm ud}$,
where $\rho$ should be less than unity so that calculating with the 
preconditioner is less costly than with the original IR part. 
The IR force $F_{\rm IR}$ is split into $F_{\rm IR}^\prime$ 
and ${\tilde F}_{\rm IR}$. The former is derived from
the preconditioner and the latter from the preconditioned action.

We employ the mass-preconditioned DDHMC (MPDDHMC) algorithm
for the run at the lightest up-down quark mass of
$\kappa_{\rm ud}=0.13781$.
With our choice of $\rho=0.9995$ 
the relative magnitudes of the force terms become
\ben   
||F_{\rm g}||:||F_{\rm UV}||:||F_{\rm IR}^\prime||:||{\tilde F}_{\rm IR}|| \approx 16:4:1:1/7.
\label{eq:force_mpddhmc}
\een
According to this result
we choose $(N_0,N_1,N_2,N_3)=(4,4,4,6)$ for the associated step sizes.
Here the choice of $N_2=4$ does not follow the criterion 
$\delta\tau_{\rm IR}^\prime ||F_{\rm IR}^\prime|| \approx 
\delta{\tilde \tau}_{\rm IR} ||{\tilde F}_{\rm IR}||$.
This is because we take account of the fluctuations of 
$||{\tilde F}_{\rm IR}||$.
The replay trick is not implemented in the runs at $\kappa_{\rm ud}=0.13781$.
For the step size for the strange quark in the UVPHMC algorithm 
we choose $\delta\tau_{\rm s}=\delta\tau_{\rm IR}^\prime$
as we observe $||F_{\rm s}||\approx ||F_{\rm IR}^\prime||$.

The inversion of $D$ during the molecular dynamics steps 
is also improved at $\kappa_{\rm ud}=0.13781$ in three ways.  
(i) We employ the chronological guess using the last 16 solutions 
to construct the initial solution vector of $D^{-1}$ 
on the full lattice\cite{chronological}.
In order to assure the reversibility we apply a stringent 
stopping condition $|Dx-b|/|b|<10^{-14}$ to the 
force calculation.
(ii) The inversion algorithm is replaced by a nested BiCGStab solver,
which consists of an inner solver
accelerated with single precision arithmetic and
with an automatic tolerance control ranging from $10^{-3}$ to $10^{-6}$,
and an outer solver with a stringent tolerance of $10^{-14}$ operated
with the double precision. The approximate solution obtained by the
inner solver works as a preconditioner for the outer solver.
(iii) We implement the deflation technique 
to make the solver robust against possible small eigenvalues 
allowed in the Wilson-type quark action.
Once the inner BiCGStab solver becomes stagnant during the inversion of $D$, 
it is automatically replaced by the GCRO-DR (Generalized Conjugate
Residual with implicit inner Orthogonalization and Deflated Restarting)
algorithm\cite{deflation}.
In our experience the GCRO-DR algorithm is important for calculating 
$D^{-1}$ but does not save the simulation time at $\kappa_{\rm ud}=0.13781$.
More details of the MPDDHMC algorithm and the improvements are given 
in Appendix~\ref{app:MPDDHMC}.

\subsection{Implementation on the PACS-CS computer}

All of the simulations reported in this article have been carried out 
on the PACS-CS parallel computer\cite{boku}.  
PACS-CS consists of 2560 nodes, each node equipped with a 
2.8GHz Intel Xeon single-core processor ({\it i.e.,} 5.6Gflops of peak 
speed) with 2 GBytes of main memory.  The nodes are arranged into a 
$16\times 16\times 10$ array and connected by a 3-dimensional 
hypercrossbar network made of a dual Gigabit Ethernet in each direction.  
The network bandwidth is 750 MBytes/sec for each node. 

The programming language is mainly Fortran 90 with Intel Fortran compiler. 
To further enhance the performance we used Intel C++ compiler
for the single precision hopping matrix multiplication routines which are
the most time consuming parts.
The Intel compiler enables us to use the Intel Streaming SIMD 
extensions 2 and 3 intrinsics directly without writing assembler language.

We employ a 256 node partition of PACS-CS to execute our $32^3\times 64$ runs. 
The sustained performance including communication overhead with our 
DDHMC code turns out to be 18\%.  
The computer time needed for one MD unit is listed in Table~\ref{tab:param}.  
  
\subsection{Efficiency of DDHMC algorithms}
\label{subsec:efficiency}

The efficiency of the DDHMC algorithm may be clarified in comparison 
with that of the HMC algorithm.
For $N_f=2$ QCD simulations with the Wilson-clover quark action,  
an empirical cost formula suggested for the HMC algorithm
based on the CP-PACS and JLQCD $N_f=2$ runs  was as follows~\cite{berlinwall}:
\begin{widetext}
\ben
{\rm cost[Tflops\cdot years]}&=&C\left[\frac{\#{\rm conf}}{1000}\right]\cdot
\left[\frac{0.6}{m_\pi/m_\rho}\right]^6\cdot
\left[\frac{L}{3{\rm ~fm}}\right]^5\cdot
\left[\frac{0.1{\rm ~fm}}{a}\right]^7
\label{eq:cost_hmc}
\een
\end{widetext}
with $C\approx 2.8$.
A strong quark mass dependence in the above formula 
$1/(m_\pi/m_\rho)^6\sim 1/m_{\rm ud}^3$ stems from three factors:
(i) the number of iterations for the quark matrix inversion increases 
as the condition number which is proportional to $1/m_{\rm ud}$, 
(ii)to keep the acceptance rate constant we should take
$\delta\tau\propto m_{\rm ud}$ for the step size in
the molecular dynamics trajectories, and (iii)
the autocorrelation time of the HMC evolution was consistent with 
an $1/m_{\rm ud}$ dependence in the CP-PACS runs\cite{cppacs_nf2}.

\begin{figure*}[h]
\vspace{13mm}
\begin{center}
\includegraphics[width=90mm,angle=0]{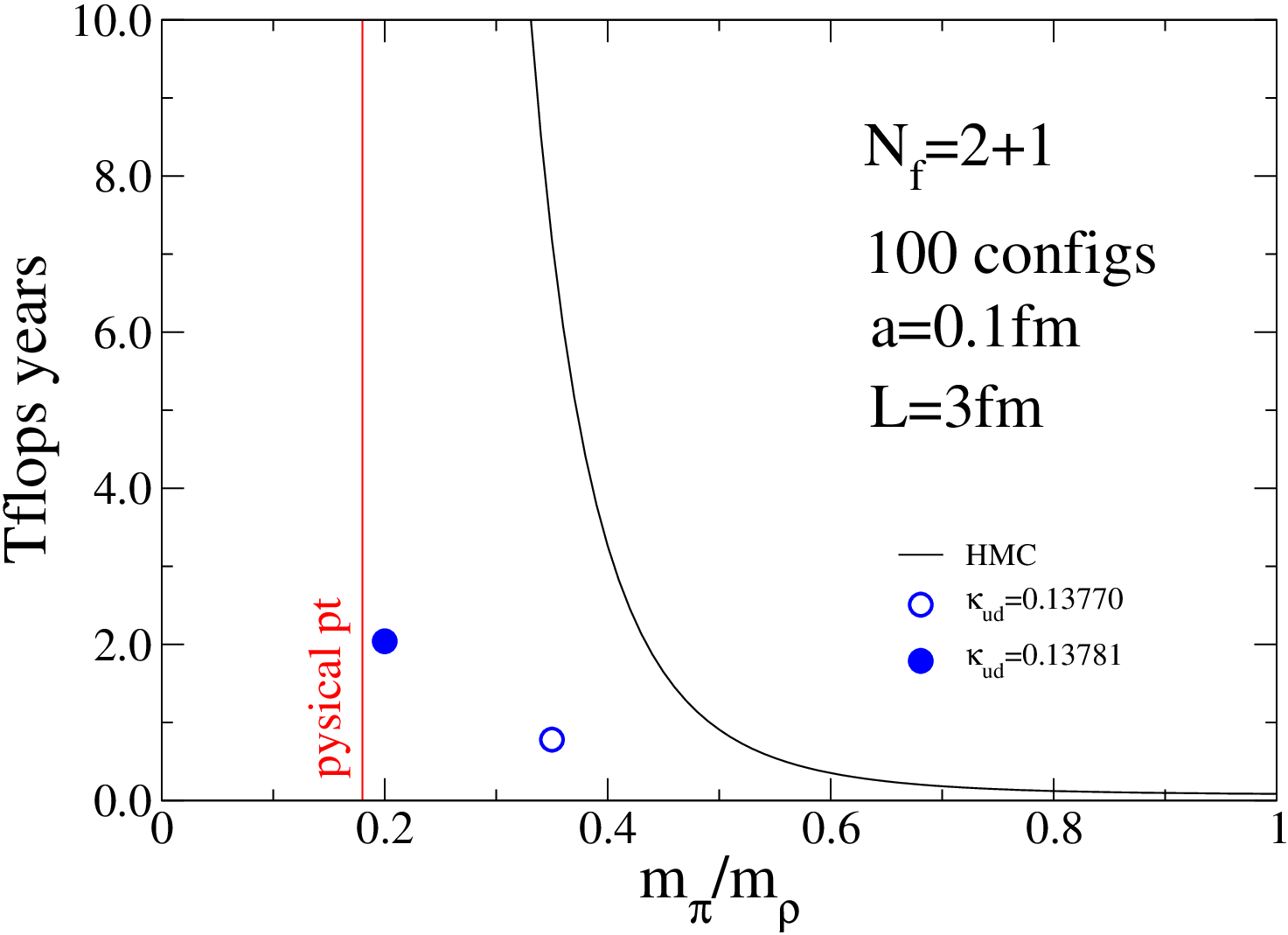}
\end{center}
\vspace{-.5cm}
\caption{Simulation cost at 
$(\kappa_{\rm ud}, \kappa_{\rm s})=(0.13770,0.13640)$ by DDHMC (blue
 open circle) and $(\kappa_{\rm ud}, \kappa_{\rm s})=(0.13781,0.13640)$ 
by MPDDHMC (blue closed circle) for 10000 trajectories.
Solid line indicates the  
cost estimate of $N_f=2+1$ QCD simulations
with the HMC algorithm at $a=0.1$~fm with $L=3$~fm
for 100 independent configurations. 
Vertical line denotes the physical point.}
\label{fig:berlinwall}
\end{figure*}

To estimate the computational cost for
$N_f=2+1$ QCD simulations with the HMC algorithm,
we assume that the strange quark contribution is given by half of
Eq.~(\ref{eq:cost_hmc}) at $m_\pi/m_\rho=0.67$ which
is a phenomenologically estimated ratio of the
strange pseudoscalar meson ``$m_{\eta_{\rm ss}}$'' and $m_\phi$:
\ben
\frac{m_{\eta_{\rm
ss}}}{m_\phi}=\frac{\sqrt{2m_K^2-m_\pi^2}}{m_\phi}\approx 0.67.\nn
\een
Since the strange quark is relatively heavy, its computational
cost occupies only a small fraction as the up-down quark mass decreases.
In Fig.~\ref{fig:berlinwall} we draw the cost formula
for the $N_f=2+1$ case as a function of $m_\pi/m_\rho$,
where we take \#conf=100, $a$=0.1~fm and $L=3$~fm in Eq.~(\ref{eq:cost_hmc})
as a representative case.
We observe a steep increase of the computational cost below
$m_\pi/m_\rho\simeq 0.5$.
At the physical point the cost expected from Eq.~(\ref{eq:cost_hmc}) 
would be $O(100)$ Tflops$\cdot$years.

Let us now see the situation with the DDHMC algorithm.
The blue open symbol in Fig.~\ref{fig:berlinwall} denotes the measured cost at
$(\kappa_{\rm ud}, \kappa_{\rm s})=(0.13770,0.13640)$, 
which is the lightest point implemented with the DDHMC algorithm.
Here we assume that we need 100 MD time separation between 
independent configurations.
We observe a remarkable reduction in the cost by a factor $20-30$ 
in magnitude.
The majority of this reduction arises from the multiple time scale
integration scheme and the GCR solver
accelerated by the SAP preconditioning
with the single-precision arithmetic.
Roughly speaking, the improvement factor is $O(10)$ for the former
and $3-4$ for the latter.
The cost of the MPDDHMC algorithm at 
$(\kappa_{\rm ud}, \kappa_{\rm s})=(0.13781,0.13640)$
is plotted by the blue closed symbol in Fig.~\ref{fig:berlinwall}. 
In this case, the reduction is mainly owing to the multiple time scale
integration scheme armored with the mass-preconditioning and 
the chronological inverter for $F_{\rm IR}^\prime$ and ${\tilde F}_{\rm IR}$.
As we already noted, the GCRO-DR solver does not accelerate the inversion
albeit it renders the solver robust against the small eigenvalues
of the Wilson-Dirac operator.

Since we find in Table~\ref{tab:param} 
that $\tau_{\rm int}[P]$ is roughly independent
of the up-down quark mass employed in the DDHMC algorithm,
the cost is expected to be proportional to $1/m_{\rm ud}^2$.
Assuming this quark mass dependence for the MPDDHMC algorithm,
we find that simulations at the physical point is feasible, 
at least for $L\approx 3$~fm lattices, with 
$O(10)$ Tflops computers, which are already available at present.

\subsection{Autocorrelations and statistical error analysis}
\label{subsec:autoc}

\begin{figure*}[h]
\vspace{13mm}
\begin{center}
\begin{tabular}{cc}
\includegraphics[width=85mm,angle=0]{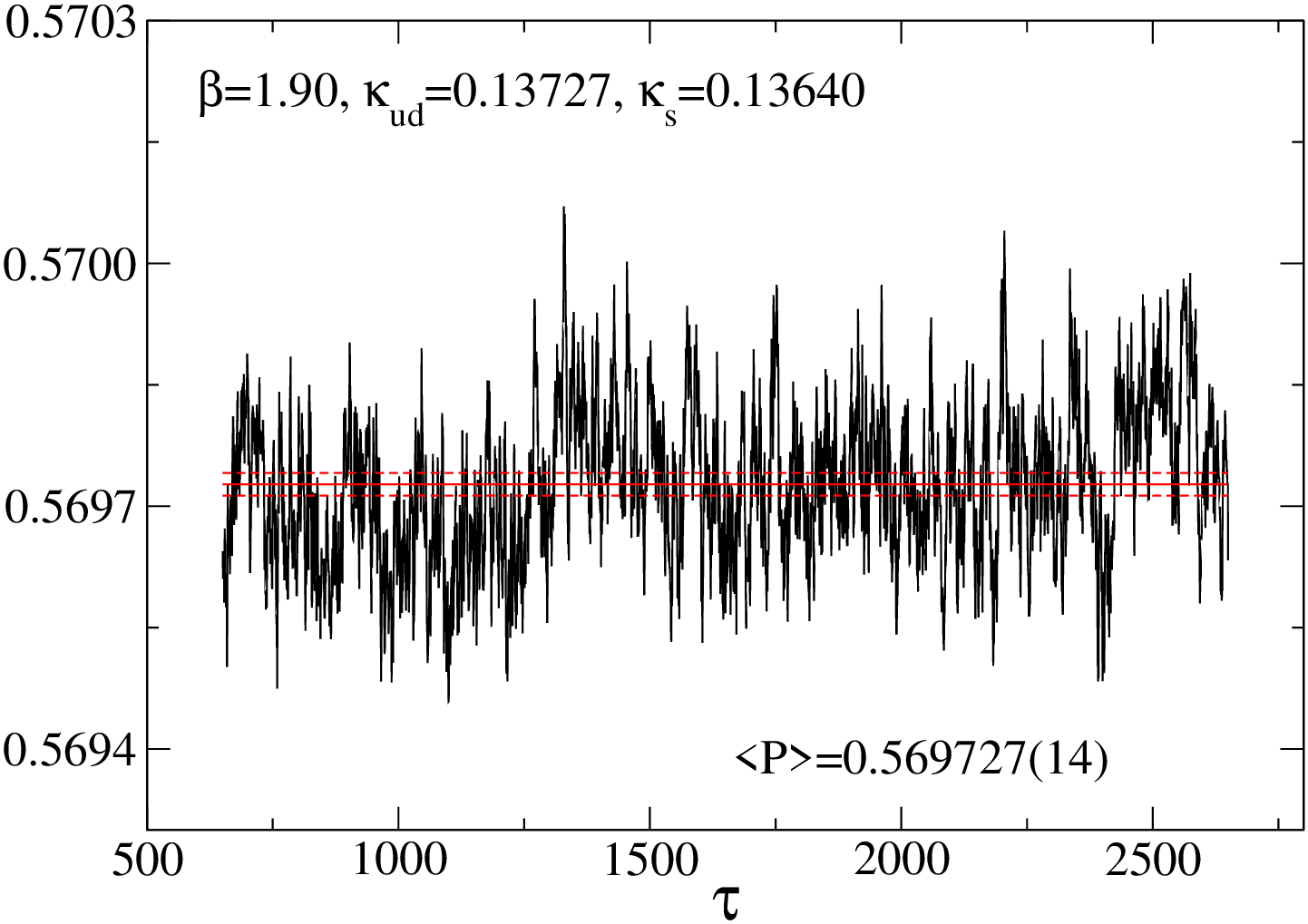}  
\hspace*{2mm}&\hspace*{2mm}
\includegraphics[width=84mm,angle=0]{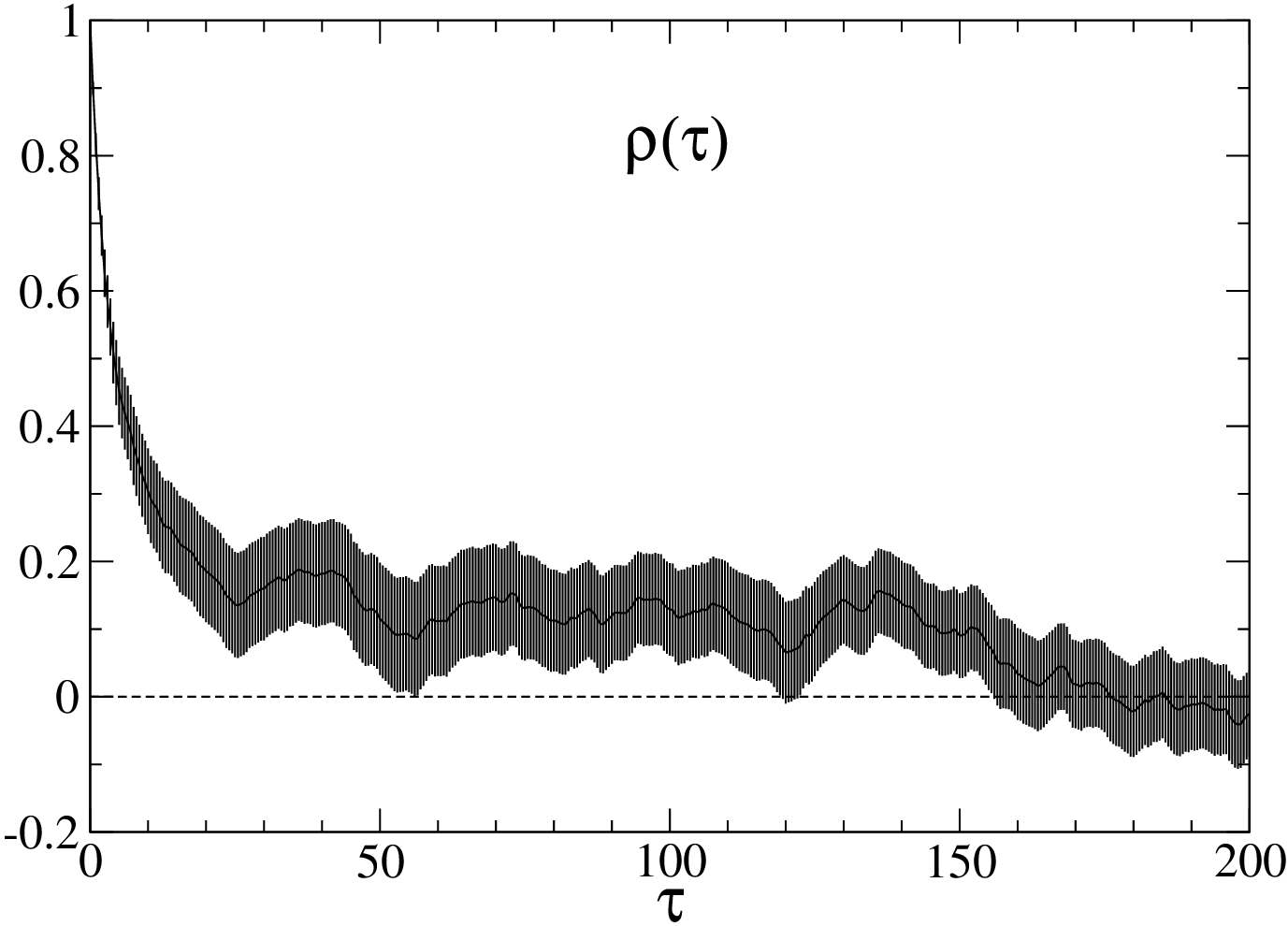}
\end{tabular}
\end{center}
\vspace{-.5cm}
\caption{Plaquette history (left) and normalized autocorrelation
function (right) for $(\kappa_{\rm ud}, \kappa_{\rm s})=(0.13727,
0.13640)$.  Horizontal lines in the left denote the average value of
the plaquette with one standard deviation error band.}
\label{fig:PLQ}
\end{figure*}

The autocorrelation function $\Gamma(\tau)$ of a time series of 
an observable ${\cal O}$ in the course of a numerical simulation is
given by
\ben
\Gamma(\tau)=\la {\cal O}(\tau_0){\cal O}(\tau_0+\tau)\ra-\la{\cal O}(\tau_0)\ra^2
\een
In Fig.~\ref{fig:PLQ} we show the plaquette history and the normalized
autocorrelation function $\rho(\tau)=\Gamma(\tau)/\Gamma(0)$ at
$\kappa_{\rm ud}=0.13727$ as an example. 
The integrated autocorrelation time
is estimated as $\tau_{\rm int}[P]=20.9(10.2)$
following the definition in Ref.~\cite{luscher}
\ben
\tau_{\rm int}(\tau)=\frac{1}{2}
+\sum_{0< \tau \le W} \rho(\tau),
\een
where the summation window $W$ is set to the first time lag $\tau$ such that
$\rho(\tau)$ becomes consistent with zero within the error bar.
In this case we find $W=119.5$.
The choice of $W$ is not critical for estimate of
$\tau_{\rm int}$ in spite of the long tail observed in Fig.~\ref{fig:PLQ}.
Extending the summation window, we find
that $\tau_{\rm int}[P]$ saturates at $\tau_{\rm int}[P]\approx 25$
beyond $W=200$, which is
within the error bar of the original estimate.

Our simulations at $\kappa_{\rm ud}=0.13700$ and $0.13727$ are fast enough 
to be executed by a single long run of 2000 MD units. 
The simulations at $\kappa_{\rm ud}\ge 0.13754$ become increasingly 
CPU time consuming so that we had to execute multiple runs in parallel .   
The data obtained from different runs are combined into 
a single extended series, for which we define the above autocorrelation
function $\Gamma(\tau)$ as if it were a single run.  
The results for $\tau_{\rm int}[P]$ are listed in 
Table~\ref{tab:param}.
Although we hardly observe any systematic quark mass dependence for
the integrated autocorrelation time,
the statistics may not be
sufficiently large to derive a definite conclusion.
  
In the physics analysis we estimate the statistical errors with the
jackknife method in order to take account of the autocorrelation.
For the simulations at $\kappa_{\rm ud}\ge 0.13754$ we apply the
jackknife analysis after combining the different runs into a single
series. The bin size dependence of the statistical error is investigated
for each physical observable. 
For a cross-check
we also carry out the bootstrap error estimation with 1000 
samples.  In all cases we find the two estimates agree for the magnitude of 
errors within 10\%.
We follow the procedure given in Appendix  B of Ref.~\cite{cppacs_nf2}
in estimating the errors for the chiral fit parameters.

\section{Measurements of hadronic observables}
\label{sec:measurement}

\subsection{Hadron masses, quark masses and decay constants}
\label{subsec:hmass}

We measure the meson and baryon correlators at the unitary
points where the valence quark masses are equal to the sea quark masses.
For the meson operators we employ
\ben
M_\Gamma^{fg}(x)={\bar q}_f(x)\Gamma q_g(x),
\een
where $f$ and $g$ denote quark flavors and $\Gamma$ are 16
Dirac matrices $\Gamma={\rm I}$, $\gamma_5$, $\gamma_\mu$, 
$i\gamma_\mu\gamma_5$ and  $i[\gamma_\mu,\gamma_\nu]/2$ 
$(\mu,\nu=1,2,3,4)$.
The octet baryon operators are given by 
\ben
{\cal O}^{fgh}_\alpha(x)=\epsilon^{abc}((q_f^a(x))^T C\gamma_5 q_g^b(x))
q_{h\alpha}^c(x),
\een 
where $a,b,c$ are color indices, $C=\gamma_4\gamma_2$ is the 
charge conjugation matrix and $\alpha=1,2$ labels the $z$-component
of the spin 1/2.
The $\Sigma$- and $\Lambda$-like octet baryons are distinguished by
the flavor structures:
\ben
\Sigma{\rm -like}\;\; &:& \;\; -\frac{{\cal O}^{[fh]g}+{\cal O}^{[gh]f}}{\sqrt{2}},\\
\Lambda{\rm -like}\;\; &:& \;\; \frac{{\cal O}^{[fh]g}-{\cal O}^{[gh]f}-2{\cal O}^{[fg]h}}
{\sqrt{6}},
\een
where $O^{[fg]h}={\cal O}^{fgh}-{\cal O}^{gfh}$.
We define the decuplet baryon operators for the four $z$-components of the
spin 3/2 as
\ben
D^{fgh}_{3/2}(x)&=&\epsilon^{abc}((q_f^a(x))^T C\Gamma_+
q_g^b(x))q_{h1}^c(x),\\
D^{fgh}_{1/2}(x)&=&
\epsilon^{abc}[((q_f^a(x))^T C\Gamma_0q_g^b(x))q_{h1}^c(x)\nn\\
&&-((q_f^a(x))^T C\Gamma_+q_g^b(x))q_{h2}^c(x)]/3,\\
D^{fgh}_{-1/2}(x)&=&
\epsilon^{abc}[((q_f^a(x))^T C\Gamma_0q_g^b(x))q_{h2}^c(x)\nn\\
&&-((q_f^a(x))^T C\Gamma_-q_g^b(x))q_{h1}^c(x)]/3,\\
D^{fgh}_{-3/2}(x)&=&\epsilon^{abc}((q_f^a(x))^T C\Gamma_-
q_g^b(x))q_{h2}^c(x),
\een
where $\Gamma_{\pm}=(\gamma_1\mp\gamma_2)/2$, $\Gamma_0=\gamma_3$ and
the flavor structures should be symmetrized.

We calculate the meson and the baryon correlators
with point and smeared sources and a local sink.
For the smeared source we employ an exponential smearing function 
$\Psi(|{\vec x}|)=A_q\exp(-B_q|{\vec x}|)$ $(q={\rm ud,s})$ with
$\Psi(0)=1$ for the ud and s quark propagators. The
parameters $A_q$ and $B_q$ are adjusted 
from a couple of configurations after the beginning of the production run
such that the pseudoscalar meson effective masses
reach a plateau as soon as possible. 
Their values are given in Table~\ref{tab:param_smear}.
The point and smeared sources allow the hadron propagators 
with nonzero spatial momentum, and we calculate them for 
${\vec p}=(0,0,0),(\pi/16,0,0),(0,\pi/16,0),(0,0,\pi/16)$.

In order to increase the statistics we calculate the hadron correlators
with four source points at $(x_0,y_0,z_0,t_0)$=$(17,17,17,1)$, $(1,1,1,9)$,
$(25,25,25,17)$, and $(9,9,9,25)$ for $\kappa_{\rm ud}\ge 0.13754$.
They are averaged on each configuration before the jackknife analysis. 
This procedure reduces the statistical errors 
by typically $20-40$\% for the vector meson and the baryon masses
and  less than 20\% for the pseudoscalar meson masses 
compared to a single source point.
For further enhancement of the signal we average zero momentum 
hadron propagators over possible spin states on each configuration: 
three polarization states for the vector meson and two (four) 
spin states for the octet (decuplet) baryons.

\begin{figure*}[h]
\vspace{13mm}
\begin{center}
\begin{tabular}{cc}
\includegraphics[width=85mm,angle=0]{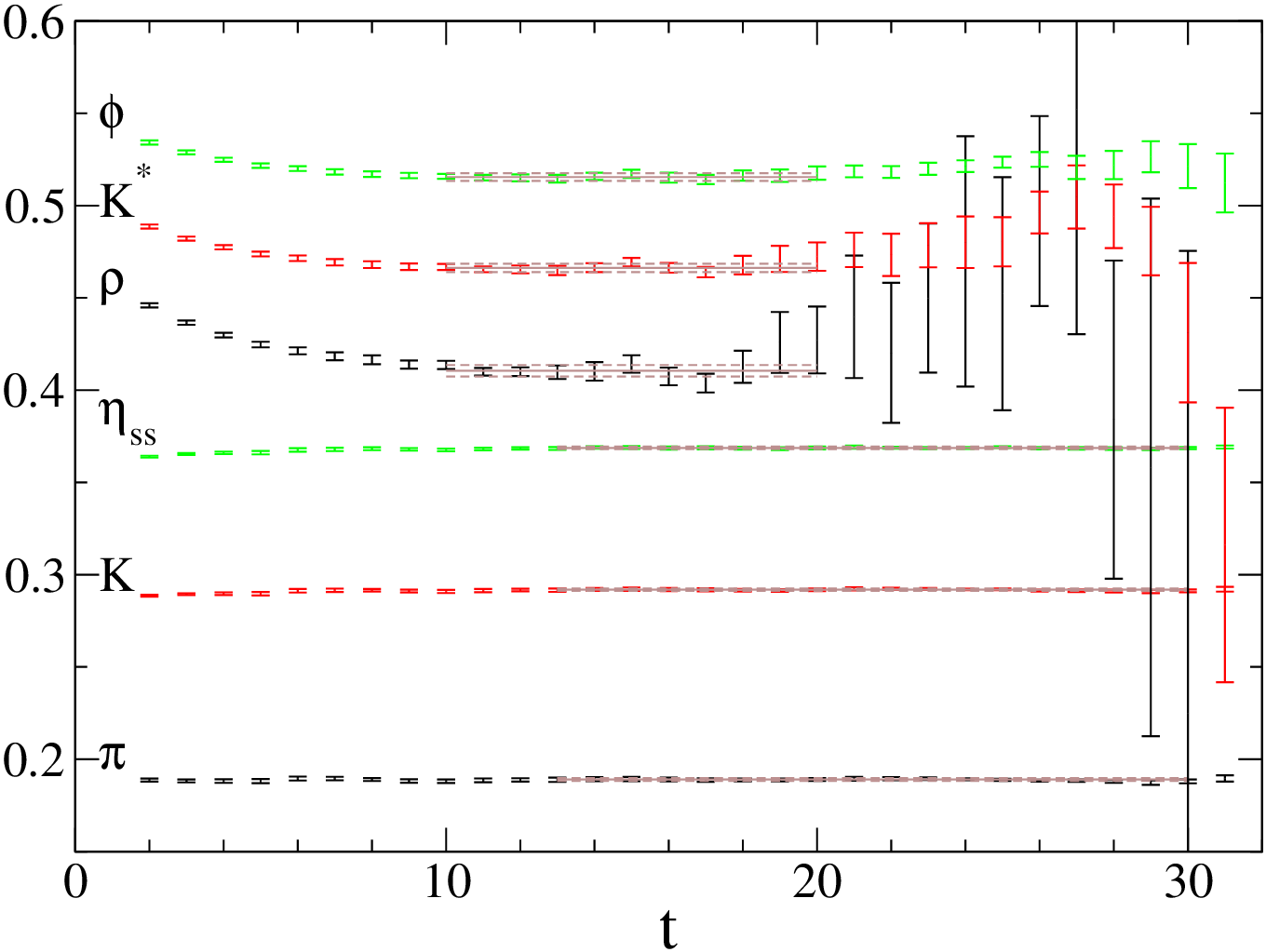} 
\hspace*{2mm}&\hspace*{2mm}
\includegraphics[width=85mm,angle=0]{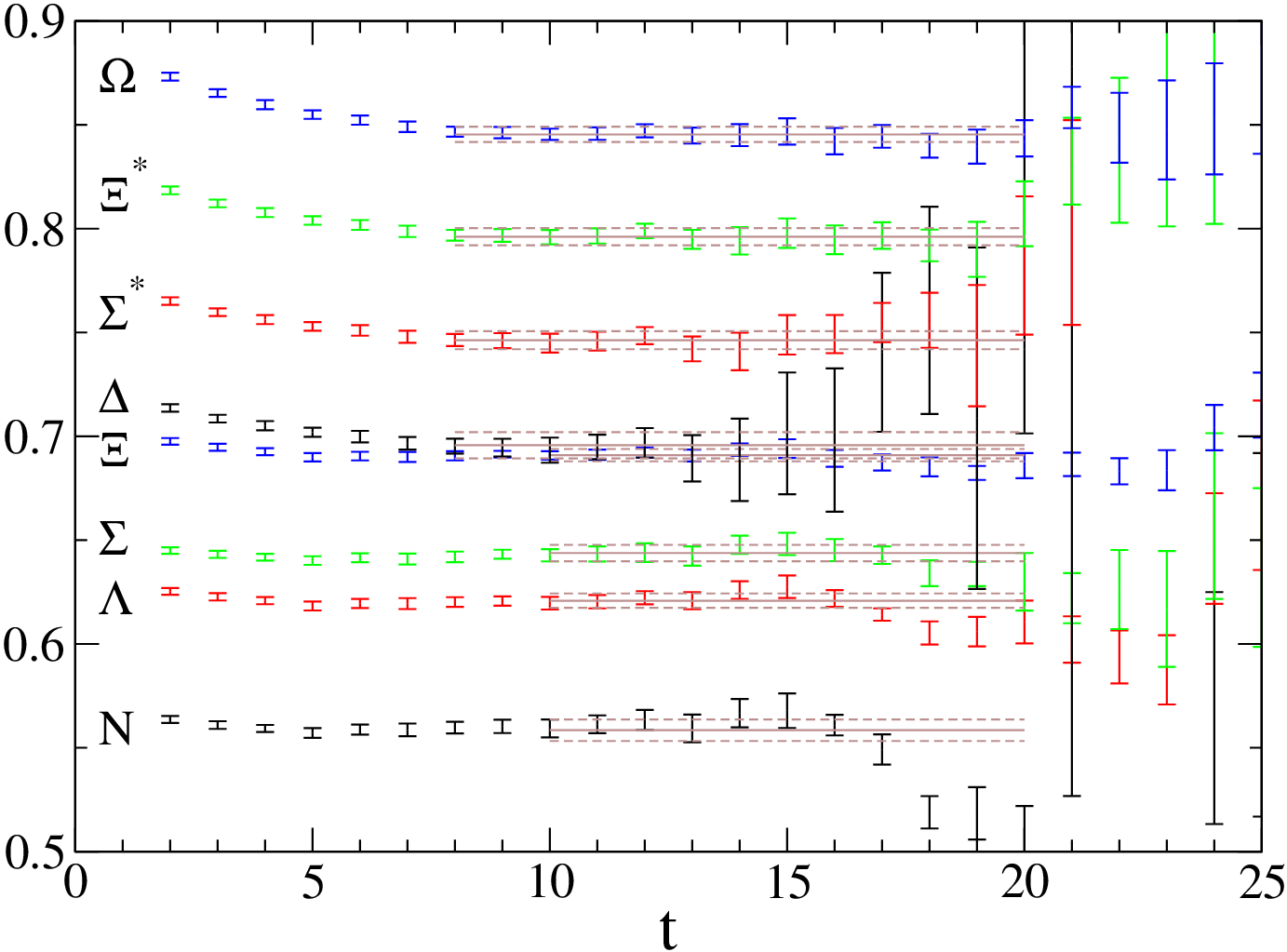}
\end{tabular}
\end{center}
\vspace{-.5cm}
\caption{Effective masses for the mesons (left) and the baryons (right) 
at $(\kappa_{\rm ud},\kappa_{\rm s})=(0.13754,0.13640)$.
Horizontal lines represent the fitting results with one standard 
deviation error band.}
\label{fig:m_eff_kud54}
\end{figure*}

\begin{figure*}[h]
\vspace{13mm}
\begin{center}
\begin{tabular}{cc}
\includegraphics[width=85mm,angle=0]{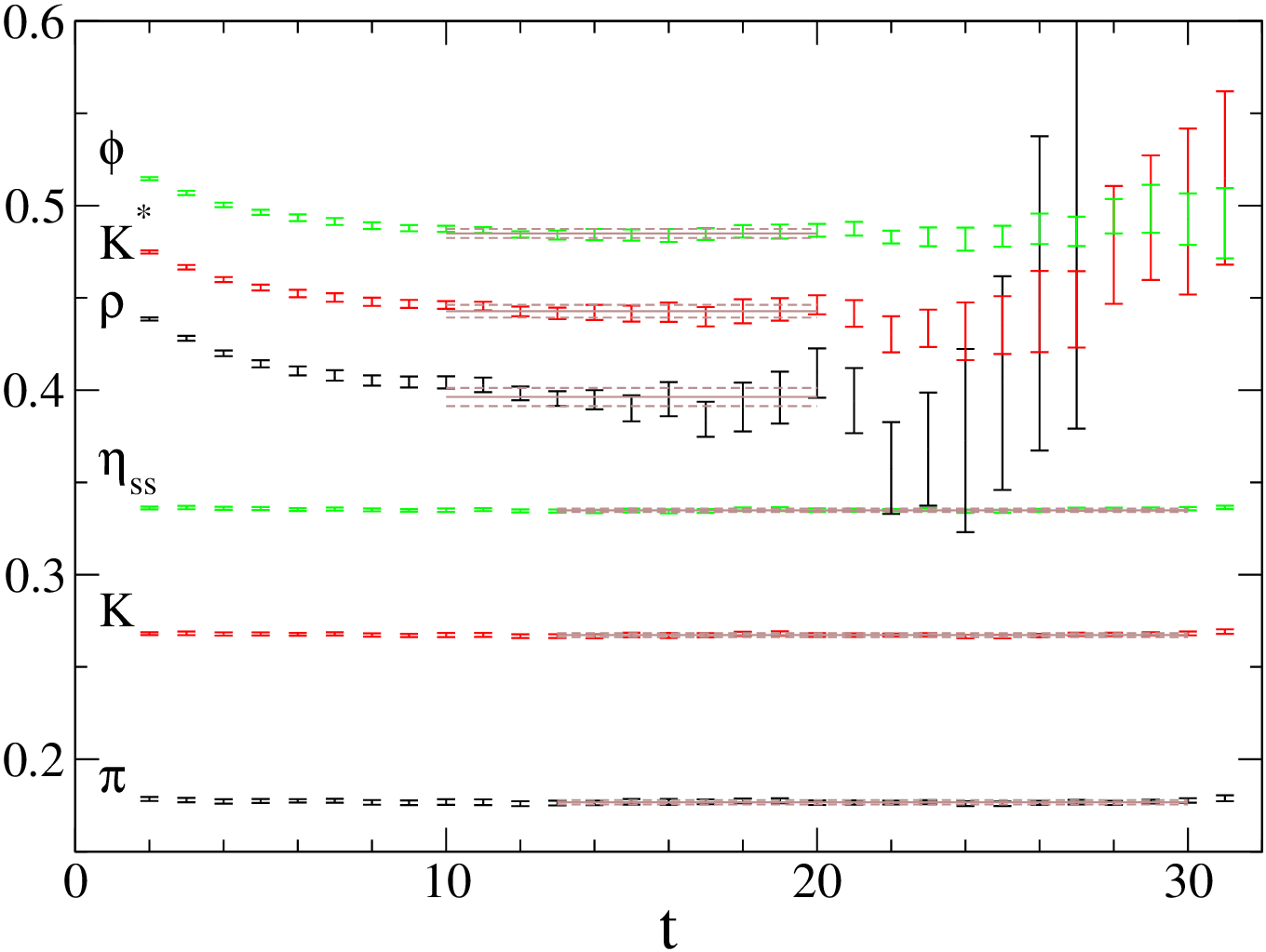} 
\hspace*{2mm}&\hspace*{2mm}
\includegraphics[width=85mm,angle=0]{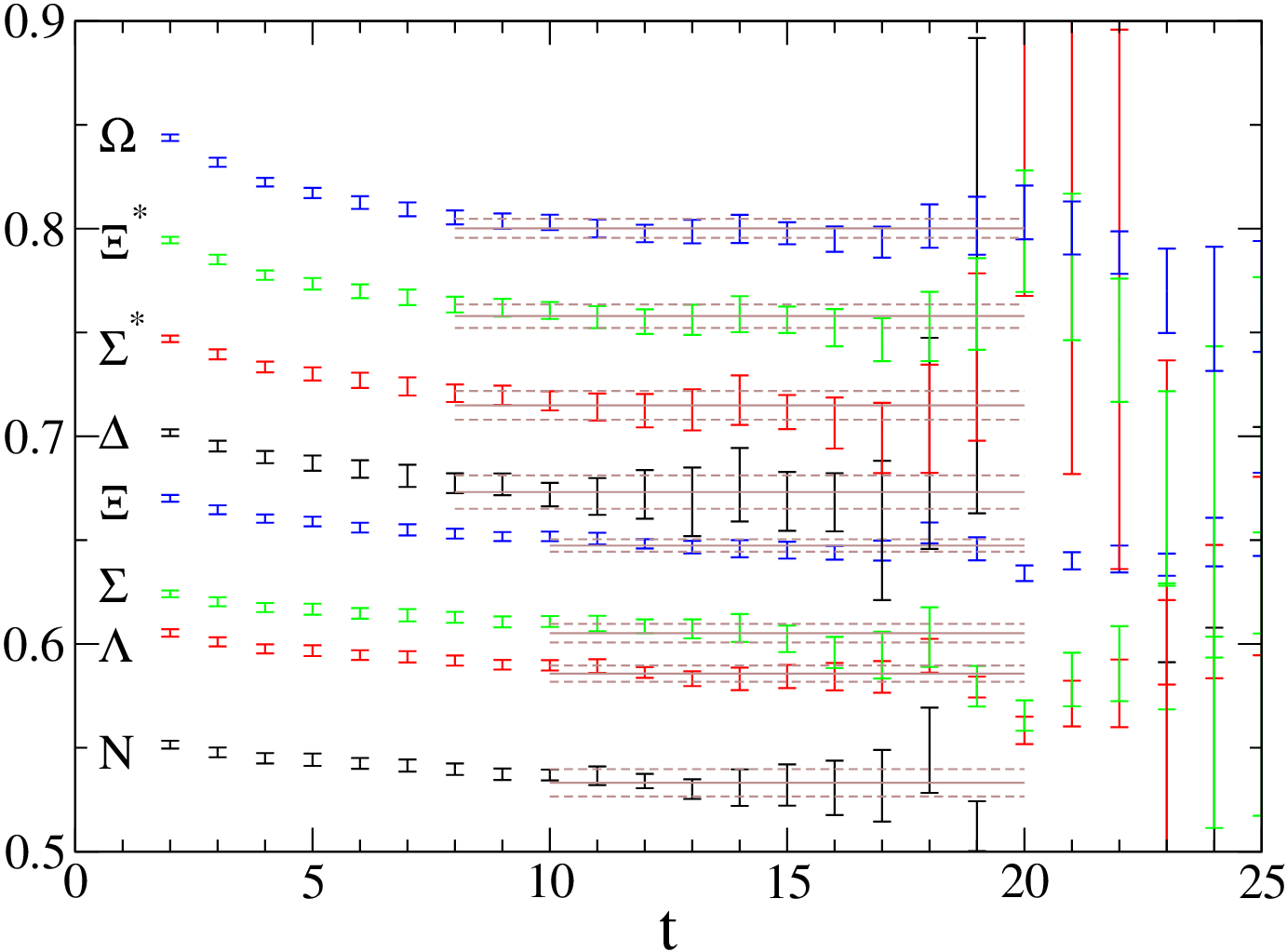}
\end{tabular}
\end{center}
\vspace{-.5cm}
\caption{Same as Fig.~\protect{\ref{fig:m_eff_kud54}}
for $(\kappa_{\rm ud},\kappa_{\rm s})=(0.13754,0.13660)$.}
\label{fig:m_eff_kud54s}
\end{figure*}

\begin{figure*}[h]
\vspace{13mm}
\begin{center}
\begin{tabular}{cc}
\includegraphics[width=85mm,angle=0]{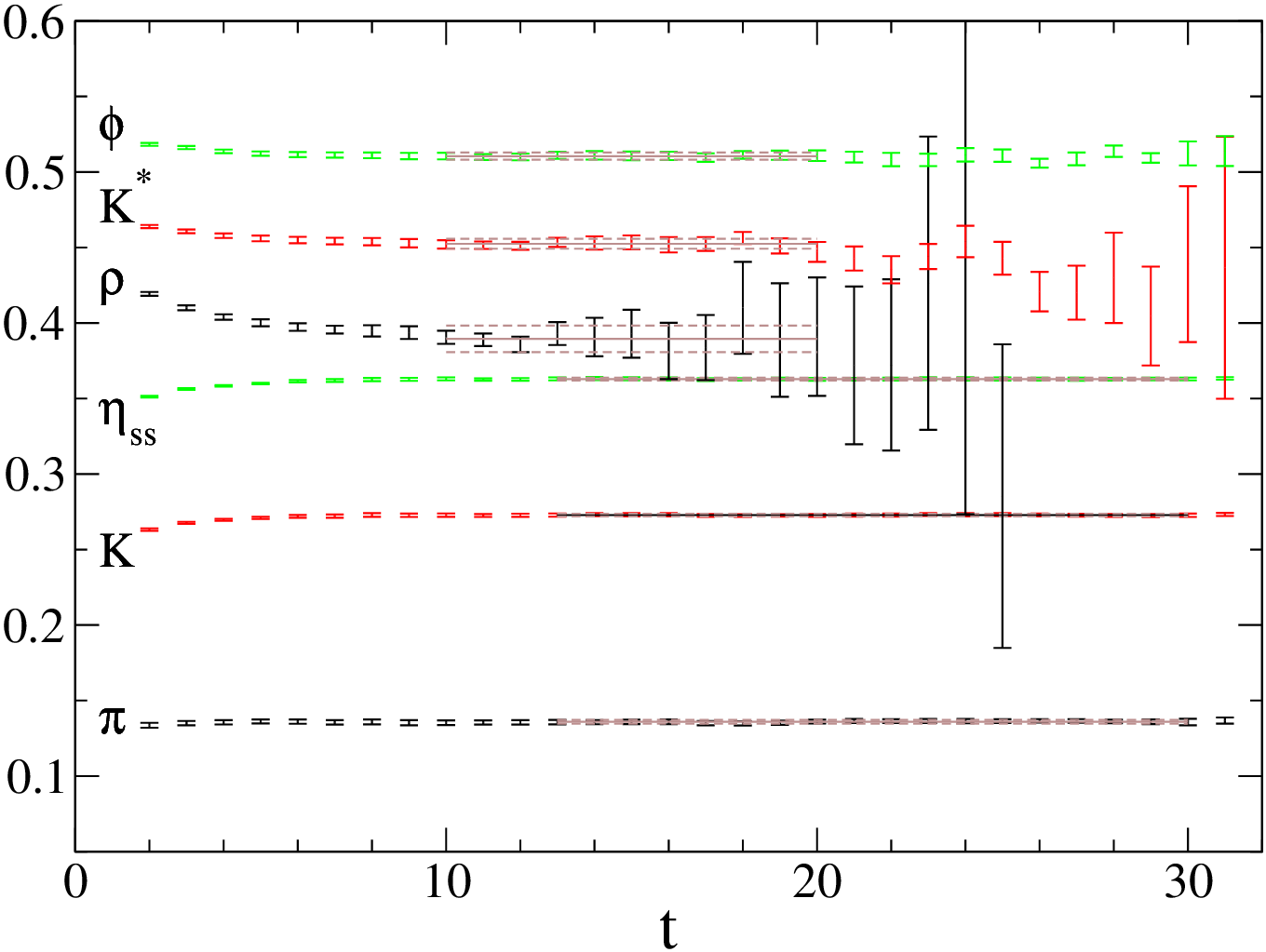} 
\hspace*{2mm}&\hspace*{2mm}
\includegraphics[width=85mm,angle=0]{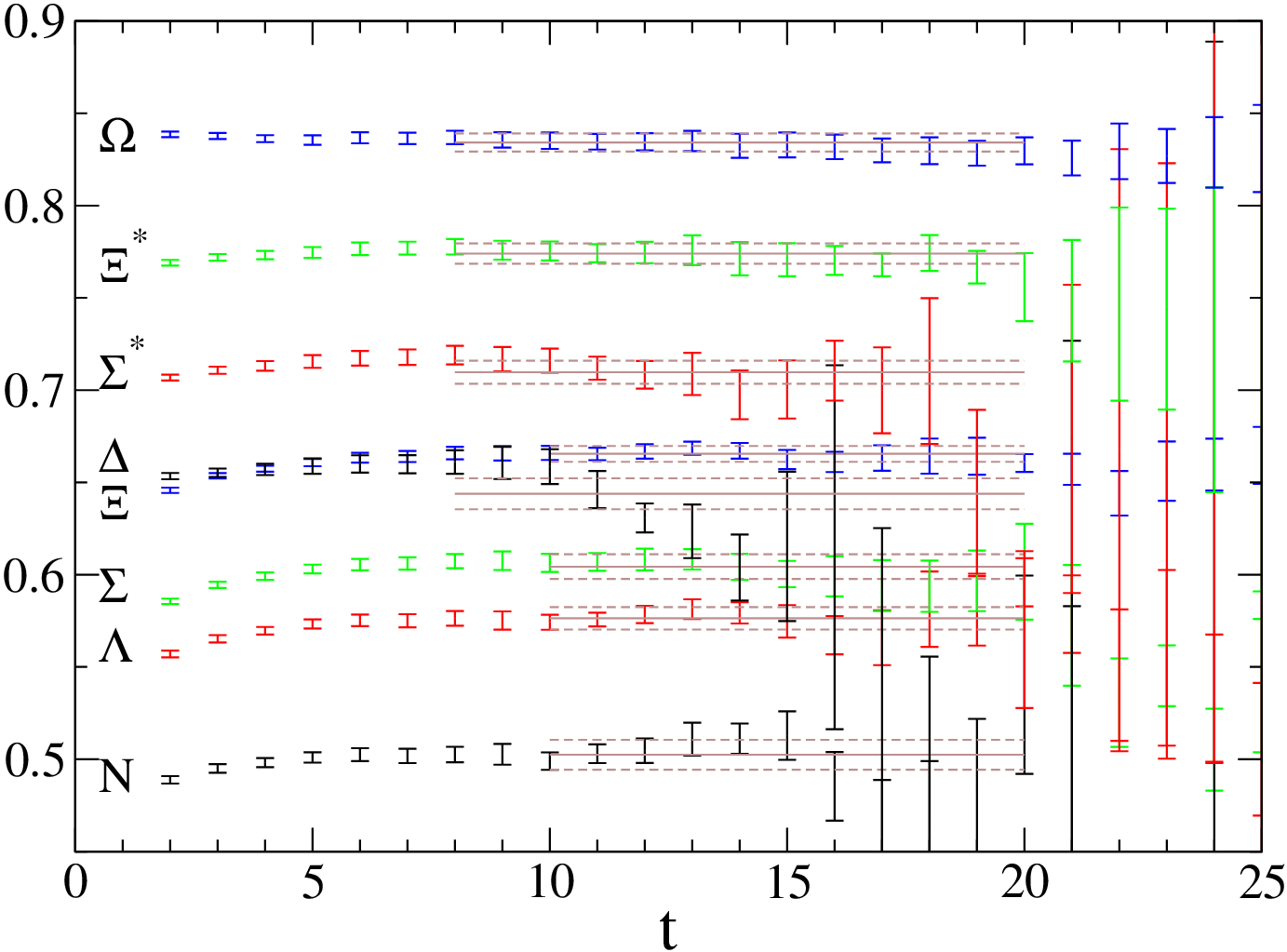}
\end{tabular}
\end{center}
\vspace{-.5cm}
\caption{Same as Fig.~\protect{\ref{fig:m_eff_kud54}}
for $(\kappa_{\rm ud},\kappa_{\rm s})=(0.13770,0.13660)$.}
\label{fig:m_eff_kud70}
\end{figure*}

\begin{figure*}[h]
\vspace{13mm}
\begin{center}
\begin{tabular}{cc}
\includegraphics[width=85mm,angle=0]{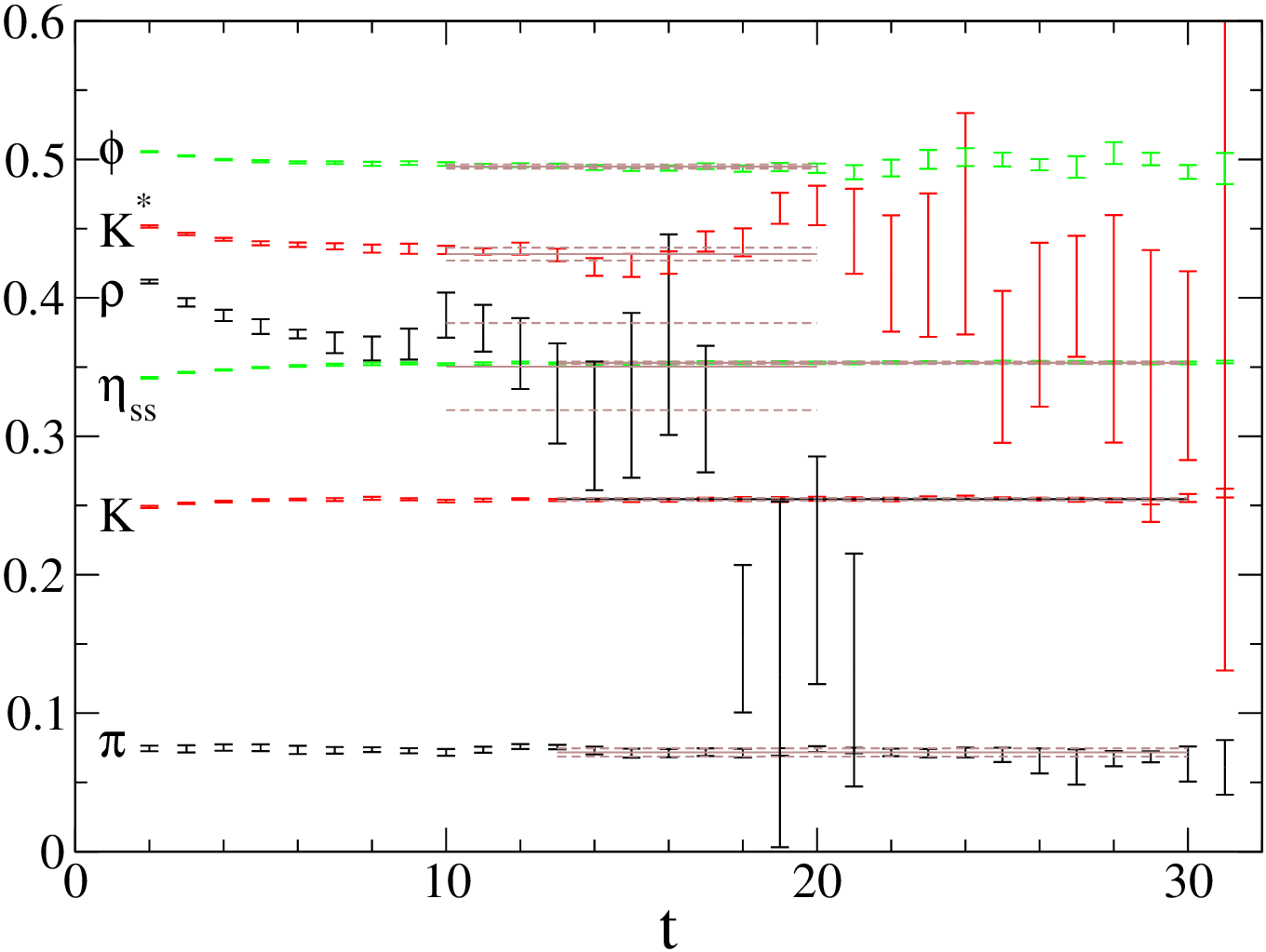} 
\hspace*{2mm}&\hspace*{2mm}
\includegraphics[width=85mm,angle=0]{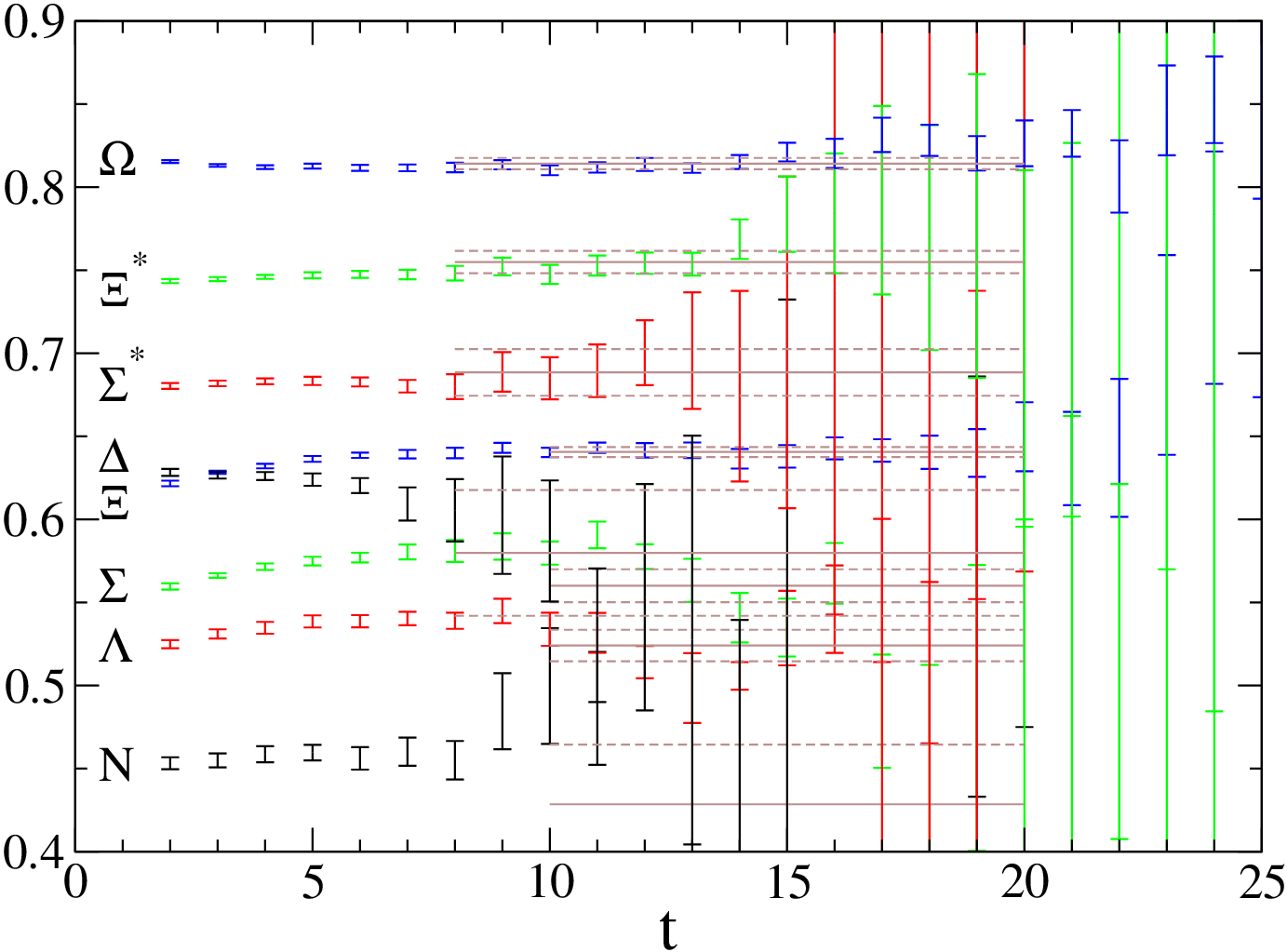}
\end{tabular}
\end{center}
\vspace{-.5cm}
\caption{Same as Fig.~\protect{\ref{fig:m_eff_kud54}}
for $(\kappa_{\rm ud},\kappa_{\rm s})=(0.13781,0.13660)$.}
\label{fig:m_eff_kud81}
\end{figure*}

We extract the meson and the baryon masses from the hadron propagators
with the point sink and the smeared source, where all the valence quark
propagators in the mesons and the baryons have the smeared sources.
Figures~\ref{fig:m_eff_kud54}$-$\ref{fig:m_eff_kud81} show effective
mass plots for the meson and the baryon propagators with the smeared
source for $\kappa_{\rm ud}\ge 0.13754$. We observe that the excited
state contributions are effectively suppressed and good plateaus start  
at small values of $t$. 

The hadron masses are extracted by uncorrelated $\chi^2$ fits 
to the propagators without
taking account of correlations between different time slices,
since we encounter instabilities for correlated fits using covariance matrix.
We assume a single hyperbolic cosine function for the mesons and a single
exponential form for the baryons. 
The lower end of the fit range $t_{\rm min}$ is determined by
investigating stability of the fitted mass. 
On the other hand, the choice of $t_{\rm max}$ 
gives little influence on the fit results as far as
the effective mass exhibits a plateau and the signal is not lost in the noise. 
We employ the same fit range $[t_{\rm min},t_{\rm max}]$ for the same 
particle type: $[13,30]$ for pseudoscalar mesons, $[10,20]$ for vector
mesons, $[10,20]$ for octet baryons and $[8,20]$ for decuplet baryons. 
These fit ranges are independent of the quark masses.
Resulting hadron masses are summarized in Table~\ref{tab:hmass}. 

\begin{figure*}[h]
\vspace{13mm}
\begin{center}
\begin{tabular}{cc}
\includegraphics[width=85mm,angle=0]{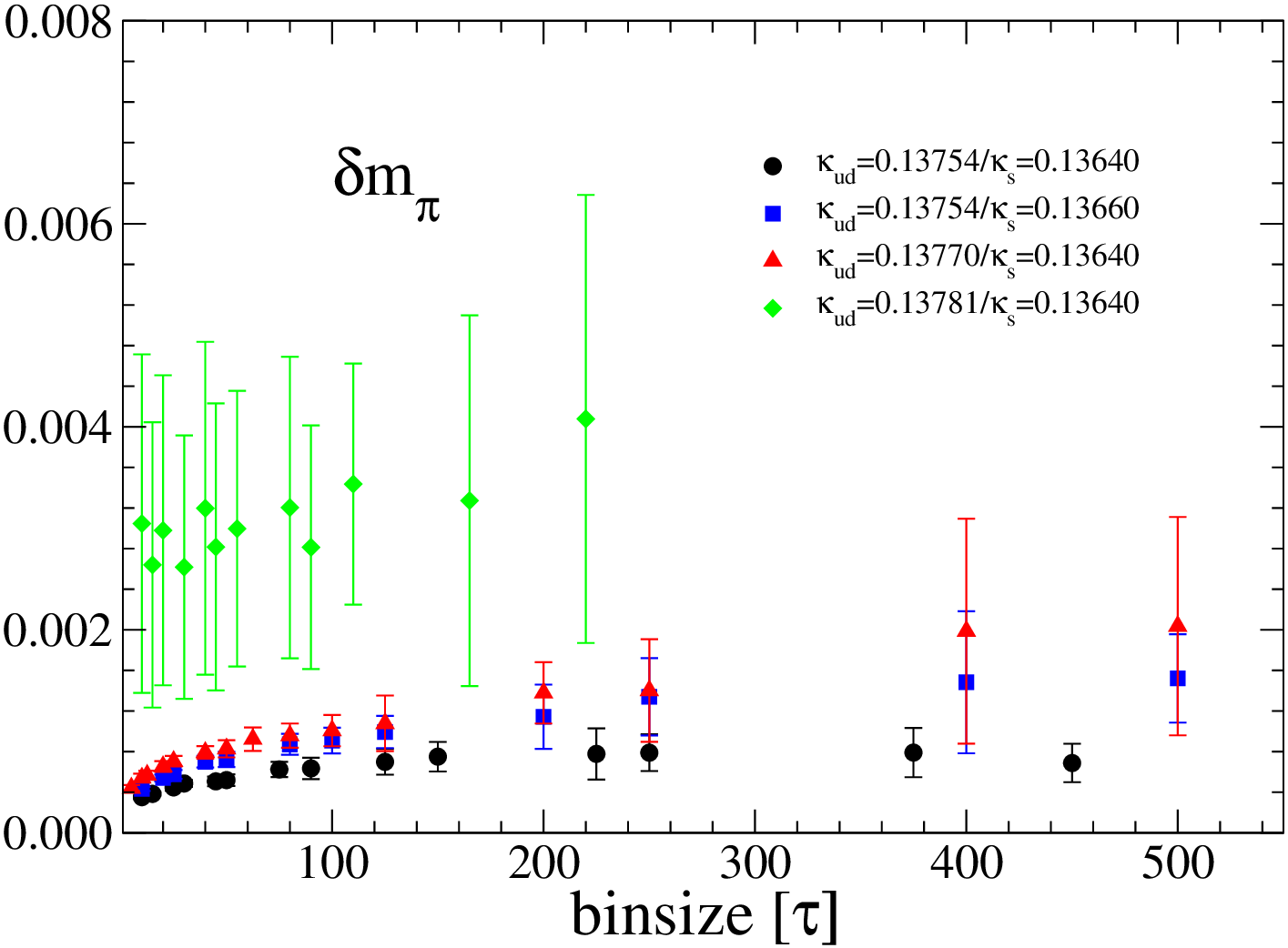} 
\hspace*{2mm}&\hspace*{2mm}
\includegraphics[width=85mm,angle=0]{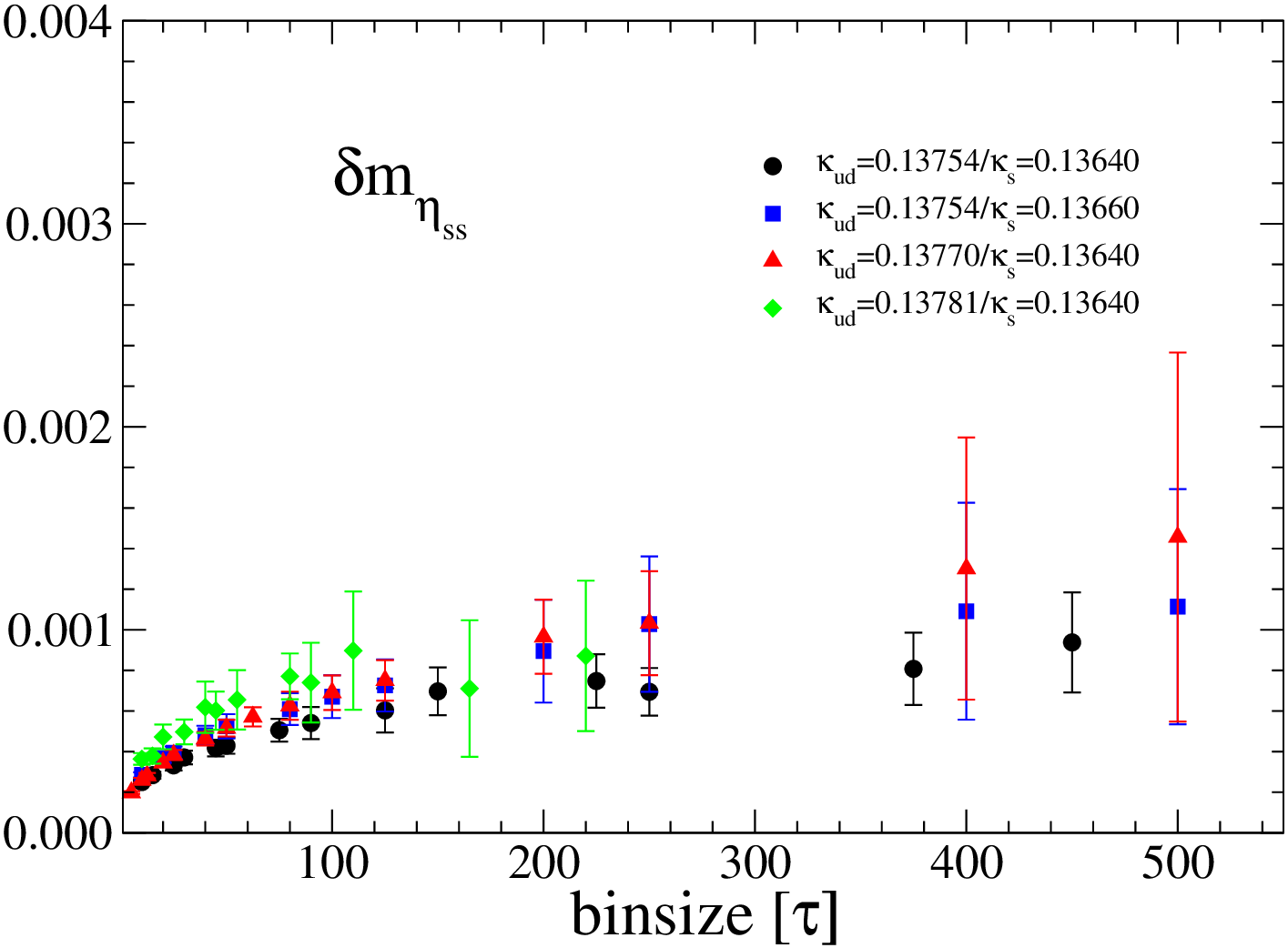} 
\end{tabular}
\end{center}
\vspace{-.5cm}
\caption{Binsize dependence of the magnitude of error for $m_\pi$ (left)
 and $m_{\eta_{\rm ss}}$ (right) at $\kappa_{\rm ud}\ge 0.13754$. 
}
\label{fig:binerr_pi}
\end{figure*}

Statistical errors are estimated with the jackknife
procedure. In Fig.~\ref{fig:binerr_pi} we show the bin size dependence of
the error for $m_\pi$ and $m_{\eta_{\rm ss}}$.  
We observe that the magnitude of error 
reaches a plateau after 100$-$200 MD time depending on the quark
mass. Since similar binsize dependences are found for 
other particle types,
we employ a binsize of 250 MD time for the jackknife analysis 
at $0.13770\ge \kappa_{\rm ud}\ge 0.13754$.  
At our lightest point $\kappa_{\rm ud}=0.13781$ with the statistics of 
990 MD units, we had to reduce the bin size to 110 MD units.

We define the bare quark mass based on the axial vector
Ward-Takahashi identity (AWI) by the ratio of matrix elements
of the pseudoscalar density $P$ and the fourth component of the
axial vector current $A_4$: 
\ben
{\bar m}^{\rm AWI}_f+ {\bar m}^{\rm AWI}_g=\frac{\langle 0 |\nabla_4
  A_4^{\rm imp} |{\rm PS} \rangle}{\langle 0| P | PS\rangle},
\een
where $|{\rm PS}\rangle$ denotes the pseudoscalar meson state
at rest and $f$ and $g$ $(f,g={\rm ud},s)$ label the flavors 
of the valence quarks.
We employ the nonperturbatively $O(a)$-improved 
axial vector current $A_4^{\rm imp}=A_4+c_A{\bar \nabla}_4 P$ 
with ${\bar \nabla}_4$ the symmetric lattice derivative, and 
$c_A=-0.03876106$ as determined in Ref.~\cite{ca}.
In practice the AWI quark mass is determined by 
\ben
{\bar m}^{\rm AWI}_f+{\bar m}^{\rm AWI}_g&=&\frac{m_{\rm PS}}{2}
\left\vert \frac{C_A^s}{C_P^s}\right\vert,
\een
where $m_{\rm PS}$, $C_A^s$ and $C_P^s$ are obtained by
applying a simultaneous $\chi^2$ fit to 
\ben
\langle A_4^{\rm imp}(t) P^s(0)\rangle=2C_A^s
\frac{\sinh(-m_{\rm PS}(t-T/2))}{\exp(m_{\rm PS}T/2)}
\label{eq:a4p_s}
\een
with a smeared source and
\ben
\langle P(t)P^s(0)\rangle=2C_P^s
\frac{\cosh(-m_{\rm PS}(t-T/2))}{\exp(m_{\rm PS}T/2)}
\label{eq:pp_s}
\een
with a smeared source, where $T$ denotes the temporal extent of the lattice.
We employ the fit range of $[t_{\rm min},t_{\rm max}]=[13,25]$ 
for the former and $[13,30]$ for the latter at all the
hopping parameters.
The renormalized quark mass in the continuum ${\overline{\rm MS}}$
scheme is defined as 
\ben
m^{\overline{\rm MS}}_f&=&\frac{Z_A}{Z_P}m^{\rm AWI}_f,
\een
with
\ben
m^{\rm AWI}_f&=&\frac{\left(1+b_A
  \frac{m_f^{\rm VWI}}{u_0}\right)}
{\left(1+b_P \frac{m_f^{\rm VWI}}{u_0} \right)}{\bar m}^{\rm AWI}_f.
\een
The renormalization factors
$Z_{A,P}$ and the improvement coefficients $b_{A,P}$ are 
perturbatively evaluated up to one-loop level\cite{z_pt,z_imp_pt,aokietal}
with the tadpole improvement. The VWI quark  masses in the $ma$
corrections are perturbatively obtained from the AWI quark masses:
\ben
\frac{m_f^{\rm VWI}}{u_0} = \frac{Z_A}{Z_P Z_m}{\bar m}^{\rm AWI}_f.
\label{eq:mq_vwi}
\een
In Table~\ref{tab:qmass} we list
the values of $m^{\overline{\rm MS}}_{\rm ud}$ and 
$m^{\overline{\rm MS}}_{\rm s}$ renormalized at the scale of $1/a$, whose 
statistical errors are provided by the jackknife analysis with 
the bin size chosen as in the hadron mass measurements.

The bare pseudoscalar meson decay constant on the lattice 
is defined by
\ben
\left\vert\langle 0|A_4^{\rm imp}|{\rm PS}\rangle\right\vert 
=f^{\rm bare}_{\rm PS}m_{\rm PS}.
\een
with $|{\rm PS}\rangle$ the pseudoscalar meson state at rest 
consisting of $f$ and $g$ valence quarks.
We evaluate $f^{\rm bare}_{\rm PS}$ from the formula
\ben
f^{\rm bare}_{\rm PS}&=&\left\vert\frac{C_A^s}{C_P^s}\right\vert
\sqrt{\frac{2\left\vert C_P^l\right\vert}{m_{\rm PS}}},
\een
where we extract $m_{\rm PS}$,  $C_A^s$ , $C_{P}^s$ and $C_P^l$
from a simultaneous fit of Eqs.~(\ref{eq:a4p_s}), (\ref{eq:pp_s}) and
\ben
\langle P(t)P^l(0)\rangle=2C_P^l
\frac{\cosh(-m_{\rm PS}(t-T/2))}{\exp(m_{\rm PS}T/2)}
\label{eq:pp_l}
\een
with a local source.
The fit ranges are $[13,25]$, $[13,30]$ and $[15,25]$, respectively,
at all the hopping parameters.
The bare decay constant $f^{\rm bare}_{\rm PS}$ is renormalized 
perturbatively with
\ben
f_{\rm PS}&=&u_0 Z_A \left(1+b_A\frac{m^{\rm VWI}_f+m^{\rm VWI}_g}{2u_0}\right)
f^{\rm bare}_{\rm PS},
\een
where $m^{\rm VWI}_f$ is estimated by Eq.~(\ref{eq:mq_vwi}). 
Table~\ref{tab:qmass} summarizes
the results for $f_{\rm PS}$ with the statistical errors
evaluated by the jackknife analysis with the bin size chosen as in the 
hadron mass measurements.

\subsection{Comparison with the previous CP-PACS/JLQCD results}
\label{subsec:comparison}

\begin{figure*}[h]
\vspace{13mm}
\begin{center}
\begin{tabular}{cc}
\includegraphics[width=85mm,angle=0]{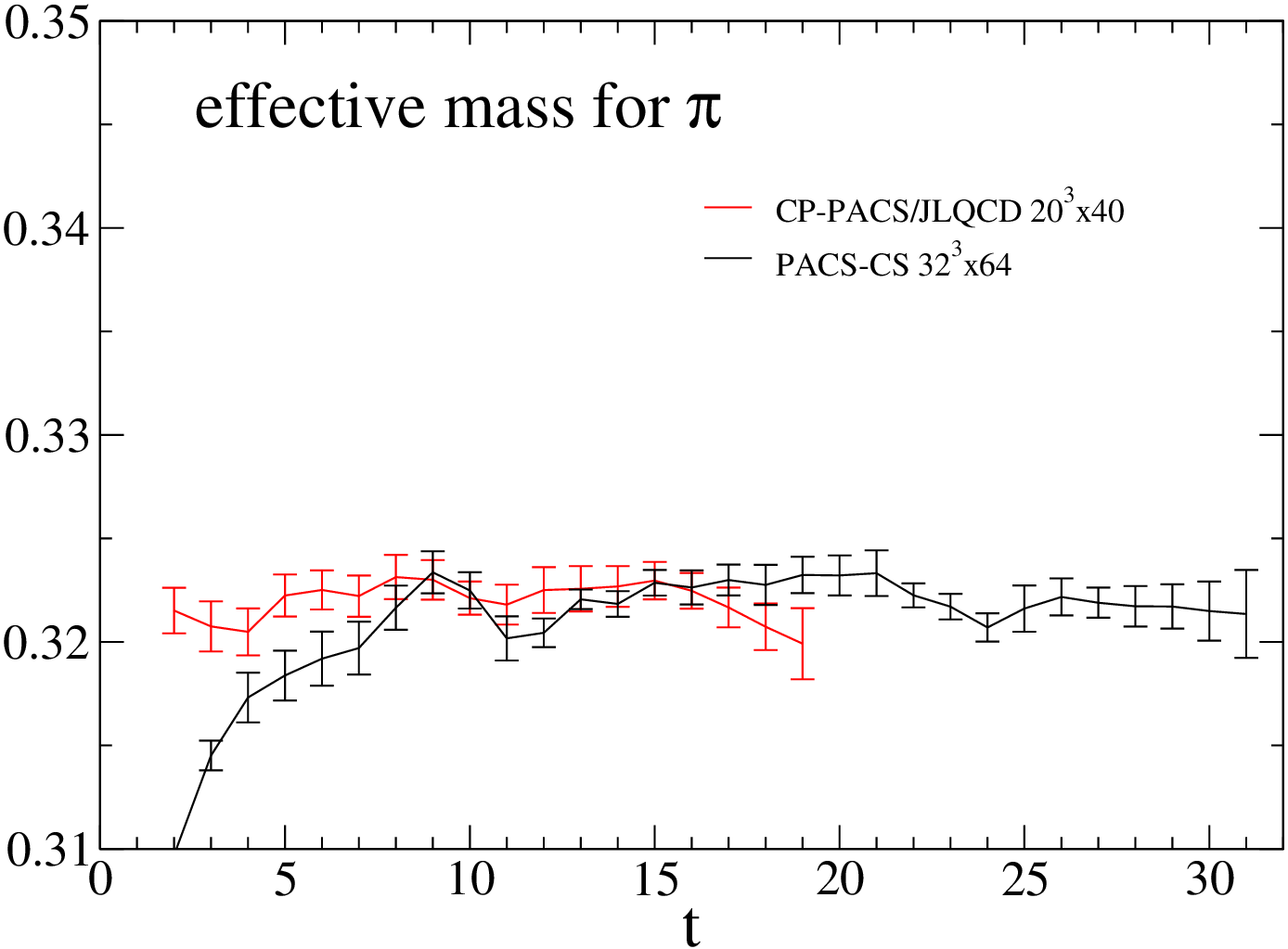} 
\hspace*{2mm}&\hspace*{2mm}
\includegraphics[width=85mm,angle=0]{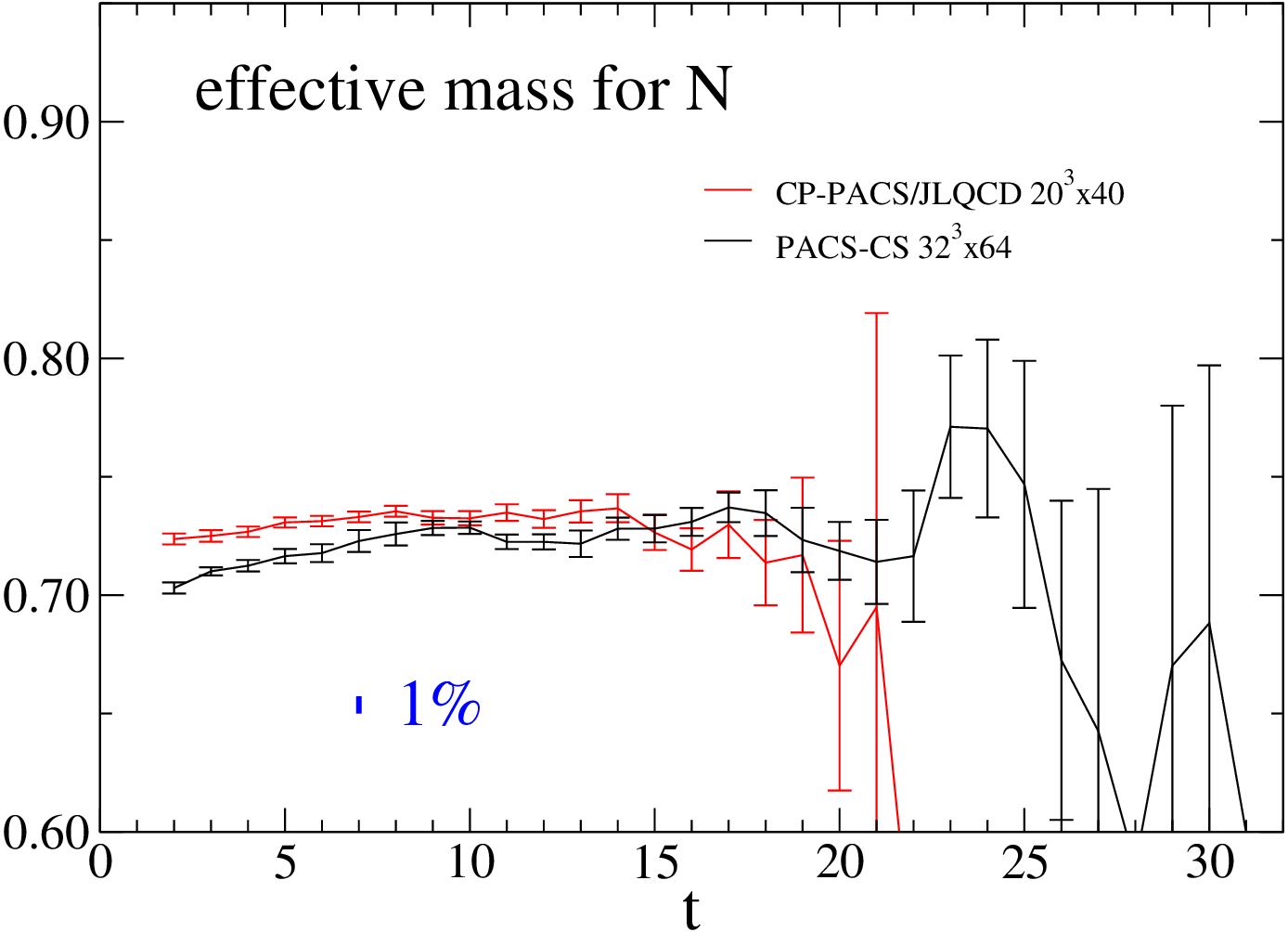}
\end{tabular}
\end{center}
\vspace{-.5cm}
\caption{Effective masses for the $\pi$ (left) and the nucleon (right) 
at $(\kappa_{\rm ud},\kappa_{\rm s})=(0.13700,0.13640)$.
Black and red symbols denote the PACS-CS and the CP-PACS/JLQCD results, 
respectively.}
\label{fig:cmpr_effm}
\end{figure*}

A comparison between the present PACS-CS results 
and those with the previous CP-PACS/JLQCD work\cite{cppacs/jlqcd1, cppacs/jlqcd2}
obtained with the same gauge and quark actions
is possible at $(\kappa_{\rm ud},\kappa_{\rm s})=(0.13700,0.13640)$,
except that the lattice sizes are different: $32^3\times 64$ for the former
and $20^3\times 40$ for the latter.
In Table~\ref{tab:comp} we list the PACS-CS and the CP-PACS/JLQCD
results for the hadron masses at 
$(\kappa_{\rm ud},\kappa_{\rm s})=(0.13700,0.13640)$.
While the results for $m_\pi$ are consistent within the errors,
we find a $1-2$\% deviation for $m_\rho$ and $m_N$. 
This is pictorially confirmed in Fig.~\ref{fig:cmpr_effm}
which shows the effective masses for the $\pi$ meson and the nucleon.
The pion effective masses  are almost degenerate from  $t=8$ to 17, while   
a slight discrepancy is observed for the nucleon results. 
The $\rho$ and  nucleon masses may be suffering  from 
finite size effects.

\section{Chiral analysis on pseudoscalar meson masses and decay constants}
\label{sec:chiral}

The analysis of chiral behavior of pseudoscalar meson masses and decay 
constants occupy an important place in lattice QCD.  
Theoretically the main points to examine are 
the presence of chiral logarithms as predicted 
by ChPT and the convergence of the ChPT series itself. 
The viability of ChPT is 
relevant also for studies of finite-size effects.  The low energy constants 
are important from phenomenological points of view.  And finally, the chiral 
analysis is required to pin down the physical point in the parameter space of the 
simulations. We begin with a discussion of a subtle point in the chiral analysis
when Wilson-clover quark action with implicit chiral symmetry breaking is employed.

\subsection{Chiral perturbation theory for $O(a)$-improved Wilson-type 
quark action }

Our simulations are carried out with a non-perturbatively $O(a)$-improved Wilson 
quark action.  At present we correct $O(ma)$ terms in the AWI quark masses 
and decay constants by one-loop perturbation theory.  These corrections 
are expected to be very small in magnitude, and hence 
leading scaling violations in 
meson masses and decay constants from our simulations can be 
taken as $O(a^2)$.  
In this case, the NLO formula of Wilson chiral perturbation theory 
for the SU(3) flavor case\cite{wchpt},  
which incorporates the leading contributions of the
implicit chiral symmetry breaking effects of the Wilson-type quarks, 
are given by
\begin{widetext}
\ben
\frac{m_{\pi}^2}{2 m_{\rm ud}}&=& B_0 \left\{
1+\mu_\pi-\frac{1}{3}\mu_\eta 
+\frac{2B_0}{f_0^2} \left(
16 m_{\rm ud}  (2L_{8}-L_5) 
+16 (2 m_{\rm ud} +m_{\rm s}) (2L_{6}-L_4)  
\right) -\frac{2H^{\prime\prime}}{f_0^2}\right\}, 
\label{eq:wchpt_mpi}
\\
\frac{m_K^2}{(m_{\rm ud} +m_{\rm s})}&=&B_0 \left\{
1+\frac{2}{3}\mu_\eta
+\frac{2B_0}{f_0^2}\left(
8(m_{\rm ud} + m_{\rm s}) (2L_{8}-L_5)
+16(2 m_{\rm ud}  +m_{\rm s}) (2L_{6}-L_4) 
\right) -\frac{2H^{\prime\prime}}{f_0^2}\right\}, 
\label{eq:wchpt_mk}
\\
f_\pi &=&f_0\left\{
1-2\mu_\pi-\mu_K
+ \frac{2B_0}{f_0^2} \left (
8 m_{\rm ud} L_5+8(2 m_{\rm ud} +m_{\rm s})L_4
\right) -\frac{2H^{\prime}}{f_0^2}\right\}, 
\label{eq:wchpt_fpi}
\\
f_K &=&f_0\left\{
1-\frac{3}{4}\mu_\pi-\frac{3}{2}\mu_K-\frac{3}{4}\mu_\eta
+\frac{2B_0}{f_0^2}\left(4(m_{\rm ud} + m_{\rm s}) L_5+8(2 m_{\rm ud} + m_{\rm s})L_4 
\right) -\frac{2H^{\prime}}{f_0^2}\right\}, 
\label{eq:wchpt_fk}
\een
\end{widetext}
where the quark masses are defined by the axial-vector Ward-Takahashi
identities: $m_{\rm ud}=m_{\rm ud}^{\rm AWI}$ and 
$m_{\rm s}=m_{\rm s}^{\rm AWI}$. 
$L_{4,5,6,8}$ are the low energy constants and 
$\mu_{\rm PS}$ is the chiral logarithm defined by 
\ben
\mu_{\rm PS}=\frac{1}{16\pi^2}\frac{{\tilde m}_{\rm PS}^2}{f_0^2}
\ln\left(\frac{{\tilde m}_{\rm PS}^2}{\mu^2}\right),
\label{eq:chlog}
\een
where 
\ben
{\tilde m}_\pi^2 &=& 2 {m_{\rm ud}}B_0,\\
{\tilde m}_K^2 &=& ({m_{\rm ud}} +m_{\rm s})B_0,\\
{\tilde m}_\eta^2 &=& \frac{2}{3}({m_{\rm ud}} +2 m_{\rm s})B_0
\een
with $\mu$ the renormalization scale.

The two additional parameters $H^{\prime\prime}$ and $H^\prime$ are associated with
the $O(a^2)$ contributions distinguishing the Wilson ChPT 
from that in the continuum.
Since these parameters are independent of the quark
masses, their contributions can be absorbed into $B_0$ and
$f_0$ by the following redefinitions:
\ben
B_0^\prime=B_0\left(1-\frac{2H^{\prime\prime}}{f_0^2}\right),\\
f_0^\prime=f_0\left(1-\frac{2H^{\prime}}{f_0^2}\right),
\een
Indeed re-expansion of the terms in the curly brackets of 
(\ref{eq:wchpt_mpi}) to (\ref{eq:wchpt_fk}) gives rise only to  
terms of form $O(m_q\cdot a^2)$ and $O(m_q\ln m_q\cdot a^2)$, which are NNLO in
the order counting of WChPT analysis and hence can be ignored. 
Thus, up to NLO, WChPT formula are equivalent to the continuum form. 
Note that the expressions in terms of the VWI
quark masses take different forms
and cannot be reduced to those of the continuum ChPT.
Hereafter we concentrate on the continuum ChPT.

\subsection{SU(3) chiral perturbation theory}

The SU(3) ChPT formula in the continuum up to NLO\cite{chpt_nf3} is given by
\begin{widetext}
\ben
\frac{m_{\pi}^2}{2 m_{\rm ud}}&=& B_0 \left\{
1+\mu_\pi-\frac{1}{3}\mu_\eta 
+\frac{2B_0}{f_0^2} \left(
16 m_{\rm ud}  (2L_{8}-L_5) 
+16 (2 m_{\rm ud} +m_{\rm s}) (2L_{6}-L_4)  
\right) \right\}, 
\label{eq:chpt_mpi}
\\
\frac{m_K^2}{(m_{\rm ud} +m_{\rm s})}&=&B_0 \left\{
1+\frac{2}{3}\mu_\eta
+\frac{2B_0}{f_0^2}\left(
8(m_{\rm ud}+m_{\rm s}) (2L_{8}-L_5)
+16(2 m_{\rm ud}  +m_{\rm s}) (2L_{6}-L_4) 
 \right)\right\}, 
\label{eq:chpt_mk}
\\
f_\pi &=&f_0\left\{
1-2\mu_\pi-\mu_K
+ \frac{2B_0}{f_0^2} \left (
8 m_{\rm ud} L_5+8(2 m_{\rm ud} +m_{\rm s})L_4
\right)\right\}, 
\label{eq:chpt_fpi}
\\
f_K &=&f_0\left\{
1-\frac{3}{4}\mu_\pi-\frac{3}{2}\mu_K-\frac{3}{4}\mu_\eta
+\frac{2B_0}{f_0^2}\left(4(m_{\rm ud}+m_{\rm s}) L_5+8(2 m_{\rm ud}+ m_{\rm s})L_4 
\right)\right\}, 
\label{eq:chpt_fk}
\een
\end{widetext}
There are six unknown low energy constants $B_0,f_0,L_{4,5,6,8}$ 
in the expressions above.
$L_{4,5,6,8}$ are scale-dependent so as to 
cancel that of the chiral logarithm given by (\ref{eq:chlog}). 
We can determine these parameters                                  
by applying a simultaneous fit to $m_\pi^2/(2 m_{\rm ud})$, 
$m_K^2/(m_{\rm ud} +m_{\rm s})$, $f_\pi$ and $f_K$.

\begin{figure*}[h]
\vspace{13mm}
\begin{center}
\begin{tabular}{cc}
\includegraphics[width=85mm,angle=0]{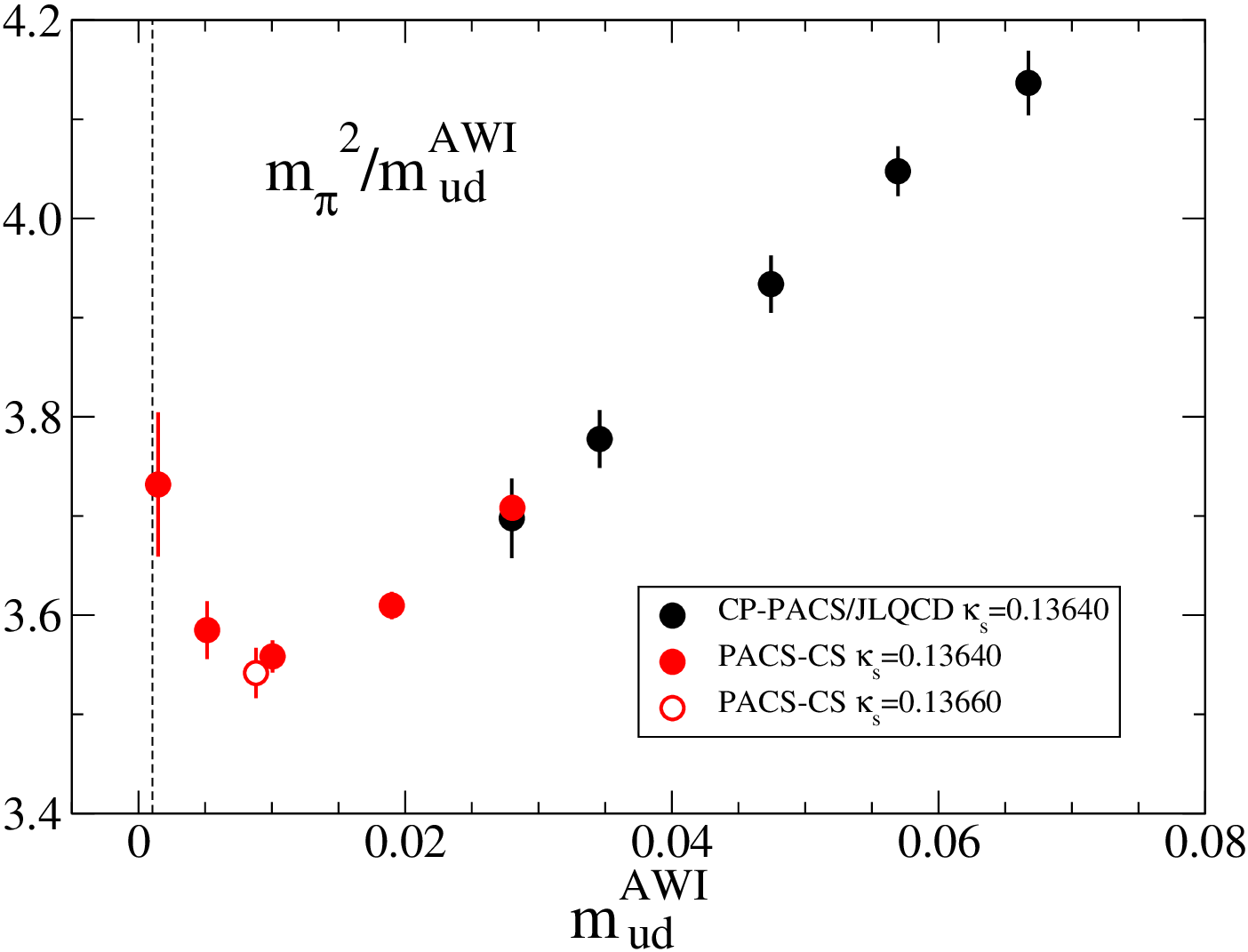}
\hspace*{2mm}&\hspace*{2mm}
\includegraphics[width=85mm,angle=0]{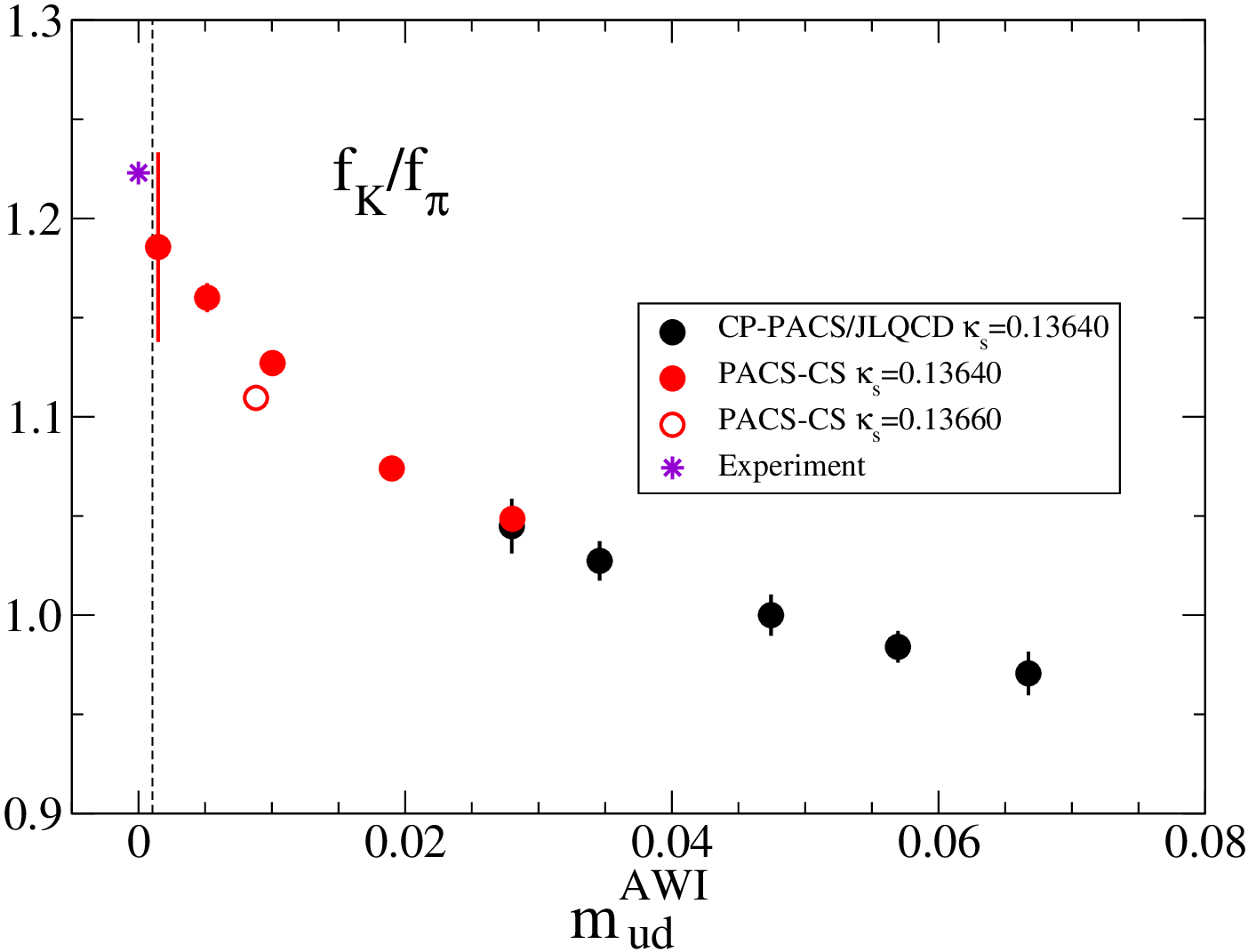}
\end{tabular}
\end{center}
\vspace{-.5cm}
\caption{Comparison of the PACS-CS (red) and the CP-PACS/JLQCD (black)
 results for $m_\pi^2/m_{\rm ud}^{\rm AWI}$ (left) and $f_K/f_\pi$
 (right) as a function of $m_{\rm ud}^{\rm AWI}$.  
Vertical line denotes the physical point and 
star symbol represents the experimental value.}
\label{fig:cmp_chlog}
\end{figure*}

In order to provide an overview of our data we plot in Fig.~\ref{fig:cmp_chlog}
a comparison of the PACS-CS (red symbols) and the CP-PACS/JLQCD 
results (black symbols) for $m_\pi^2/m_{\rm ud}^{\rm AWI}$ and $f_K/f_\pi$ 
as a function of $m_{\rm ud}^{\rm AWI}$.  
The two data sets show a smooth connection at
$\kappa_{\rm ud}=0.13700$ ($m_{\rm ud}^{\rm AWI}=0.028$). 
More important is the fact that an almost linear quark mass dependence
of the CP-PACS/JLQCD results in heavier quark mass region
changes into a convex behavior,
both for $m_\pi^2/m_{\rm ud}^{\rm AWI}$ and $f_K/f_\pi$, 
as $m_{\rm ud}^{\rm AWI}$ is diminished in the PACS-CS results. 
This is a characteristic feature expected from 
the ChPT prediction in the small quark mass region
due to the chiral logarithm.
This curvature drives up the ratio $f_K/f_\pi$ toward the 
experimental value as the physical point is approached.

Having confirmed signals for the presence of the chiral logarithm, 
we apply the SU(3) ChPT formulae (\ref{eq:chpt_mpi})$-$(\ref{eq:chpt_fk})
to our results.  We choose the four simulation points at 
$\kappa_{\rm ud}\ge 0.13754$. In Fig.~\ref{fig:cmp_chlog} these four 
points lie to the left and around the turning point of the curvature.  
They also correspond to the region where the $\rho$ meson 
mass satisfies the condition $m_\rho > 2m_\pi$, and hence lie to the left 
of the threshold singularity in the complex energy plane 
for the $\rho$ meson.
The heaviest pion mass at $(\kappa_{\rm ud},\kappa_{\rm s})=(0.13754,0.13640)$
is about 410~MeV with the use of $a^{-1}=2.176(31)$ GeV determined below.
The measured bare AWI quark masses, but corrected for the $O(ma)$ corrections
at one-loop perturbation theory,  
are used for ${m_{\rm ud}}$ and $m_{\rm s}$ 
in Eqs.~(\ref{eq:chpt_mpi})$-$(\ref{eq:chpt_fk}).

We present the fit results for the low energy constants 
in Table~\ref{tab:fit_su3chpt}.  
The results are quoted both without (w/o FSE) 
and with (w/ FSE) finite-size corrections in the ChPT formulae 
(see Sec~\ref{subsec:fse}).   We also list 
the phenomenological estimates with the experimental
inputs~\cite{colangelo05,amoros01}, and the results obtained 
by recent 2+1 flavor lattice QCD calculations~\cite{rbcukqcd08,milc07}.  
The renormalization scale is set to be 770~MeV. The MILC results
for the low energy constants quoted at the scale of $m_\eta$ 
are converted according to Ref.~\cite{chpt_nf3}
\begin{widetext}
\ben
L_4(\mu)&=&L_4(m_\eta)-\frac{1}{256\pi^2}
\ln\left(\frac{\mu^2}{m_\eta^2}\right),\\
L_5(\mu)&=&L_5(m_\eta)-\frac{3}{256\pi^2}
\ln\left(\frac{\mu^2}{m_\eta^2}\right),\\
(2L_6-L_4)(\mu)&=&(2L_6-L_4)(m_\eta)-\left(\frac{2}{9}\right)\frac{1}{256\pi^2}
\ln\left(\frac{\mu^2}{m_\eta^2}\right),\\
(2L_8-L_5)(\mu)&=&(2L_8-L_5)(m_\eta)+\left(\frac{4}{3}\right)\frac{1}{256\pi^2}
\ln\left(\frac{\mu^2}{m_\eta^2}\right)
\een   
\end{widetext}
with $\mu$ the renormalization scale.
For $L_4$ and $L_5$ governing the behavior of $f_\pi$ and $f_K$, 
we find that all the results are compatible. 
On the other hand, some discrepancies are observed for the results 
of $2L_6-L_4$ and $2L_8-L_5$ contained in the ChPT formulae 
for $m_\pi^2$ and $m_K^2$.

For later convenience we convert the SU(3) low energy constants
$B_0,f_0,L_{4,5,6,8}$ to the SU(2) low energy constants $B,f,
l_{3,4}$ defined by
\ben
\frac{m_\pi^2}{2{m_{\rm ud}}}&=&B\left\{1+\mu_\pi(B_0\rightarrow B,
f_0\rightarrow f)+4\frac{{\bar m}_\pi^2}{f^2}l_3\right\},\nn\\
\label{eq:su2chpt_mpi}\\
f_\pi&=&f\left\{1- 2\mu_\pi(B_0\rightarrow B, f_0\rightarrow f)
+2\frac{{\bar m}_\pi^2}{f^2}l_4\right\}\nn\\
\label{eq:su2chpt_fpi}
\een
with ${\bar m}_\pi^2=m_{\rm ud}B$.
The NLO relations are given by\cite{chpt_nf3}
\ben
B&=&B_0\left(1-\frac{1}{3}{\bar \mu}_\eta+
\frac{32 {\bar m}^2_{K}}{f_0^2}(2L_6-L_4) \right),\\
f&=&f_0\left(1-{\bar \mu}_K+
\frac{16 {\bar m}^2_{K}}{f_0^2}L_4 \right),\\
l_3&=&-8L_4-4L_5+16L_6+8L_8-\frac{1}{18}{\bar \nu}_\eta,\\
l_4&=&8L_4+4L_5-\frac{1}{2}{\bar \nu}_K,
\een
where ${\bar \mu}_{K,\eta}$ and ${\bar \nu}_{K,\eta}$ are defined by
\ben
{\bar \mu}_{K,\eta}=\frac{{\bar m}^2_{K,\eta}}{16\pi^2f_0^2}
\ln\left(\frac{{\bar m}^2_{K,\eta}}{\mu^2}\right),\\
{\bar \nu}_{K,\eta}=\frac{1}{32\pi^2}
\left(\ln\left(\frac{{\bar m}^2_{K,\eta}}{\mu^2}\right)+1\right)
\een
with 
\ben
{\bar m}^2_{K}&=&m_{\rm s}B_0,\\
{\bar m}^2_{\eta}&=&\frac{4}{3}m_{\rm s}B_0,
\een
and ${\bar l}_i$ ($i=3,4$) are defined at the renormalization scale
$\mu=m_\pi=139.6$ MeV\cite{chpt_nf2}:
\ben
l_i&=&\frac{\gamma_i}{32\pi^2}\left({\bar l}_i+\ln\frac{m^2_\pi}{\mu^2}\right)
\een
with
\ben
\gamma_3&=&-\frac{1}{2},\\
\gamma_4&=&2.
\een
In Table~\ref{tab:lec_su2_conv} we summarize the results for 
the SU(2) low energy constants obtained by the conversion from the
SU(3) low energy constants.
The vacuum condensations are defined by
\ben
\langle {\bar u}u\rangle_0 &\equiv & \langle {\bar
u}u\rangle\vert_{m_{\rm ud}=m_{\rm s}=0}=-\frac{1}{2}f_0^2 B_0,\\ 
\langle {\bar u}u\rangle &\equiv & \langle {\bar
u}u\rangle\vert_{m_{\rm ud}=0,m_{\rm s}=m_{\rm s}^{\rm physical}}
=-\frac{1}{2}f^2B.
\een
These quantities are perturbatively renormalized
at the scale of 2 GeV.

\begin{figure*}[h]
\vspace{13mm}
\begin{center}
\begin{tabular}{cc}
\includegraphics[width=80mm,angle=0]{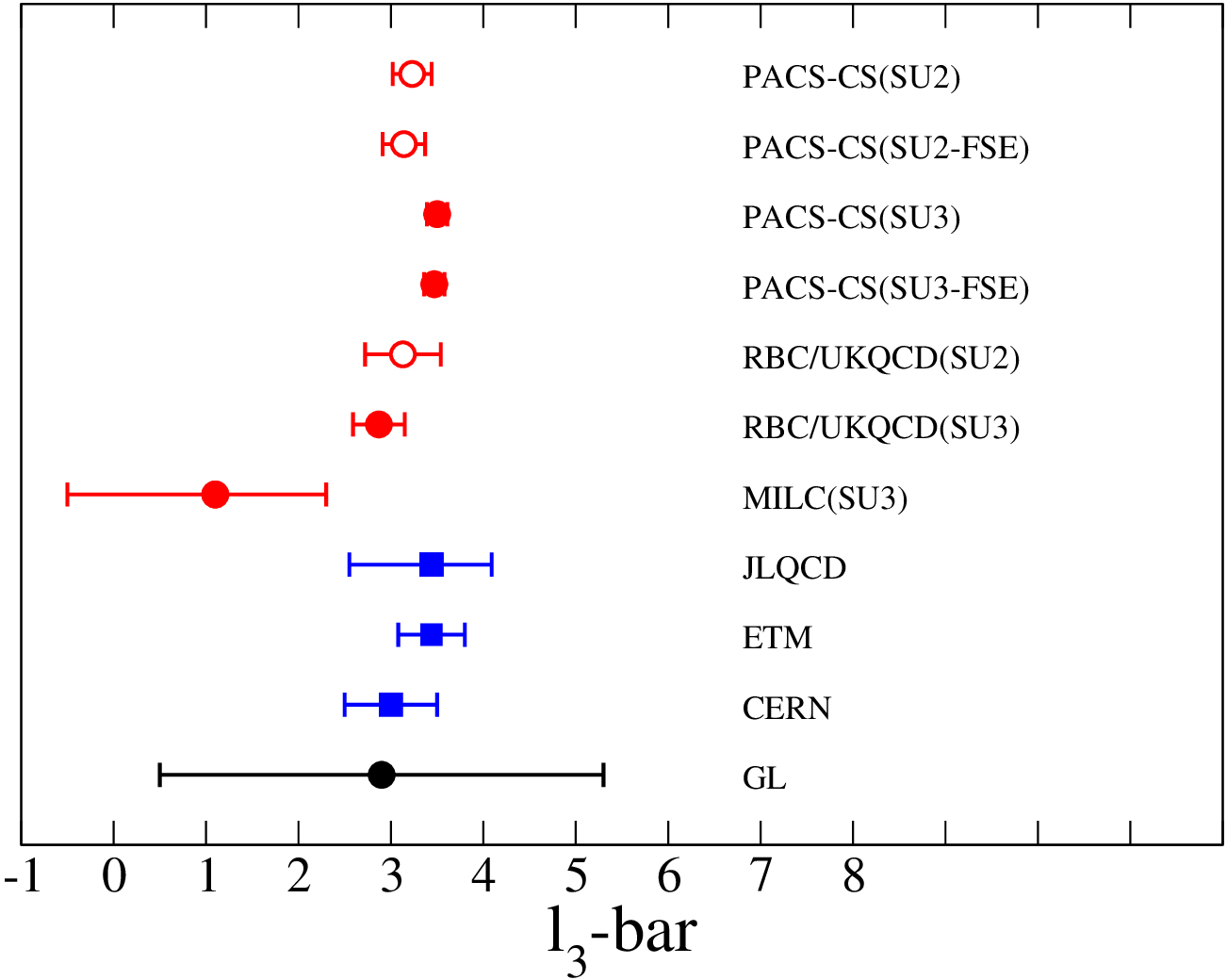}
\hspace*{5mm}&\hspace*{5mm}
\includegraphics[width=80mm,angle=0]{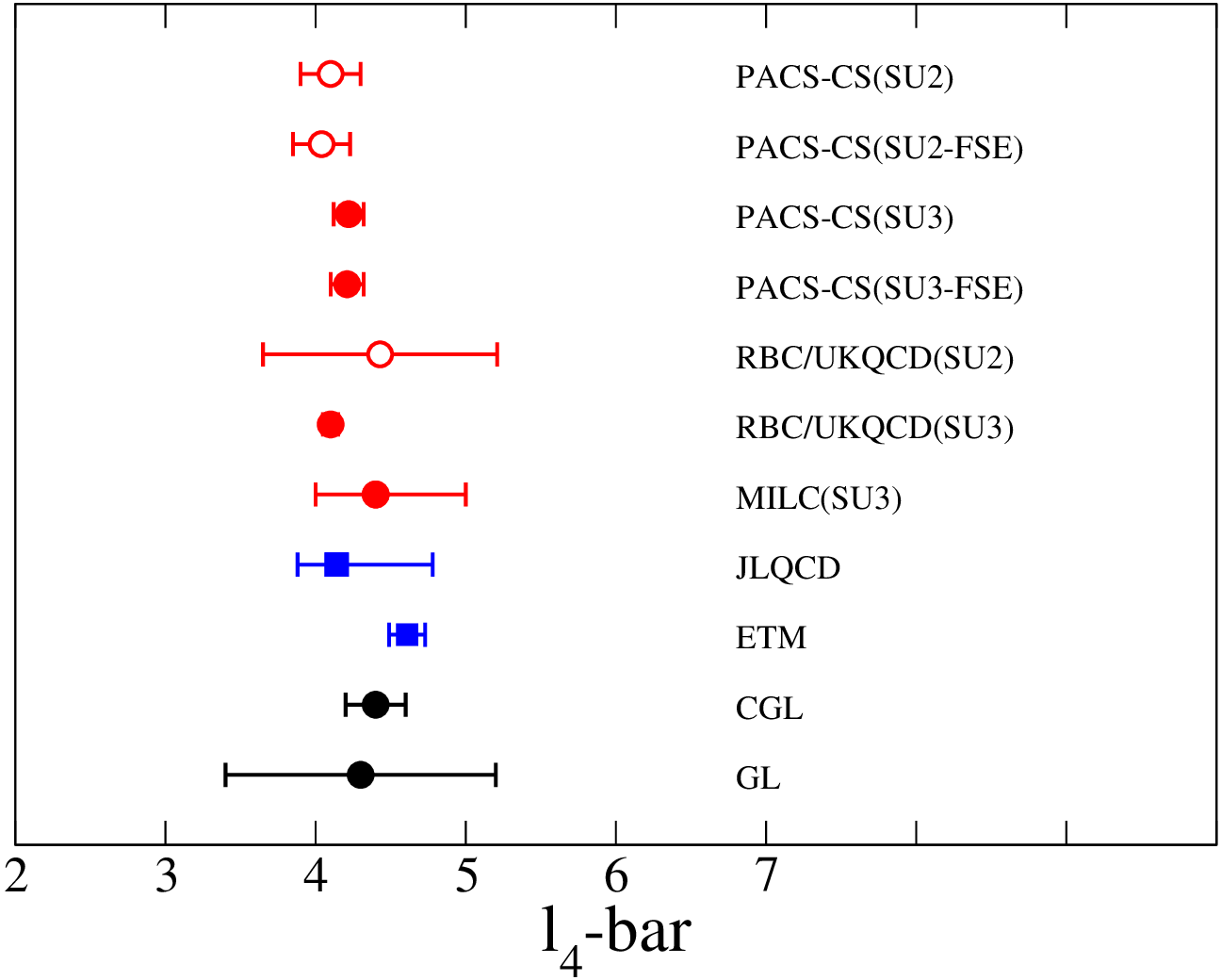}
\end{tabular}
\end{center}
\vspace{-.5cm}
\caption{Comparison of the results for ${\bar l}_3$ and ${\bar l}_4$. 
Black symbols denote the phenomenological estimates. Blue ones are
for 2 flavor lattice results. Red closed (open) symbols represent
the results for the SU(3) (SU(2)) ChPT analyses 
in the 2+1 flavor dynamical simulations.  
See text and Table~\protect{\ref{tab:l_34}} for details.}
\label{fig:l_34}
\end{figure*}

In Fig.~\ref{fig:l_34} we compare our results for 
${\bar l}_{3,4}$ with 
those obtained by other groups 
whose numerical values are listed in Table~\ref{tab:l_34}. 
Black symbols denote the
phenomenological estimates, blue symbols represent
the results obtained by the SU(2) ChPT fit on 
2 flavor dynamical configurations and red closed (open) symbols
are for those obtained by the SU(3) (SU(2)) ChPT fit
on 2+1 flavor dynamical configurations.
For ${\bar l}_3$ all the results 
reside between 3.0 and 3.5, except for the MILC result which is 
sizably smaller and marginally consistent with others within a large error. 
On the other hand, we find a good consistency
among the results for ${\bar l}_4$.

\begin{figure*}[h]
\vspace{13mm}
\begin{center}
\begin{tabular}{cc}
\includegraphics[width=87mm,angle=0]{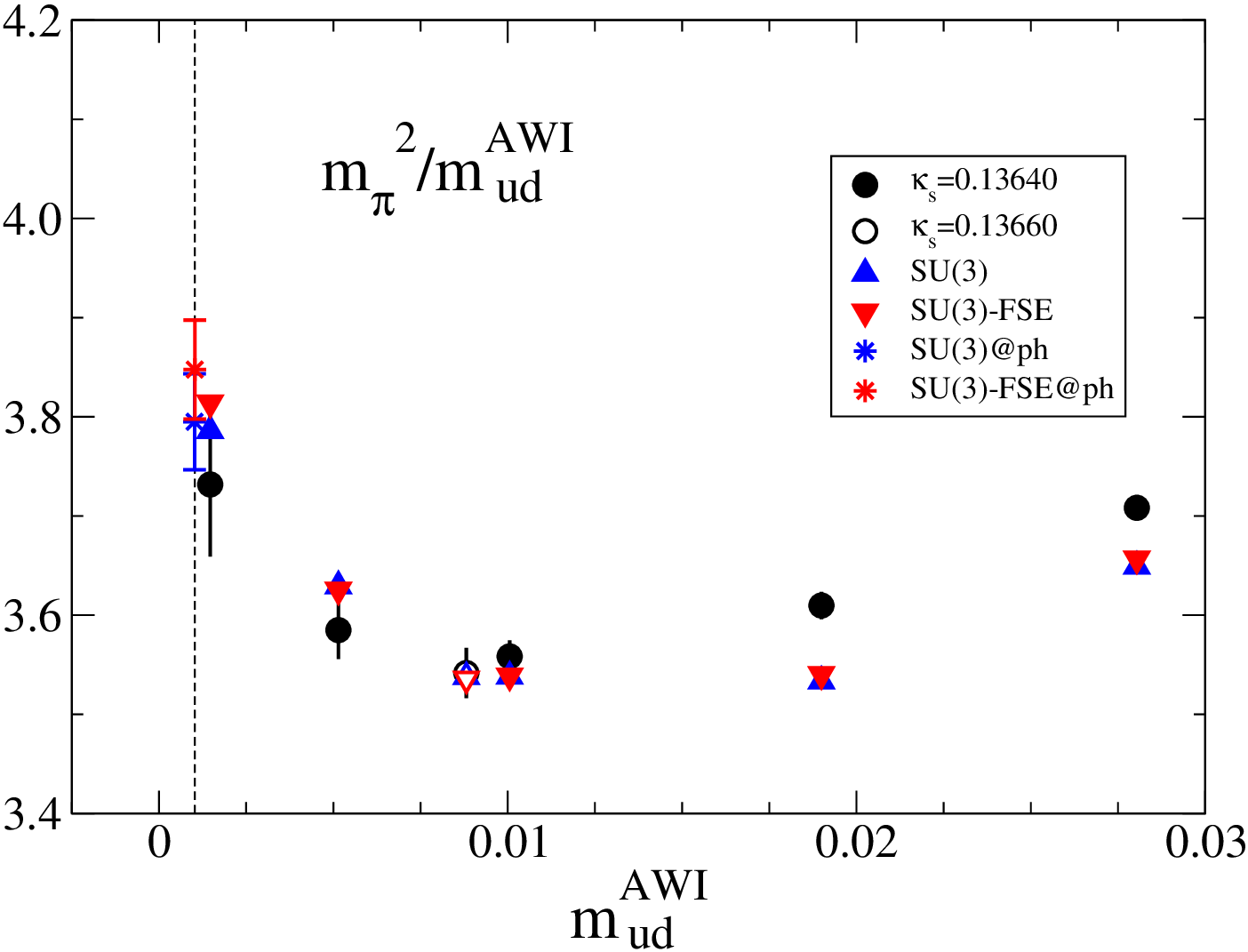}
\hspace*{2mm}&\hspace*{2mm}
\includegraphics[width=85mm,angle=0]{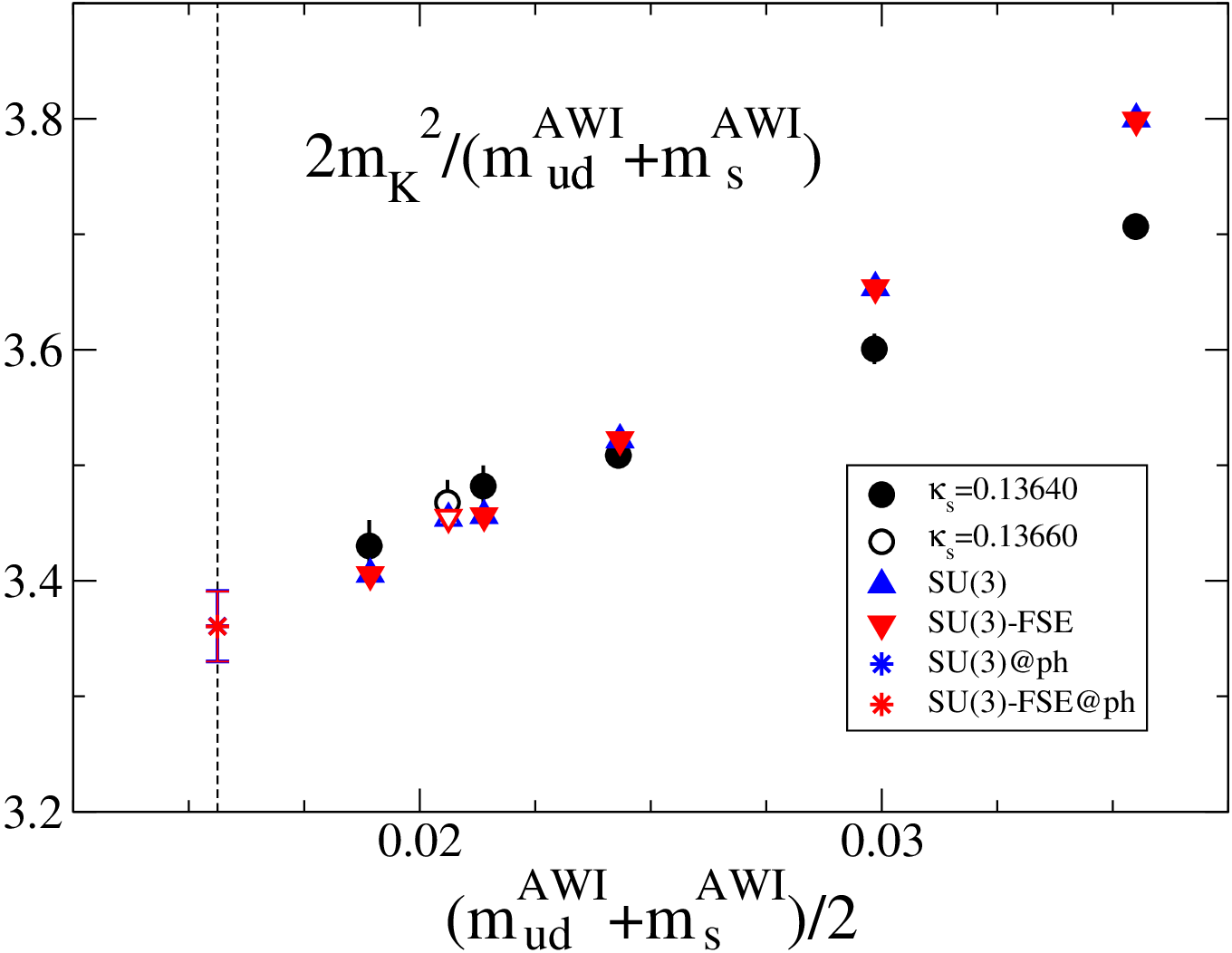}
\end{tabular}
\end{center}
\vspace{-.5cm}
\caption{SU(3) ChPT fit for $m_\pi^2/m_{\rm ud}^{\rm AWI}$ (left)
and $2 m_K^2/(m_{\rm ud}^{\rm AWI}+m_{\rm s}^{\rm AWI})$  (right).}
\label{fig:su3fit_mps}
\end{figure*}

\begin{figure*}[h]
\vspace{13mm}
\begin{center}
\begin{tabular}{cc}
\includegraphics[width=85mm,angle=0]{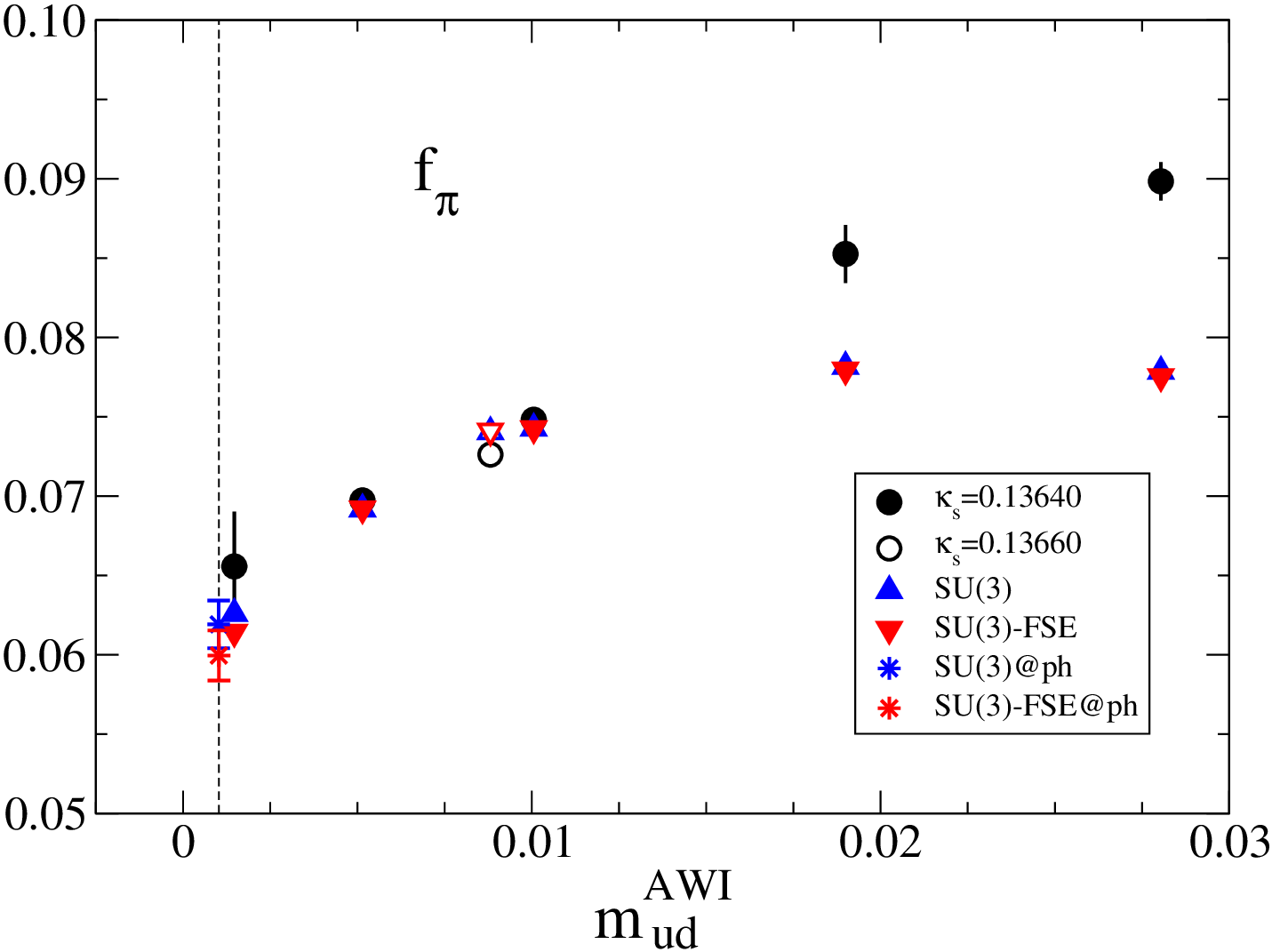}
\hspace*{2mm}&\hspace*{2mm}
\includegraphics[width=85mm,angle=0]{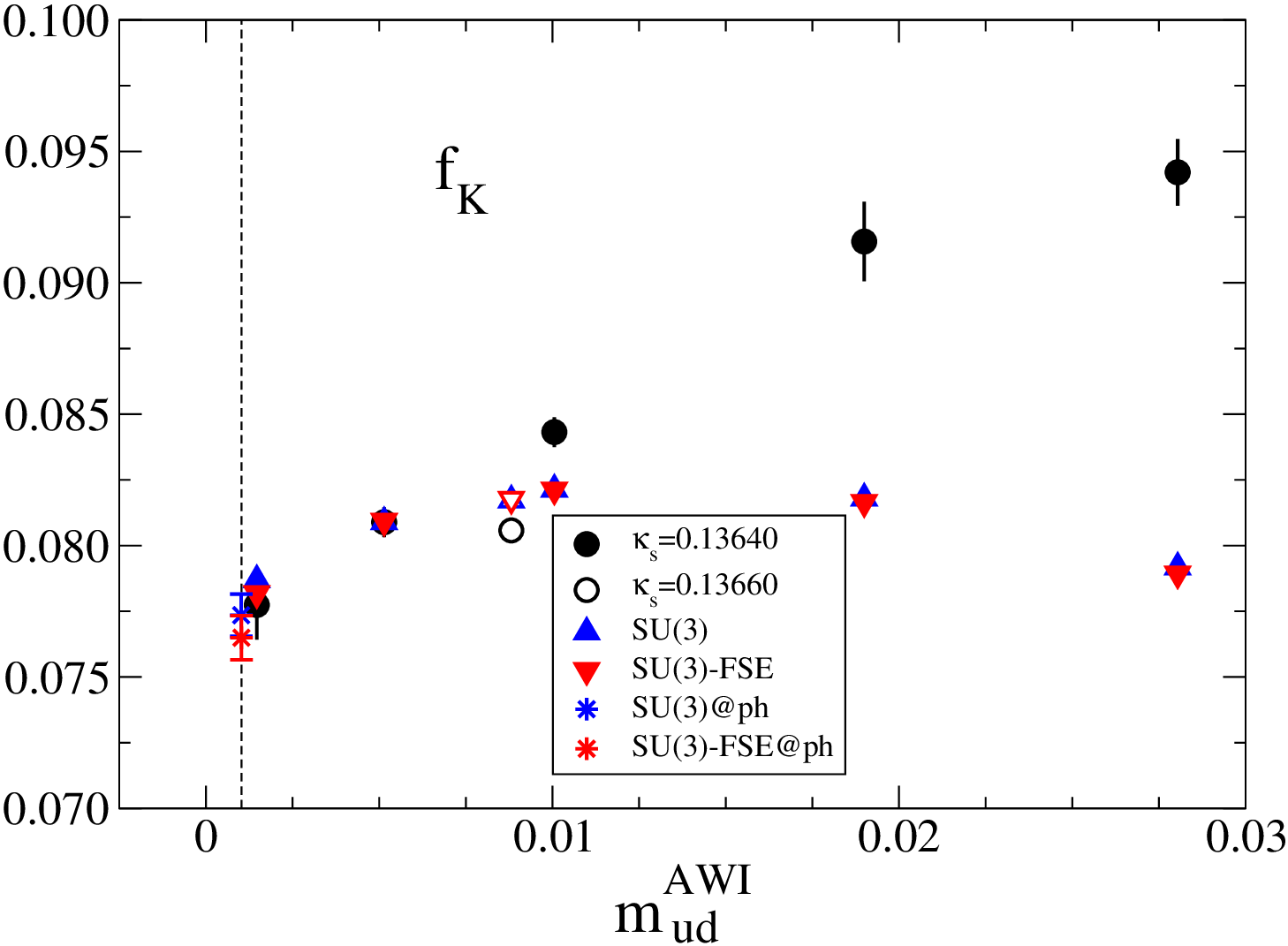}
\end{tabular}
\end{center}
\vspace{-.5cm}
\caption{SU(3) ChPT fit for $f_\pi$ (left) and $f_K$ (right).}
\label{fig:su3fit_fps}
\end{figure*}

We have found that the SU(3) ChPT fit gives reasonable 
values for the low energy constants.  However, we are concerned with a rather
large value of $\chi^2$/dof=4.2(2.9) (see Table~\ref{tab:fit_su3chpt}).
Figures~\ref{fig:su3fit_mps} and \ref{fig:su3fit_fps}
show how well the
data for $m_\pi^2/m_{\rm ud}$,  $2m_K^2/(m_{\rm ud}+m_{\rm s})$, $f_\pi$
and $f_K$ are described by the SU(3) ChPT up to NLO. 
The filled and open circles are our data, and the fit results are plotted 
by blue triangles.
We note in passing that , 
for the Wilson-clover quark action, $m_s^{\rm AWI}$ 
varies at $O(a^2)$ as $m_{\rm ud}$ varies even if $\kappa_{\rm s}$ is 
held fixed. Thus we are not able to draw a line with a fixed value 
for $m_s^{\rm AWI}$.
The blue star symbols represent the extrapolated values at the
physical point whose determination will be explained below
in Sec.~\ref{sec:physicalpt}. 

The points around $m_{\rm ud}^{\rm AWI}\approx 0.01$ corresponds to 
$(\kappa_{\rm ud},\kappa_{\rm s})=(0.13754,0.13640)$ and $(0.13754,0.13660)$. 
Marked deviations between circles and triangles show that the SU(3) ChPT poorly 
accounts for the strange quark mass dependence of $f_\pi$ and $f_K$.
This flaw is mainly responsible for the large value of $\chi^2$/dof.

\begin{figure*}[h]
\vspace{13mm}
\begin{center}
\begin{tabular}{cc}
\includegraphics[width=85mm,angle=0]{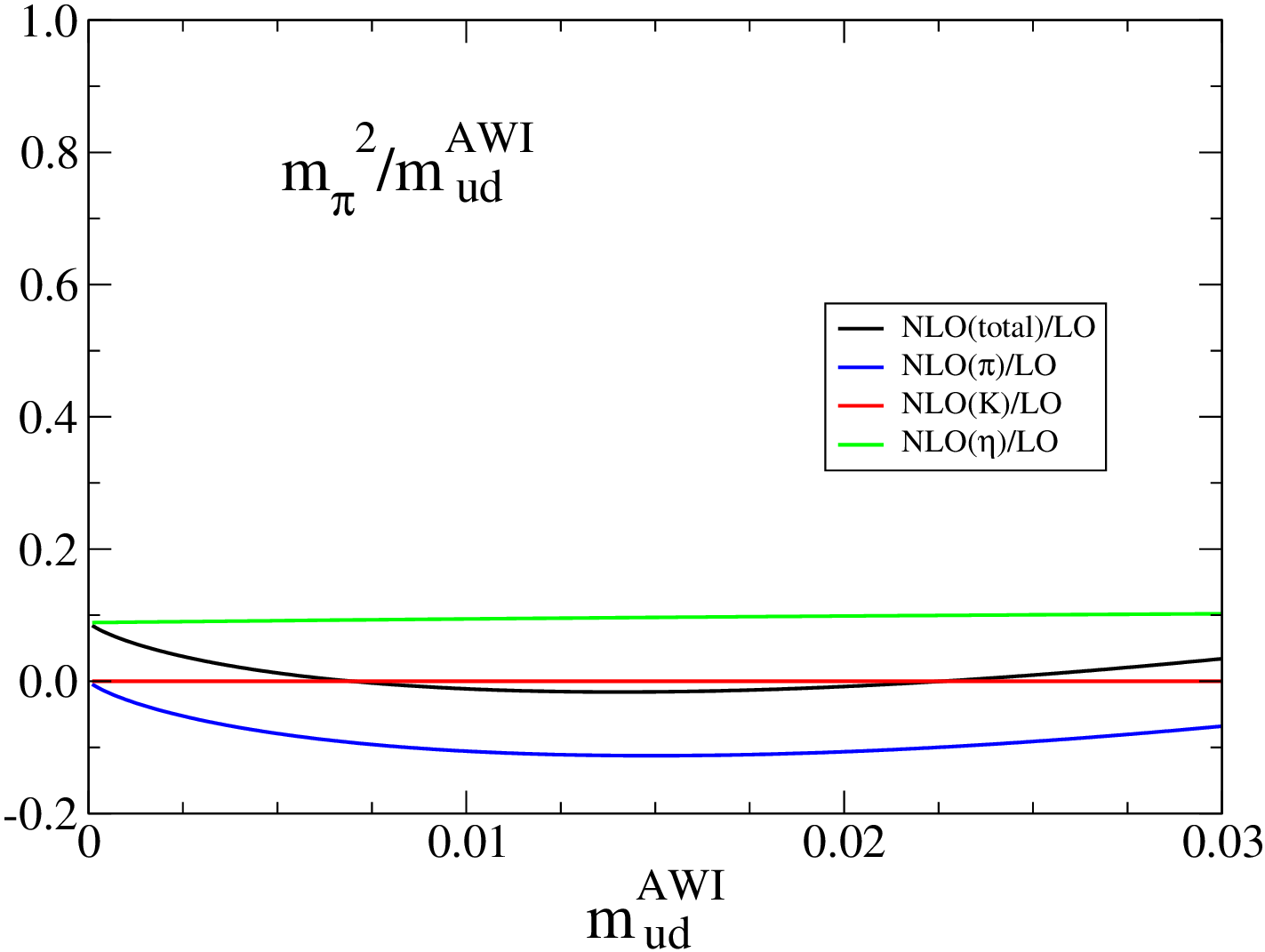}
\hspace*{2mm}&\hspace*{2mm}
\includegraphics[width=85mm,angle=0]{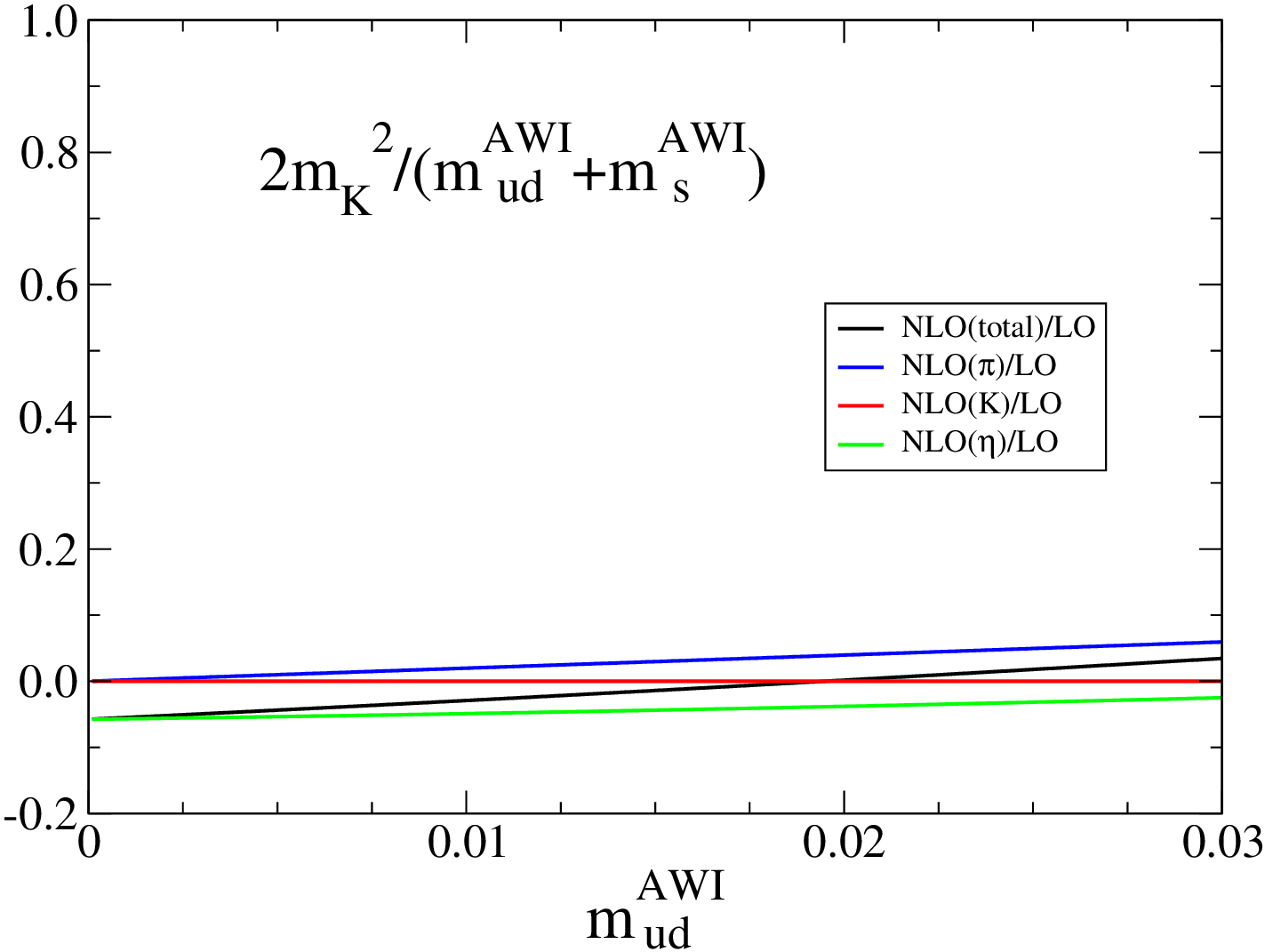}
\end{tabular}
\end{center}
\vspace{-.5cm}
\caption{Ratio of the NLO contribution to the LO one in the SU(3) ChPT
 fit for $m_\pi^2/m_{\rm ud}^{\rm AWI}$ (left) and 
$2 m_K^2/(m_{\rm ud}^{\rm AWI}+m_{\rm s})$ (right).}
\label{fig:su3nlo2lo_mps}
\end{figure*}

\begin{figure*}[h]
\vspace{13mm}
\begin{center}
\begin{tabular}{cc}
\includegraphics[width=85mm,angle=0]{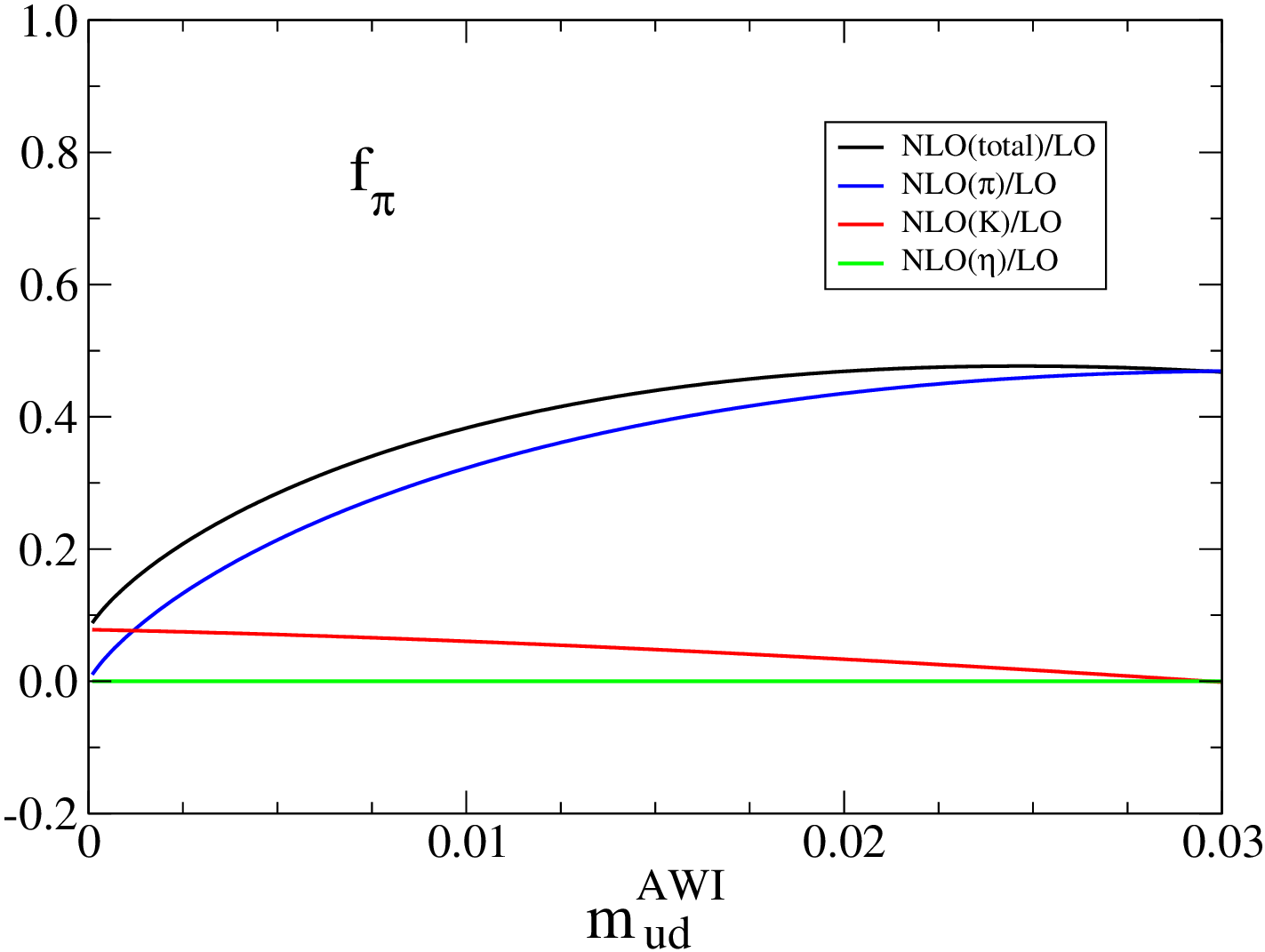}
\hspace*{2mm}&\hspace*{2mm}
\includegraphics[width=85mm,angle=0]{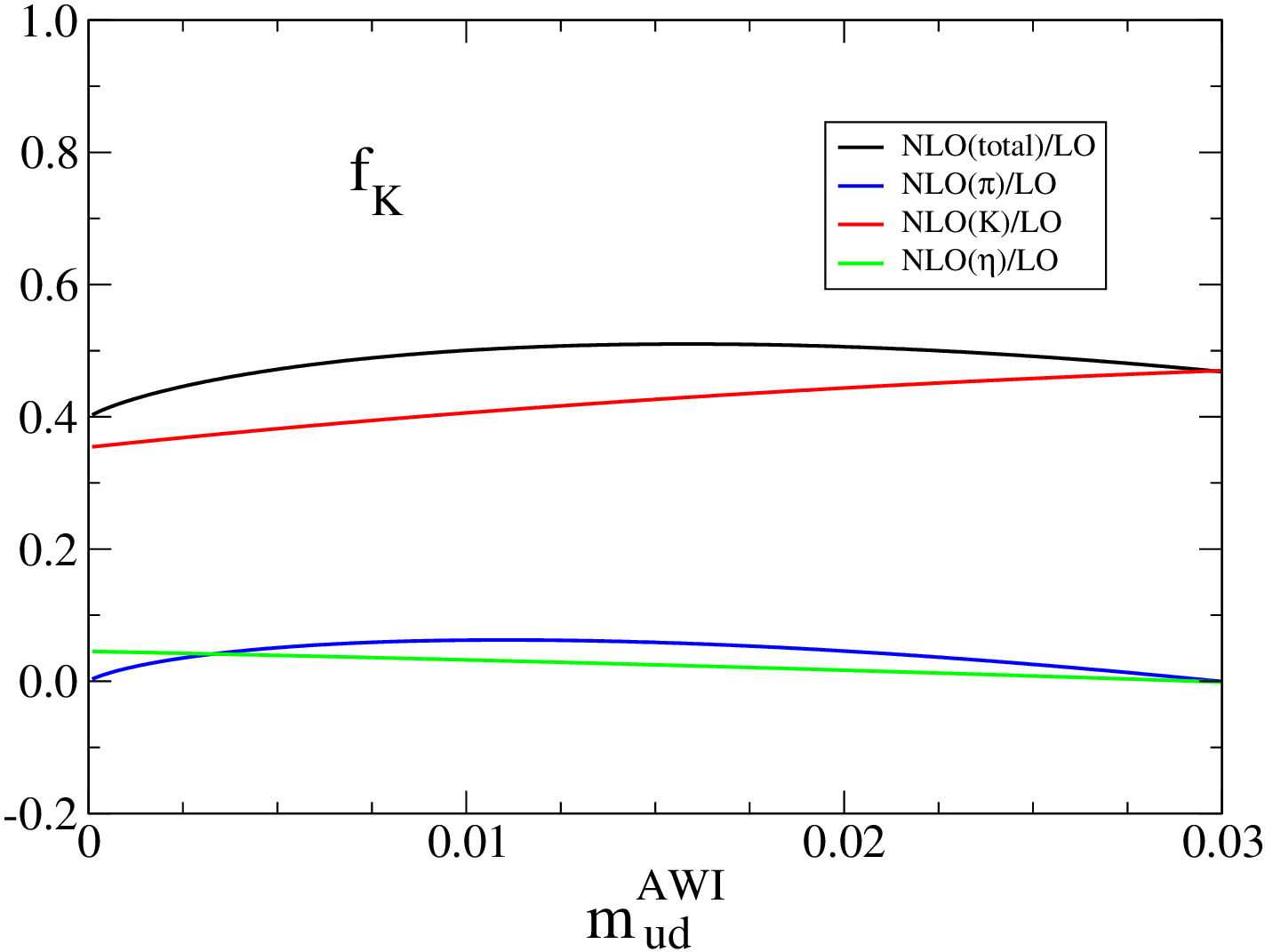}
\end{tabular}
\end{center}
\vspace{-.5cm}
\caption{Ratio of the NLO contribution to the LO one in the SU(3) ChPT
 fit for $f_\pi$ (left) and $f_K$ (right).}
\label{fig:su3nlo2lo_fps}
\end{figure*}

In order to investigate the origin of discrepancy between the data and the fit 
more closely,  we draw the relative magnitude of the NLO contribution to the 
LO one for $m_\pi^2/m_{\rm ud}$,  $2m_K^2/(m_{\rm ud}+m_{\rm s})$, $f_\pi$ and $f_K$
as a function of $m_{\rm ud}^{\rm AWI}$ in Figs.~\ref{fig:su3nlo2lo_mps} 
and \ref{fig:su3nlo2lo_fps}.  The strange quark 
mass is fixed at the physical value, and the contributions from $\pi$, $K$ and 
$\eta$ loops are separately drawn. 
The relative magnitudes are at most 10\% for
$m_\pi^2/m_{\rm ud}$ and $2m_K/(m_{\rm ud}+m_{\rm s})$.  We find, however, 
significant NLO contributions for the decay constants. 
For $f_\pi$ the relative magnitude rapidly increases from 10\% at 
$m_{\rm ud}=0$ to 40\% at around $m_{\rm ud}=0.01$. 
The situation is worse for $f_K$ for which the NLO contribution is about 40\% of 
the LO one even at $m_{\rm ud}=0$, most of which arises from the $K$ loop.

\subsection{SU(2) chiral perturbation theory}

The bad convergences of the chiral expansions 
for $f_\pi$ and $f_K$ tell us that the strange quark
mass is not light enough to be appropriately treated by the NLO SU(3)
ChPT. There are two alternative choices for further chiral
analysis. One is to extend SU(3) ChPT to NNLO, and the other is to use 
SU(2) ChPT with the aid of an analytic expansion for the strange
quark contribution around the physical strange quark mass.

The former method, which has been employed by the MILC collaboration 
in an incomplete fashion\cite{milc07}, is very demanding: 
we cannot determine the additional low energy constants at NNLO without 
significantly increasing the data points.  There is in addition no guarantee
that the expansion is controlled at NNLO. 
We therefore consider that the latter route is more natural.
This alternative was employed by the RBC/UKQCD collaboration\cite{rbcukqcd08}. 
Since they had data only at a single strange quark mass, they could not study 
the strange quark mass dependence.  This we shall do with our data 
thanks to the second choice of 
the strange quark mass at $\kappa_{\rm ud}=0.13754$.

For $m_\pi$ and $f_\pi$ the SU(2) ChPT formulae of (\ref{eq:su2chpt_mpi}) 
and (\ref{eq:su2chpt_fpi}) are employed.  The low energy constants $B$
and $f$ are functions of the strange quark mass.  Assuming that we run 
simulations close enough around the physical point for the strange quark mass 
so that a linear expansion in $m_{\rm s}$ is sufficient, we write 
$B=B_s^{(0)}+m_{\rm s}B_s^{(1)}$ and $f=f_s^{(0)}+m_{\rm s}f_s^{(1)}$, 
where it should be noted that $B_s^{(0)}\ne B_0$ and $f_s^{(0)}\ne f_0$.

For the kaon sector we treat the $K$ mesons as matter fields 
in the isospin $1/2$ linear representation, and couple pions 
in SU(2) invariant ways (see, {\it e.g.}, Ref.~\cite{roessl}).
For $m_K$ and $f_K$ this leads to the following fit formulae: 
\ben
m_K^2&=&\alpha_m+\beta_m m_{\rm ud}+\gamma_m m_{\rm s},
\label{eq:su2chpt_mk}\\
f_K&=&{\bar f} \left\{1+\beta_f m_{\rm ud} -\frac{3}{4}\frac{{\tilde m}_\pi^2}{16\pi^2 f^2}
\ln\left(\frac{{\tilde m}_\pi^2}{\mu^2}\right) \right\}
\label{eq:su2chpt_fk}
\een
with ${\bar f}={\bar f}_s^{(0)}+m_{\rm s}{\bar f}_s^{(1)}$. 
In these formulae, the linear expansion in $m_{\rm s}$ should be regarded 
as that around the physical strange quark mass.

We apply a simultaneous fit to $m_\pi$, $f_\pi$ and $f_K$ 
employing the formulae of Eqs.~(\ref{eq:su2chpt_mpi}), (\ref{eq:su2chpt_fpi})
and (\ref{eq:su2chpt_fk}).  The kaon mass 
$m_K^2$ is independently fitted according to Eq.~(\ref{eq:su2chpt_mk}). 
Calling the four data points corresponding to $\kappa_{\rm ud}\geq 0.13754$
as Range I, the fit results for $B,f,{\bar l}_3,{\bar l}_4$ at the physical 
strange quark mass are presented in
Table~\ref{tab:fit_su2chpt} and Fig.~\ref{fig:l_34} both without and with 
finite-size corrections. 
We find that they are consistent with
those obtained by the NLO conversion from the SU(3) low energy constants
given in Table~\ref{tab:lec_su2_conv}.
Although our result for $\la {\bar u}u\ra$ is about 50\% smaller than that 
of RBC/UKQCD, the difference comes from estimates of 
the renormalization factor: we use one-loop perturbation while they 
employ the nonperturbative RI-MOM scheme.
This is verified by the observation that
the value of $f$ and the renormalization-free quantities $m_{\rm ud} B$
and $m_{\rm s} B$ show consistency
between our results and those of RBC/UKQCD.

\begin{figure*}[h]
\vspace{13mm}
\begin{center}
\includegraphics[width=87mm,angle=0]{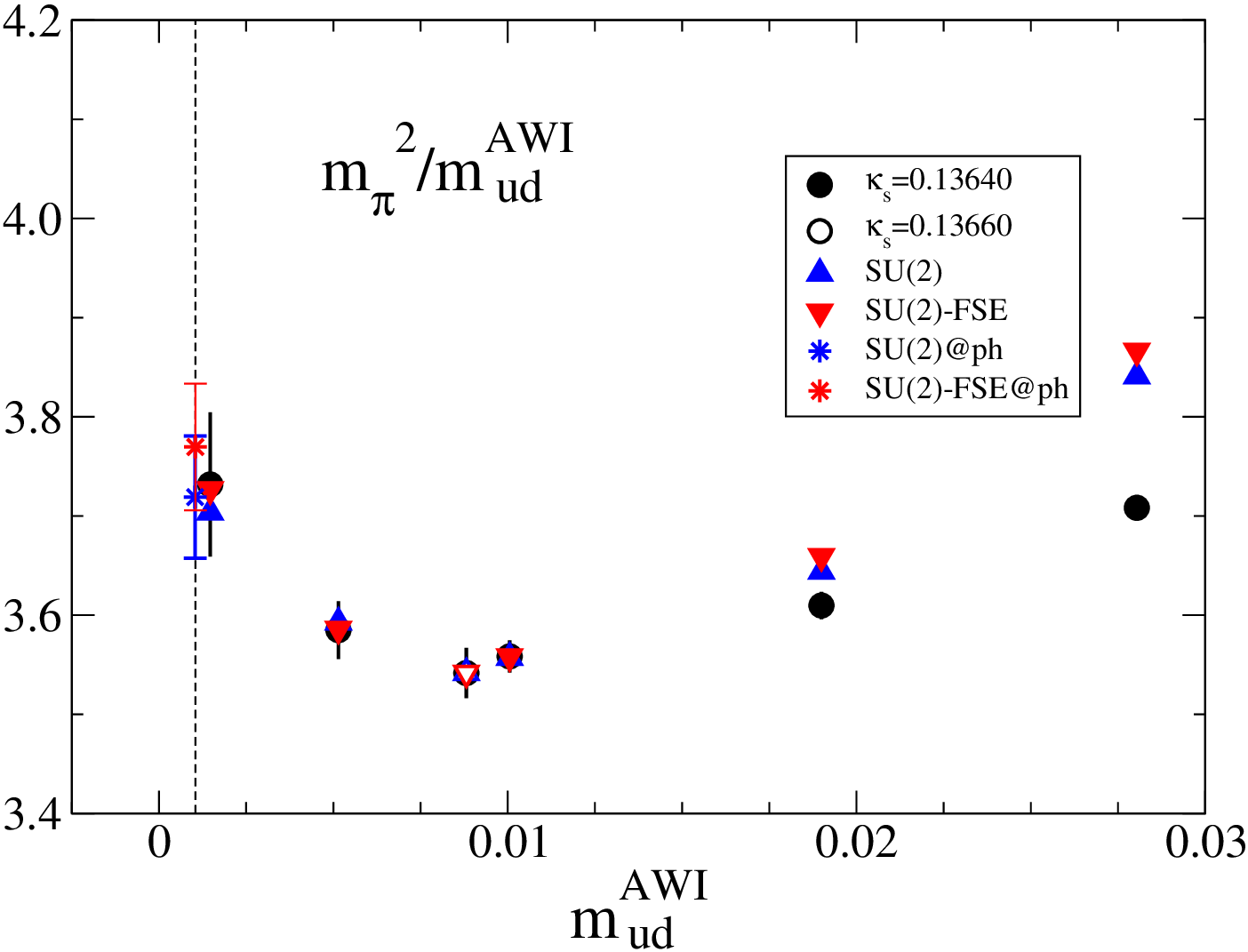}
\end{center}
\vspace{-.5cm}
\caption{SU(2) ChPT fit for $m_\pi^2/m_{\rm ud}^{\rm AWI}$.}
\label{fig:su2fit_mpi}
\end{figure*}

\begin{figure*}[h]
\vspace{13mm}
\begin{center}
\begin{tabular}{cc}
\includegraphics[width=85mm,angle=0]{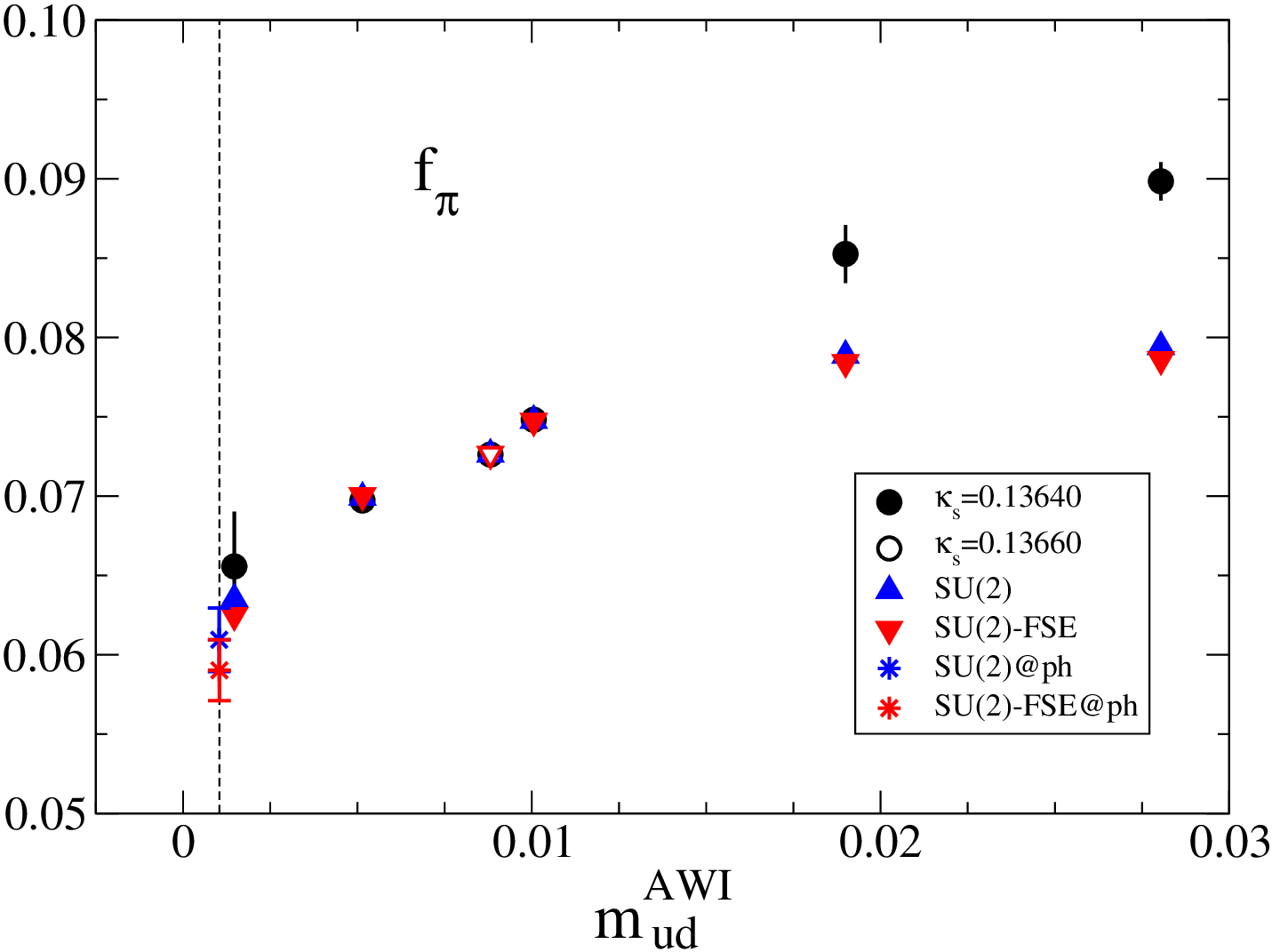}
\hspace*{2mm}&\hspace*{2mm}
\includegraphics[width=85mm,angle=0]{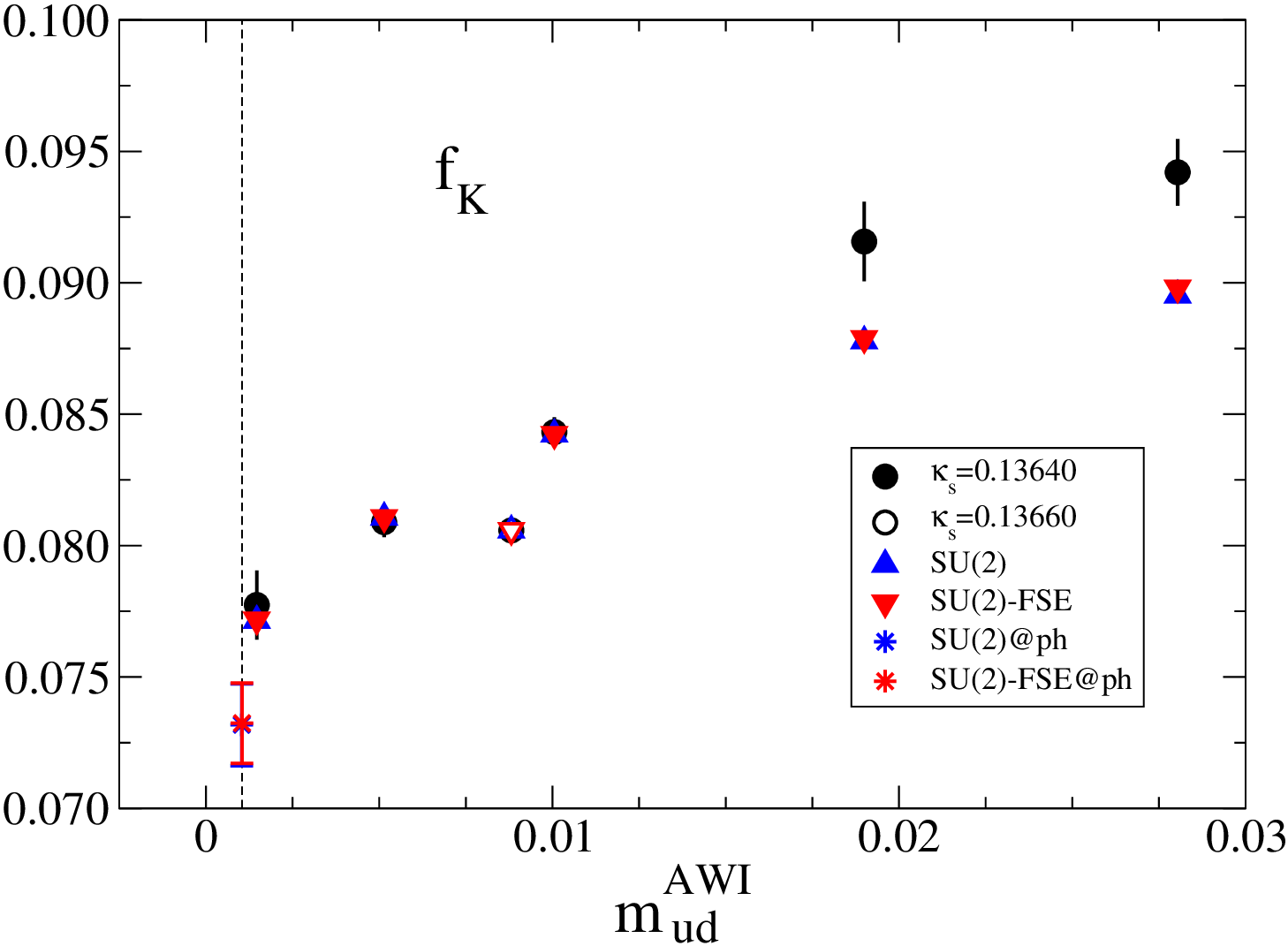}
\end{tabular}
\end{center}
\vspace{-.5cm}
\caption{SU(2) ChPT fit for $f_\pi$ (left) and $f_K$ (right).}
\label{fig:su2fit_fps}
\end{figure*}

\begin{figure*}[h]
\vspace{13mm}
\begin{center}
\includegraphics[width=90mm,angle=0]{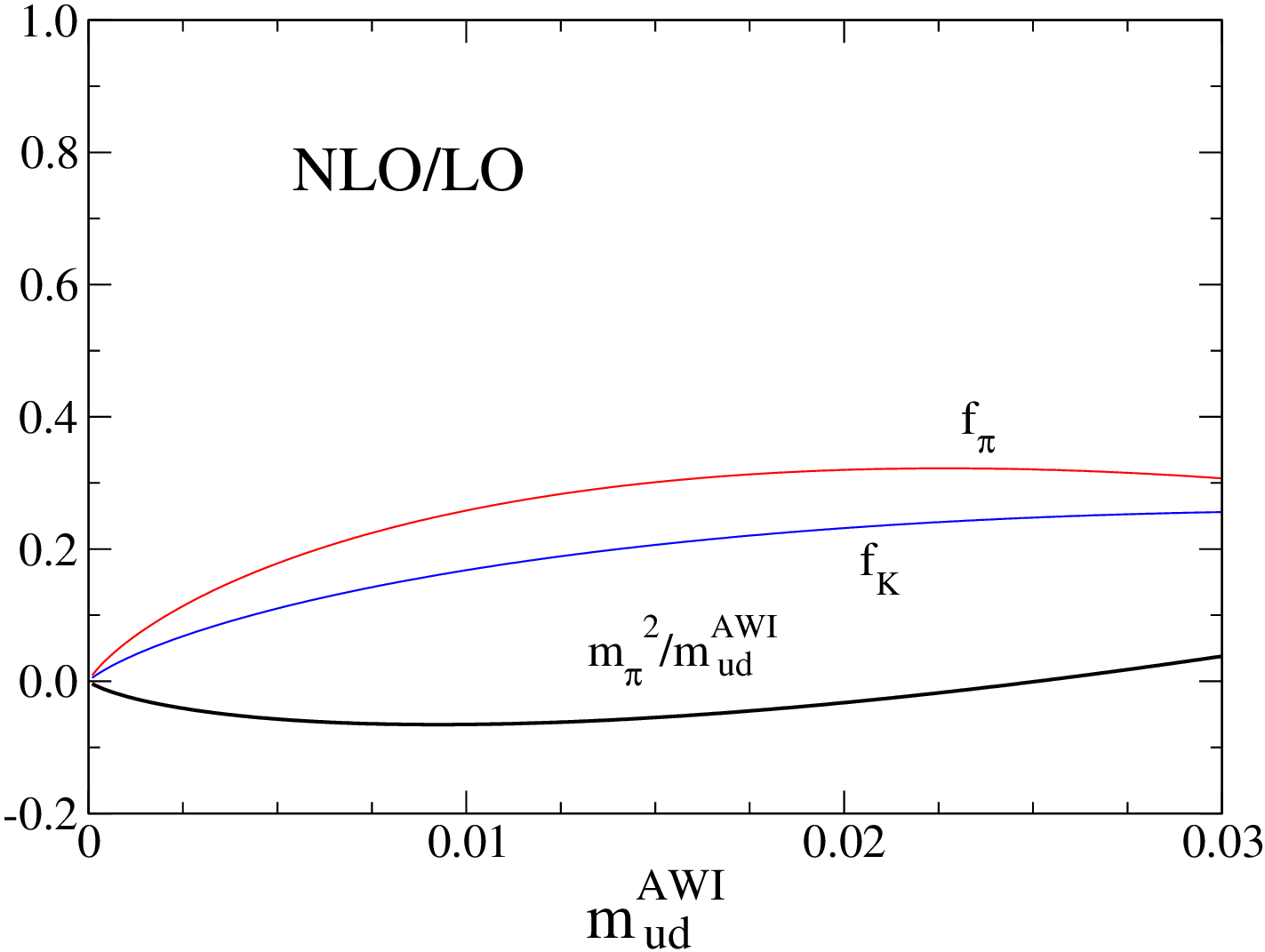}
\end{center}
\vspace{-.5cm}
\caption{Ratio of the NLO contribution to the LO one in the SU(2) ChPT
 fit for $m_\pi^2/m_{\rm ud}^{\rm AWI}$, $f_\pi$ and $f_K$.}
\label{fig:su2nlo2lo}
\end{figure*}

Figures~\ref{fig:su2fit_mpi} and \ref{fig:su2fit_fps} show that 
the quark mass dependences of $m_\pi^2/m_{\rm ud}^{\rm AWI}$, 
$f_\pi$ and $f_K$
are reasonably described by the SU(2) ChPT formulae of
(\ref{eq:su2chpt_mpi}), (\ref{eq:su2chpt_fpi}) 
and (\ref{eq:su2chpt_fk}). The resulting  $\chi^2$/dof is 0.33(72),
which is an order of magnitude smaller than in the SU(3) case. 
In Fig.~\ref{fig:su2nlo2lo} we illustrate the relative magnitude of  
the NLO contribution against the LO value for 
$m_\pi^2/m_{\rm ud}$, $f_\pi$ and $f_K$ as a function of $m_{\rm ud}^{\rm AWI}$
fixing the strange quark mass at the physical value. 
The convergences for $f_\pi$ and $f_K$ are clearly better than the SU(3) case.

In order to investigate the stability of the fit, we try two additional 
choices of the data sets for the SU(2) ChPT fit: 
Range II ($\kappa_{\rm ud}$,$\kappa_{\rm s}$)=(0.13781,0.13640),(0.13770,0.13640),
(0.13754,0.13640),(0.13754,0.13660),(0.13727,0.13640) includes one more data 
at a heavier pion mass added to Range I, and Range III 
($\kappa_{\rm ud}$,$\kappa_{\rm s}$)=(0.13770,0.13640),
(0.13754,0.13640),(0.13754,0.13660),(0.13727,0.13640) removes the point with 
the lightest pion mass from Range II.  
The results for $B,f,{\bar l}_3,{\bar l}_4$ and corresponding 
$\chi^2$/dof are given in Table~\ref{tab:fit_su2chpt}.
While inclusion of the data at $\kappa_{\rm ud}=0.13727$
increases the value of $\chi^2$/dof, the results for $B,f,{\bar l}_3,{\bar
l}_4$ are consistent among the three cases within the error bars.

\subsection{Finite size effects based on chiral perturbation theory}
\label{subsec:fse}

\begin{figure*}[h]
\vspace{13mm}
\begin{center}
\includegraphics[width=90mm,angle=0]{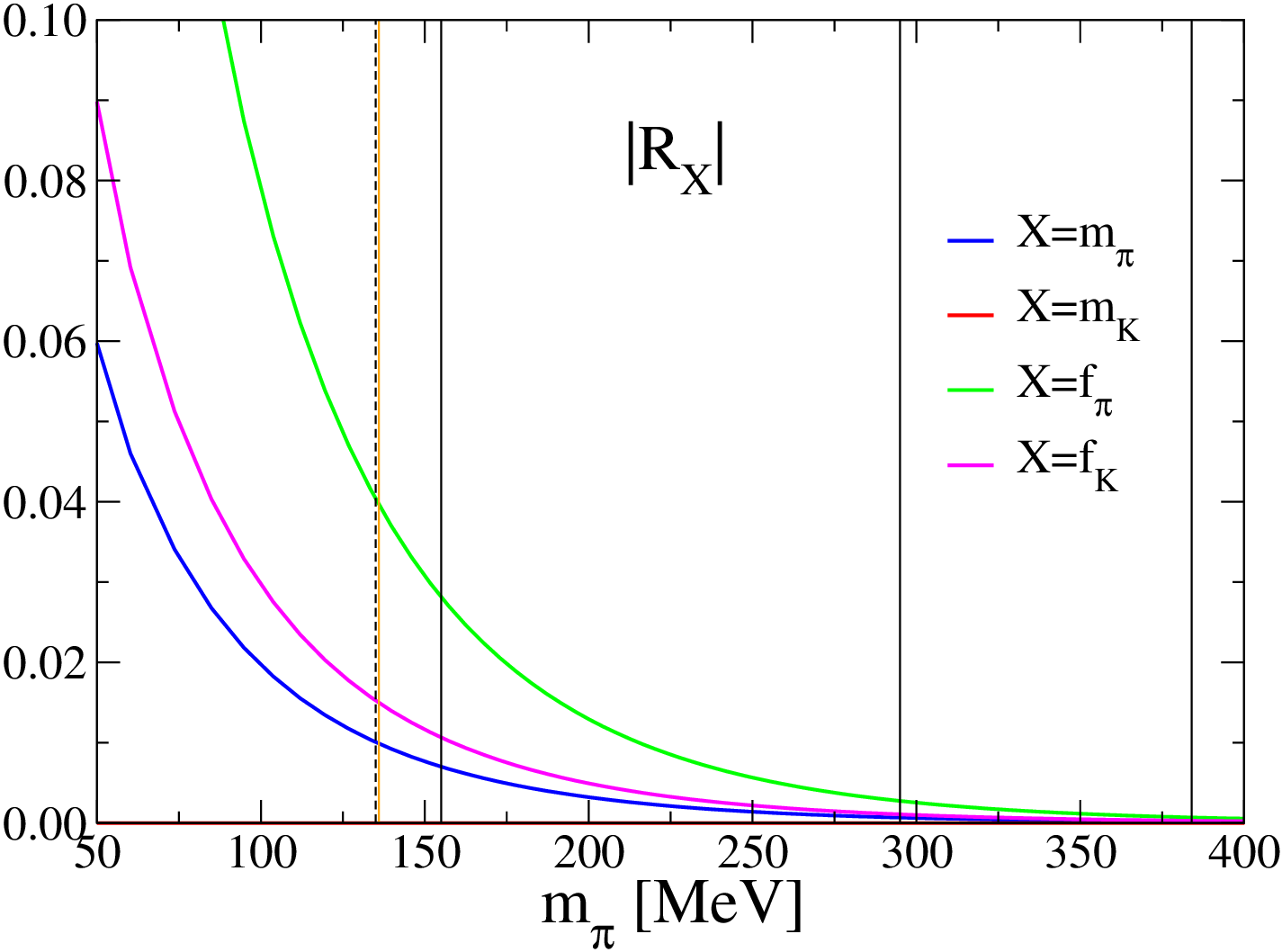} 
\end{center}
\vspace{-.5cm}
\caption{$\vert R_X\vert$ ($R_{m_{\rm PS}}>0$ 
and $R_{f_{\rm PS}}<0$) for 
$X=m_\pi,m_K,f_\pi,f_K$ with $L=2.9$ fm as a function of $m_\pi$ at the
physical strange quark mass based on the NLO SU(3) ChPT. 
Dotted vertical line denotes the physical point and the solid ones are
 for our simulation points. Orange vertical line represents the pion
 mass with $m_\pi L=2$.}
\label{fig:fse_su3}
\end{figure*}

We evaluate finite-size effects based on the NLO formulae of ChPT.
In the case of SU(3) ChPT the finite size effects defined by 
$R_X=(X(L)-X(\infty))/X(\infty)$ for $X=m_\pi,m_K,f_\pi,f_K$ are given 
by~\cite{colangelo05}:
\ben
R_{m_\pi}
&=&\frac{1}{4}\xi_{\pi}\tilde g_1(\lambda_\pi)-\frac{1}{12}\xi_{\eta}\tilde g_1(\lambda_\eta),\\
R_{m_K} 
&=&\frac{1}{6}\xi_{\eta}\tilde g_1(\lambda_\eta),\\
R_{f_\pi} 
&=&-\xi_{\pi}\tilde g_1(\lambda_\pi)-\frac{1}{2}\xi_{K}\tilde g_1(\lambda_K),\\
R_{f_K} 
&=&-\frac{3}{8}\xi_{\pi}\tilde g_1(\lambda_\pi)-\frac{3}{4}\xi_{K}\tilde g_1(\lambda_K)
-\frac{3}{8}\xi_{\eta}\tilde g_1(\lambda_\eta)\nn\\
\een
with
\ben
\xi_{\rm PS} &\equiv& \frac{2m_{\rm PS}^2}{(4\pi f_\pi)^2},\\
\lambda_{\rm PS} &\equiv& m_{\rm PS} L, \\
\tilde g_1(x)&=&\sum_{n=1}^{\infty}\frac{4m(n)}{{\sqrt n }x} K_1({\sqrt n} x), 
\een
where $K_1$ is the Bessel function of the second kind and $m(n)$
denotes the multiplicity of the partition $n=n_x^2+n_y^2+n_z^2$. 
The authors in Ref.~\cite{colangelo05} expect that the above formulea
are valid for $m_\pi L>2$, in which our simulation points reside. 
In Figs.~\ref{fig:su3fit_mps} and \ref{fig:su3fit_fps}
we also plotted the ChPT fit results including finite size effects.
The results are almost degenerate with the fit results 
without finite size effects except at the lightest simulation point 
at $\kappa_{\rm ud}=0.13781$ and the
extrapolated values at the physical point. This feature is understood
by looking at Fig.~\ref{fig:fse_su3} where we plot the magnitude of $R_X$ for 
$X=m_\pi,m_K,f_\pi,f_K$ with $L=2.9$ fm as a function of $m_\pi$ 
keeping the strange quark mass fixed at the physical value.
The expected finite size effects are less than 2\% for $m_{\rm PS}$ and
$f_{\rm PS}$ at our simulation points. For $m_{\rm PS}$ this is true 
even at the physical point, while the value of $f_\pi$ is 
decreased by 4\% due to the finite size effects. 

\begin{figure*}[h]
\vspace{13mm}
\begin{center}
\includegraphics[width=90mm,angle=0]{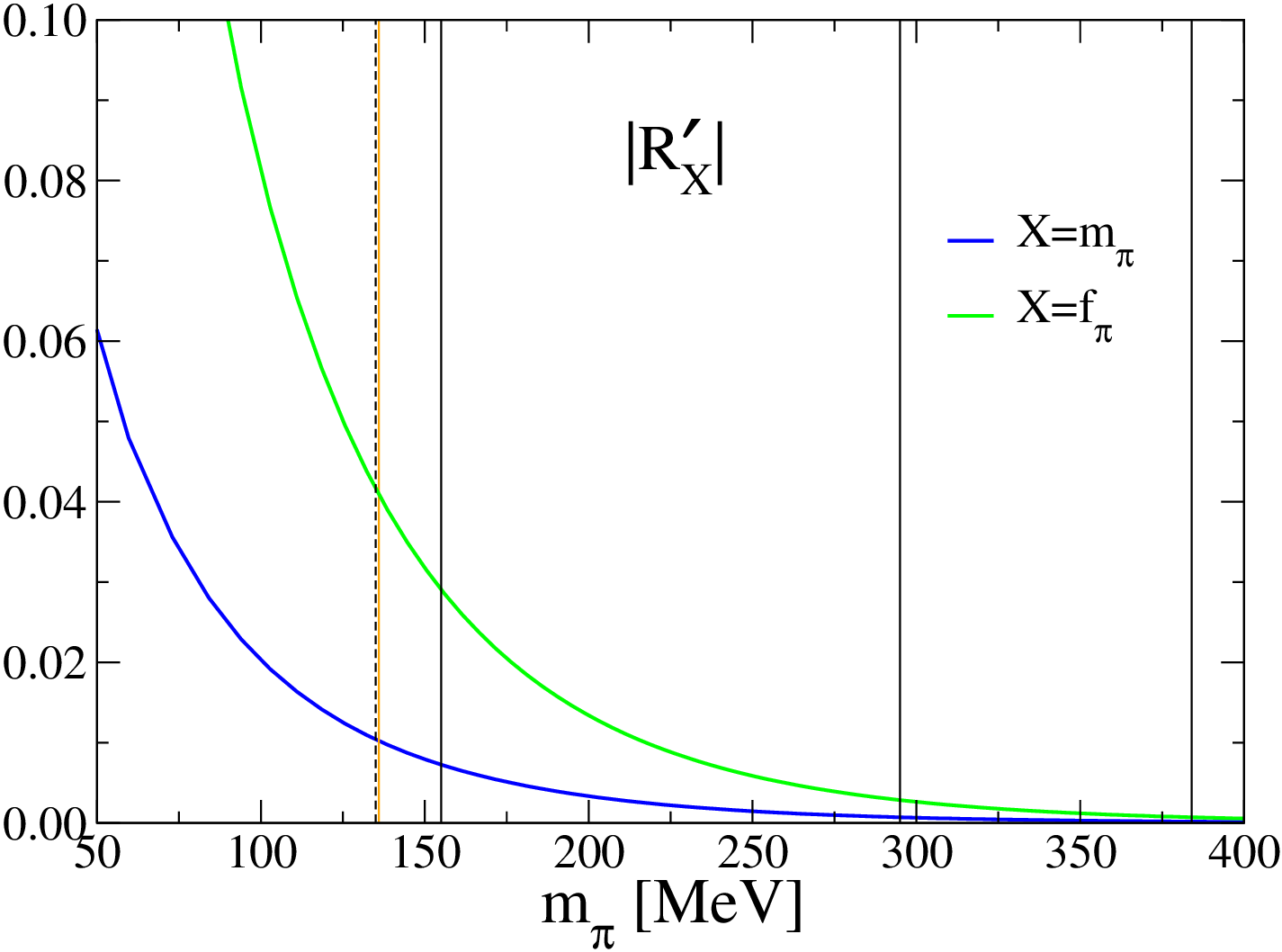} 
\end{center}
\vspace{-.5cm}
\caption{$\vert R^\prime_X\vert$ ($R^\prime_{m_{\rm PS}}>0$ 
and $R^\prime_{f_{\rm PS}}<0$) for 
$X=m_\pi,f_\pi$ with $L=2.9$ fm as a function of $m_\pi$ 
based on the NLO SU(2) ChPT. Dotted vertical line denotes 
the physical point and the solid ones are
 for our simulation points. Orange vertical line represents the pion
 mass with $m_\pi L=2$.}
\label{fig:fse_su2}
\end{figure*}

We can repeat the above study for the SU(2) case. 
The NLO formulae for $m_\pi$ and $f_\pi$ are given by\cite{colangelo05}
\ben
R^\prime_{m_\pi}
&=&\frac{1}{4}\xi_{\pi}\tilde g_1(\lambda_\pi),\\
R^\prime_{f_\pi} 
&=&-\xi_{\pi}\tilde g_1(\lambda_\pi).
\een
In Figs.~\ref{fig:su2fit_mpi} and \ref{fig:su2fit_fps}
we hardly detect finite size effects for $m_{\rm ud}^{\rm AWI}>0.001$.   
Figure~\ref{fig:fse_su2} shows $R_X^\prime$ for 
$X=m_\pi,f_\pi$ with $L=2.9$ fm as a function of $m_\pi$.
The situation is similar to the SU(3) case: although finite size
effects increase as $m_\pi$ decreases, their magnitudes are at most 2\%
for $m_\pi$ and 4\% for $f_\pi$
even at the physical point, which is easily expected 
by comparing the expressions of $R$ and $R^\prime$.

Let us add a cautionary note that the finite-size formulae analyzed 
here lose viability when $m_\pi L$ becomes too small.  Precisely at 
what values of $m_\pi L$ this takes place is not well controlled 
theoretically, however.  Direct simulations on a larger lattice 
is required to pin down the actual magnitude of finite-size effects at the 
physical point.  The need for such calculations are even more for baryons 
whose sizes are larger than mesons.

\section{Results at the physical point}
\label{sec:physicalpt}

We need three physical inputs to determine the up-down and 
the strange quark masses and the lattice cutoff.
We choose $m_\pi$, $m_K$ and $m_\Omega$.  The choice of 
$m_\Omega$ has both theoretical and practical advantages:
the $\Omega$ baryon is stable in the strong interactions
and its mass, being composed of three strange quarks, 
is determined with good precision with small finite 
size effects. 

\begin{figure*}[h]
\vspace{13mm}
\begin{center}
\begin{tabular}{cc}
\includegraphics[width=85mm,angle=0]{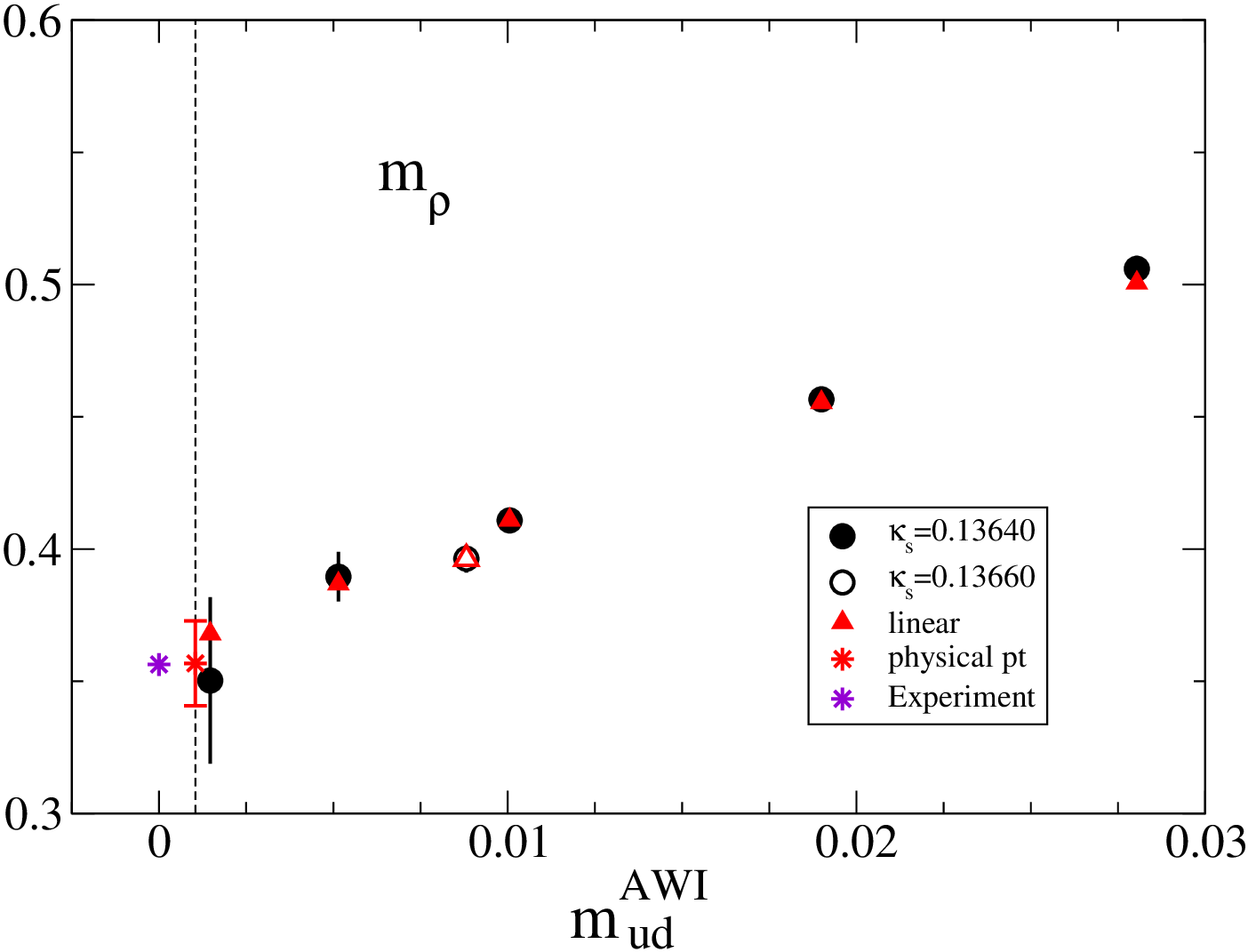}
\hspace*{2mm}&\hspace*{2mm}
\includegraphics[width=85mm,angle=0]{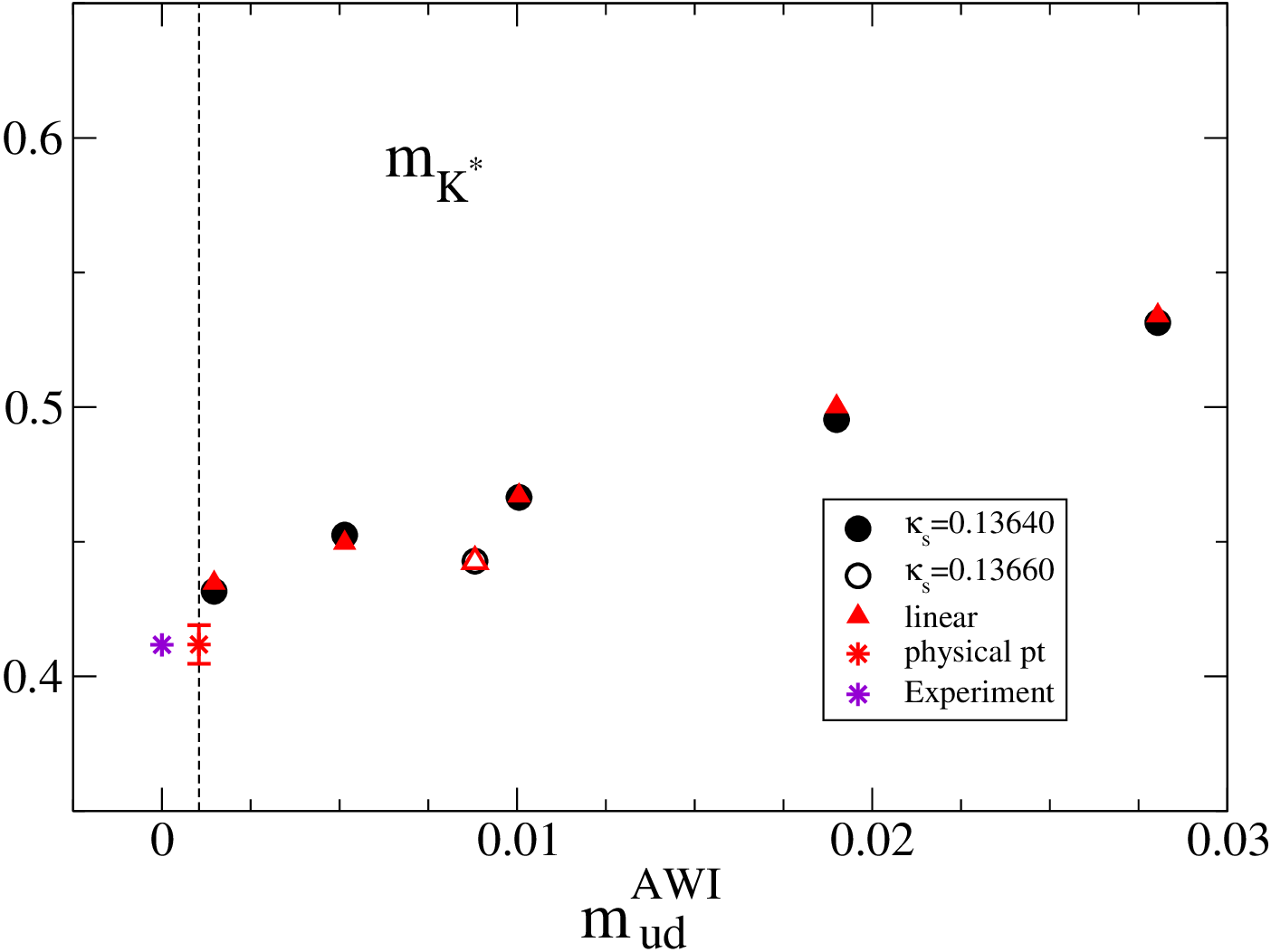}\\
\vspace*{7mm}& \\
\multicolumn{2}{c}
{\includegraphics[width=85mm,angle=0]{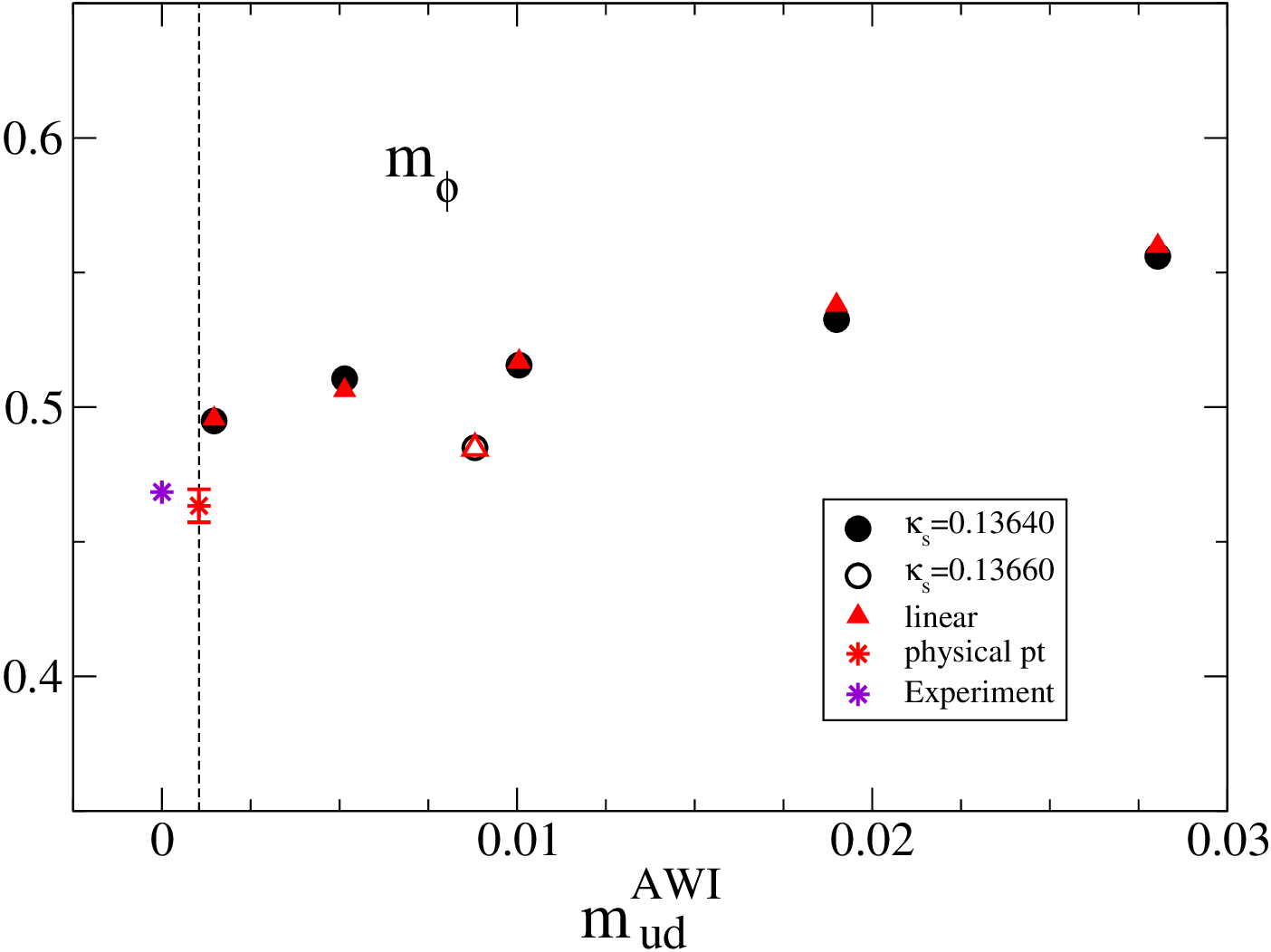}} 
\end{tabular}
\end{center}
\vspace{-.5cm}
\caption{Linear chiral extrapolation for the vector meson masses
 in lattice units. Red triangles represent the fit results
at the measured values of $m_{\rm ud}^{\rm AWI}$. 
Red star symbol denotes the extrapolated value at the physical point.
Experimental value in lattice units is also plotted 
at $m_{\rm ud}^{\rm AWI}=0$ for comparison.}
\label{fig:chexp_vector}
\end{figure*}

\begin{figure*}[h]
\vspace{13mm}
\begin{center}
\begin{tabular}{cc}
\includegraphics[width=85mm,angle=0]{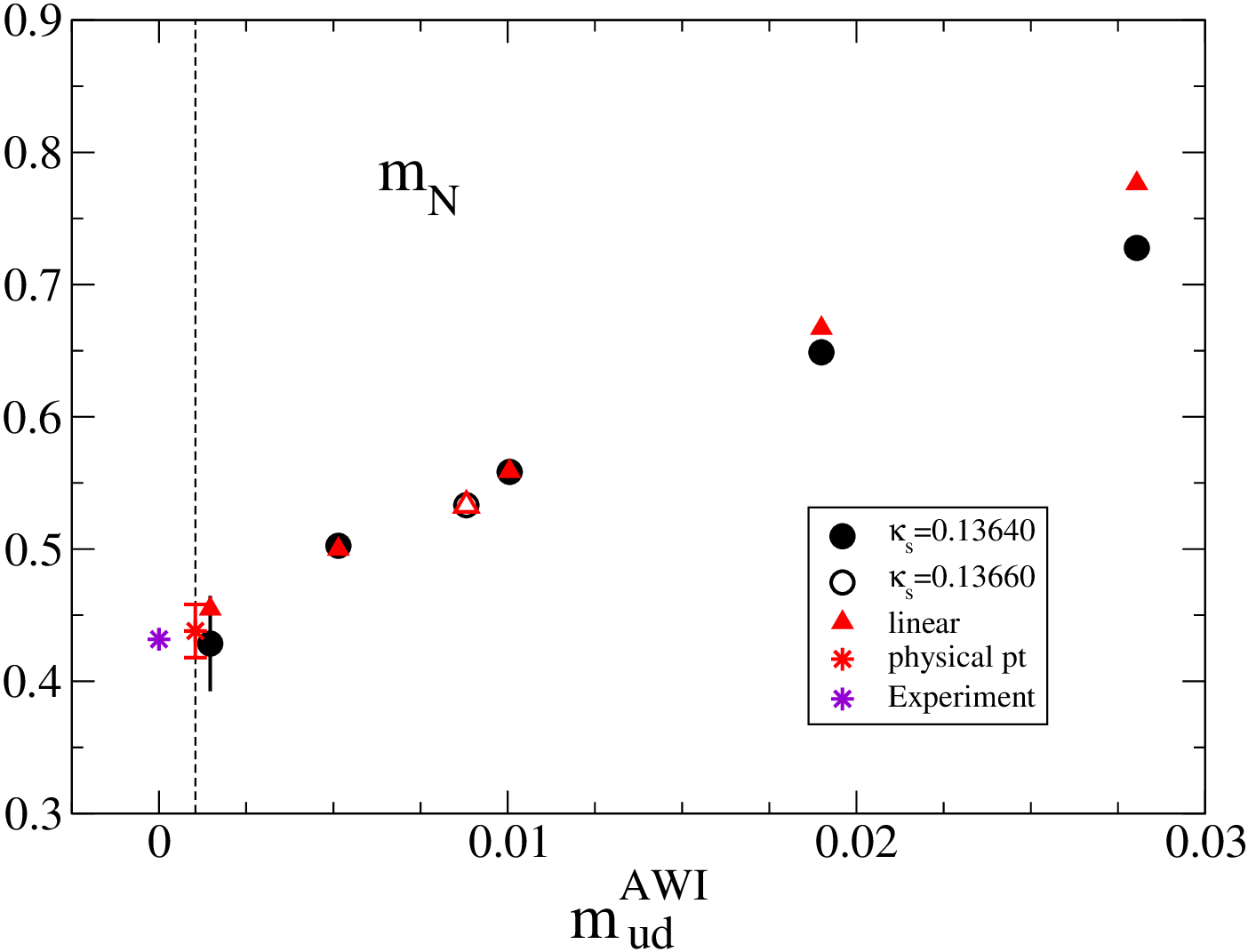}
\hspace*{2mm}&\hspace*{2mm}
\includegraphics[width=85mm,angle=0]{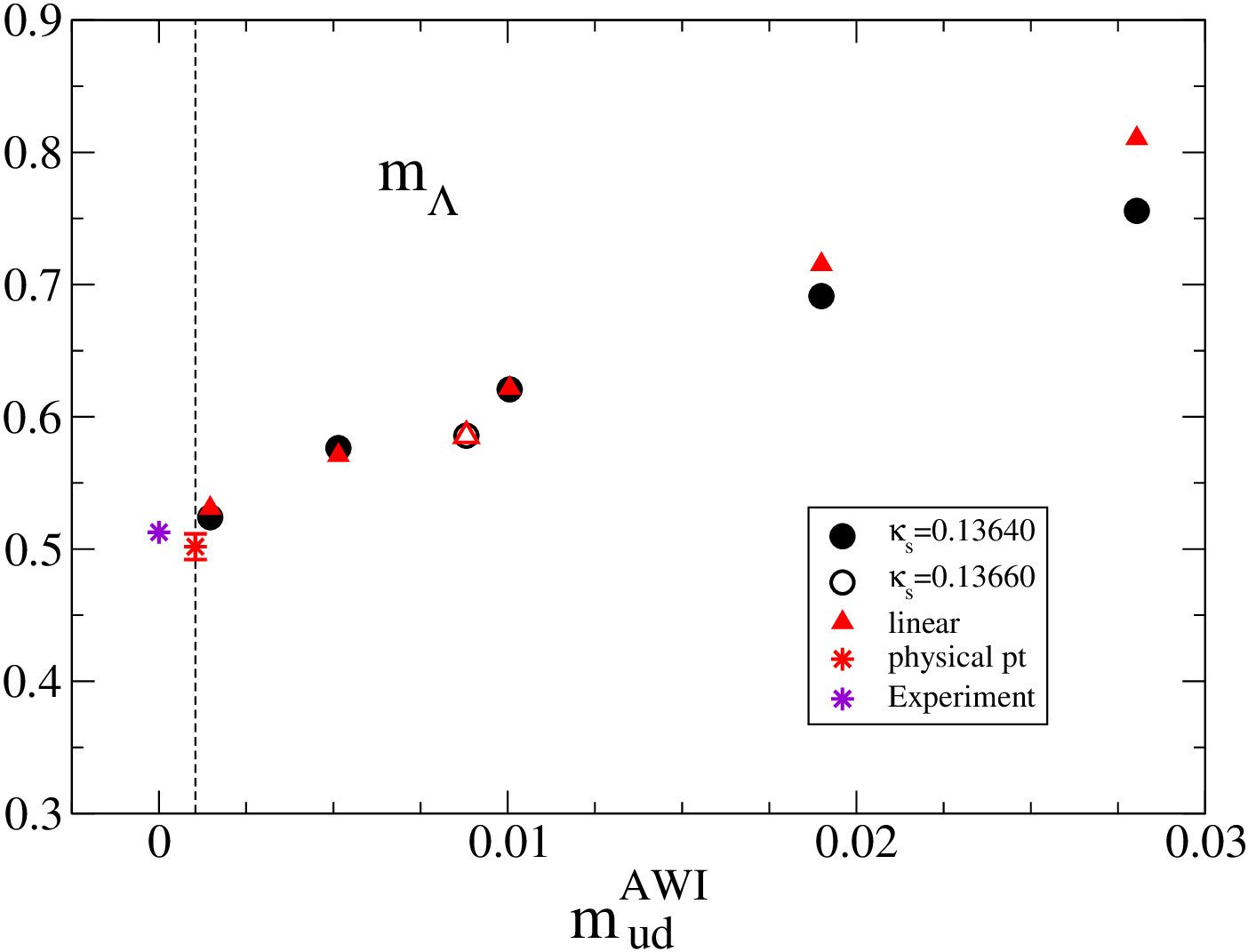}\\
\vspace*{7mm}& \\
\includegraphics[width=85mm,angle=0]{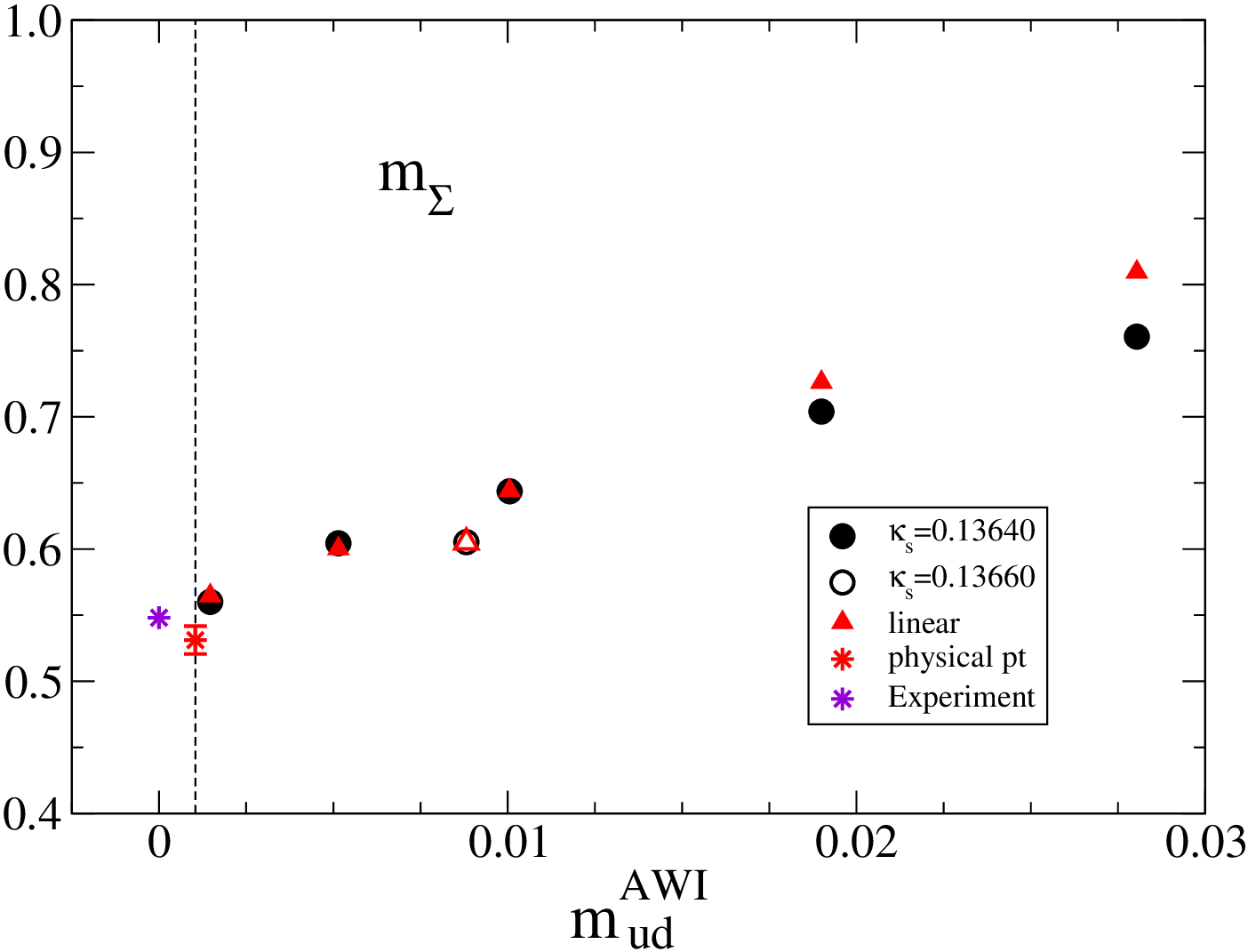}
\hspace*{2mm}&\hspace*{2mm}
\includegraphics[width=85mm,angle=0]{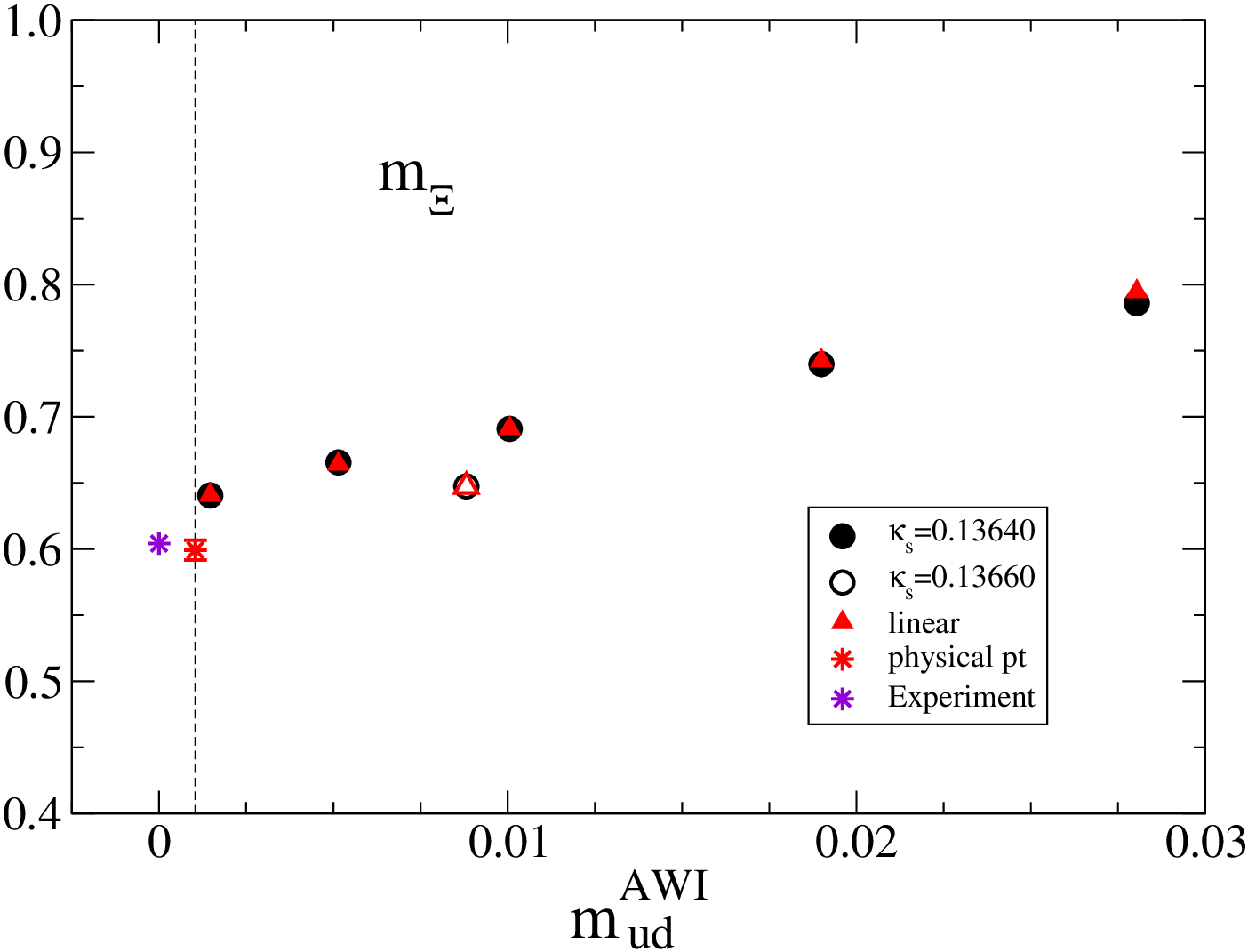}
\end{tabular}
\end{center}
\vspace{-.5cm}
\caption{Same as Fig.~\protect{\ref{fig:chexp_vector}} for the octet baryons.}
\label{fig:chexp_octet}
\end{figure*}

\begin{figure*}[h]
\vspace{13mm}
\begin{center}
\begin{tabular}{cc}
\includegraphics[width=85mm,angle=0]{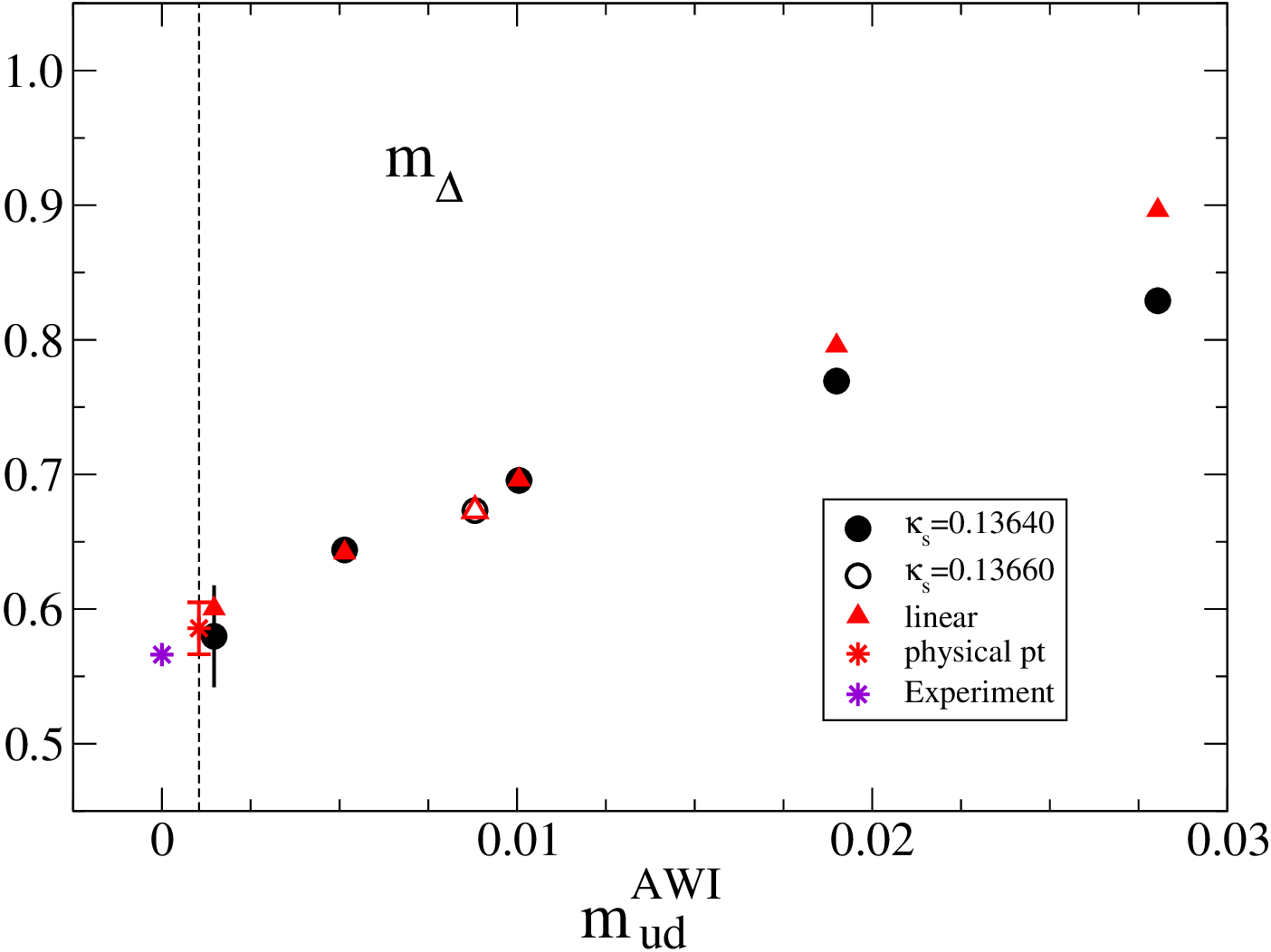}
\hspace*{2mm}&\hspace*{2mm}
\includegraphics[width=85mm,angle=0]{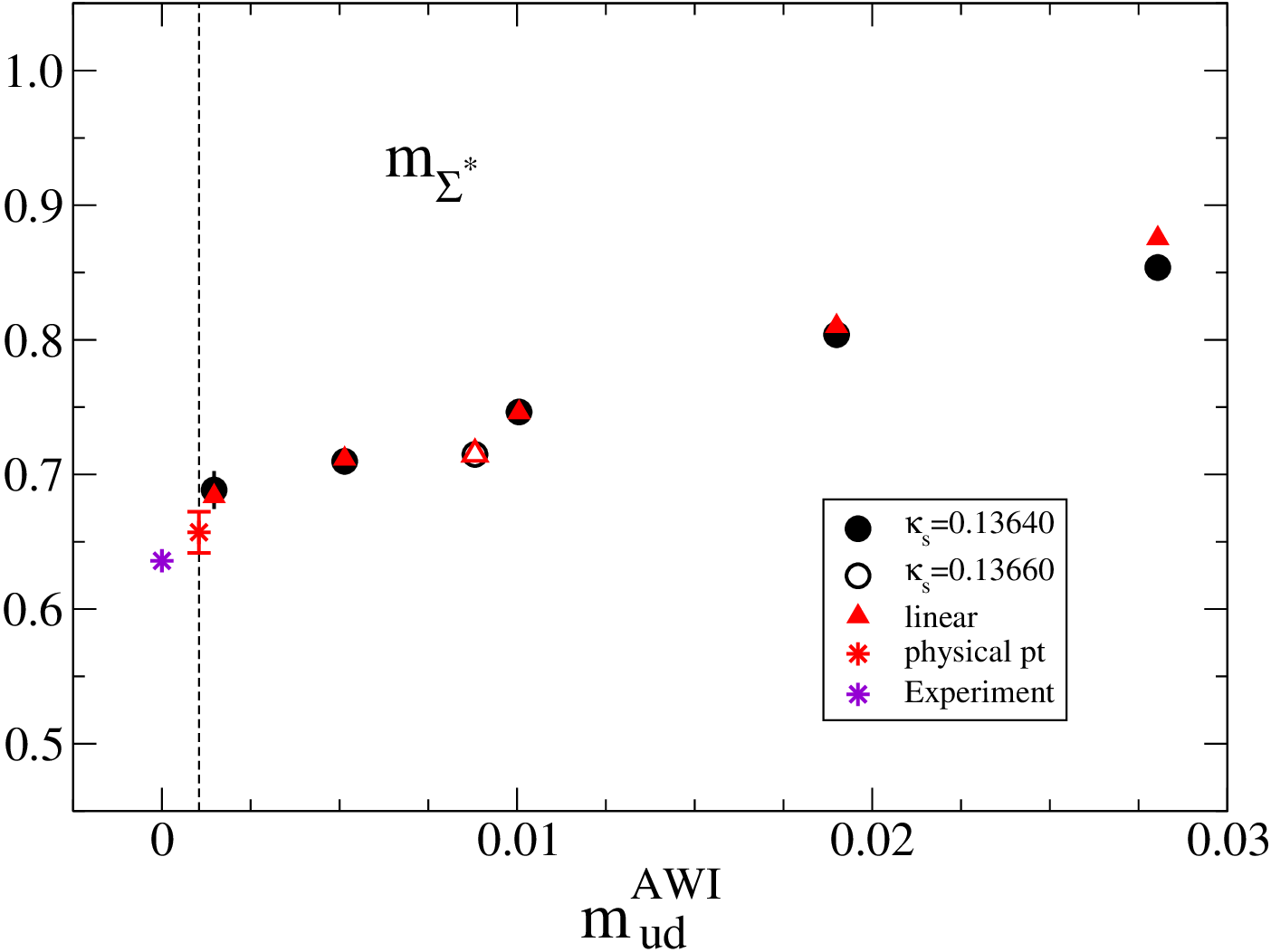}\\
\vspace*{7mm}& \\
\includegraphics[width=85mm,angle=0]{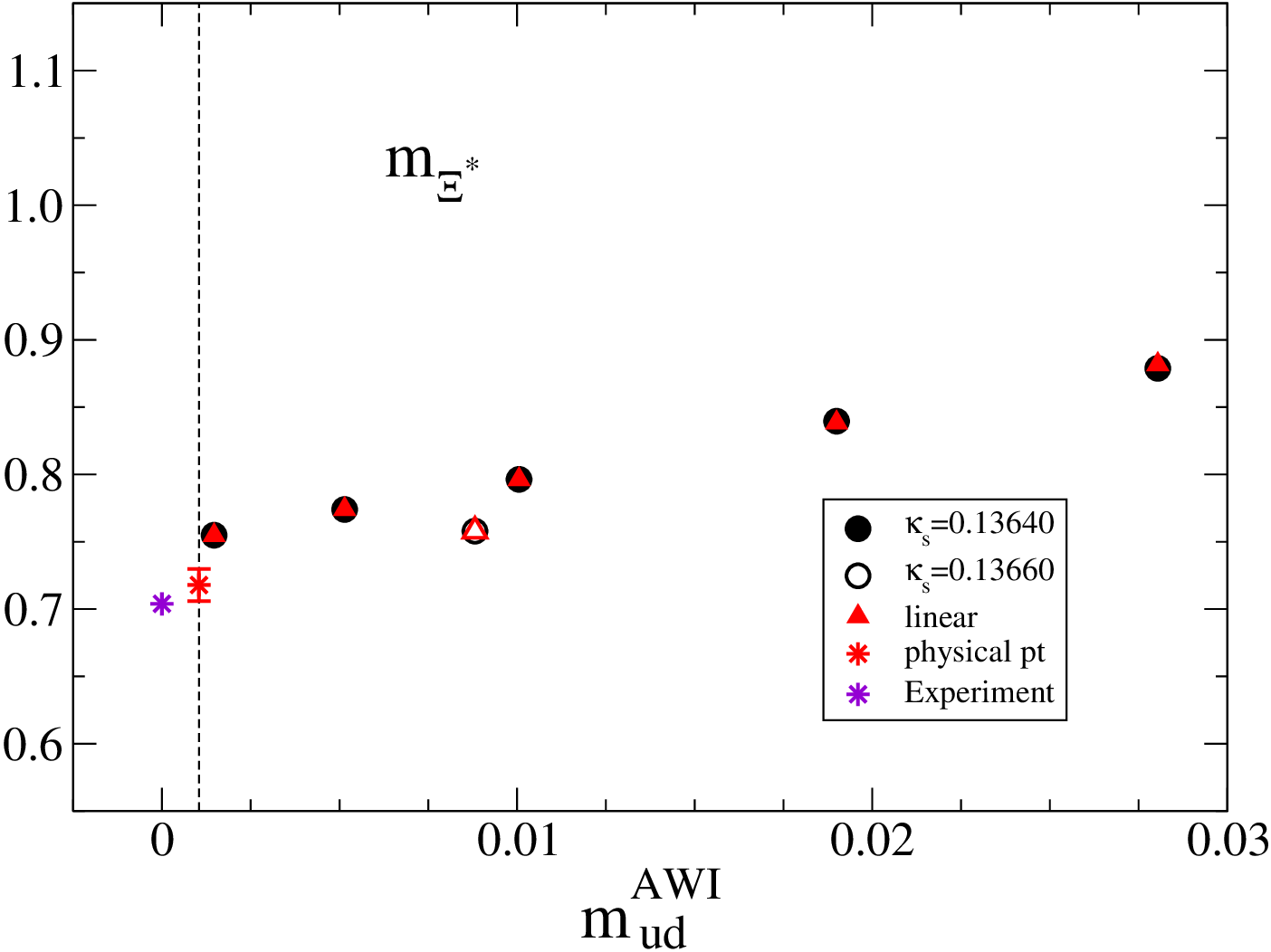}
\hspace*{2mm}&\hspace*{2mm}
\includegraphics[width=85mm,angle=0]{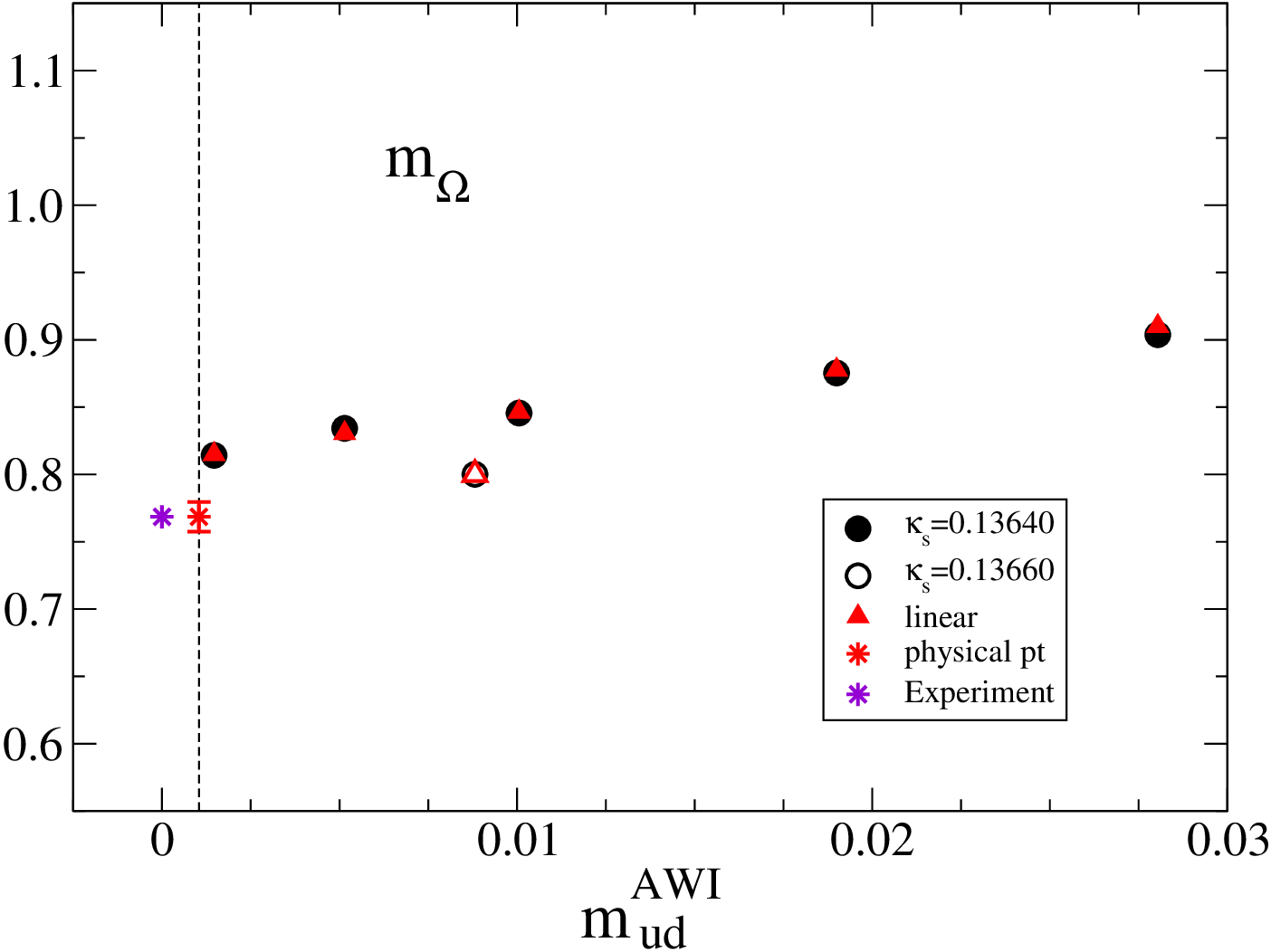}
\end{tabular}
\end{center}
\vspace{-.5cm}
\caption{Same as Fig.~\protect{\ref{fig:chexp_vector}} 
for the decuplet baryons.}
\label{fig:chexp_decuplet}
\end{figure*}

For the pseudoscalar meson sector, we employ SU(2) chiral expansion 
as explained in the previous Section.
For the vector mesons and the baryons we use a simple linear formula
$m_{\rm had}=\alpha_h+\beta_h m_{\rm ud}^{\rm AWI}+\gamma_h m_{\rm s}^{\rm
AWI}$, employing the data set in the same range   
$\kappa_{\rm ud}\ge 0.13754$ as for the pseudoscalar meson sector.
We do not rely on heavy meson effective 
theory (HMET)\cite{hmet} or heavy baryon ChPT (HBChPT)\cite{hbchpt}
since they show very poor convergences even 
at the physical point\cite{convergence}.
In Figs.~\ref{fig:chexp_vector},  \ref{fig:chexp_octet} and 
\ref{fig:chexp_decuplet},
we show linear chiral extrapolations
of the vector meson, the octet and the decuplet baryon masses,
respectively. Blue symbols represent the fit results at the measured values of
$m_{\rm ud}^{\rm AWI}$. The extrapolated values at the 
physical point are also denoted by blue star symbols, which should be
compared with the experimental values plotted at $m_{\rm ud}^{\rm AWI}=0$. 

Since the linear fit is applied to the data set at $\kappa_{\rm ud}\ge
0.13754$, blue symbols at  $\kappa_{\rm ud}< 0.13754$ express the
predictions from the fit results.
We observe that the quark mass dependence of $m_\Omega$ 
is remarkably well described by the linear function, which assures
that $m_\Omega$ is a good quantity for the physical input in the sense
that its chiral behavior is easily controlled.

The results for the physical quark masses and the lattice cutoff are listed in
Table~\ref{tab:physicalpt_qmass},
where the errors are statistical. They are provided
with and without the finite size corrections based on the NLO SU(2) ChPT
analyses. Both results are almost degenerate.
We find that our quark masses
are smaller than the estimates in the recent 2+1 flavor 
lattice QCD calculations\cite{rbcukqcd08,milc07}.  We note, however, 
that we employ the perturbative renormalization factors at one-loop level 
which should contain an uncertainty.  A nonperturbative calculation 
of the renormalization factor is in progress using the Schr\"odinger 
functional scheme. 

In Table~\ref{tab:physicalpt_qmass} we also present
the results for the pseudoscalar meson decay constants
at the physical point
using the physical quark masses and the cutoff determined above, 
which should be compared with the experimental values 
$f_\pi=130.7$~MeV, $f_K=159.8$~MeV, $f_K/f_\pi = 1.223$\cite{pdg}.
We observe a good consistency within the error of 2--3\%.  The
ratio is 3\% smaller than the experimental value in the case of the
SU(2) ChPT fit with the finite size corrections. 
A nonperturbative calculation of $Z_A$ is also in progress. 

\begin{figure*}[h]
\vspace{13mm}
\begin{center}
\begin{tabular}{cc}
\includegraphics[width=90mm,angle=0]{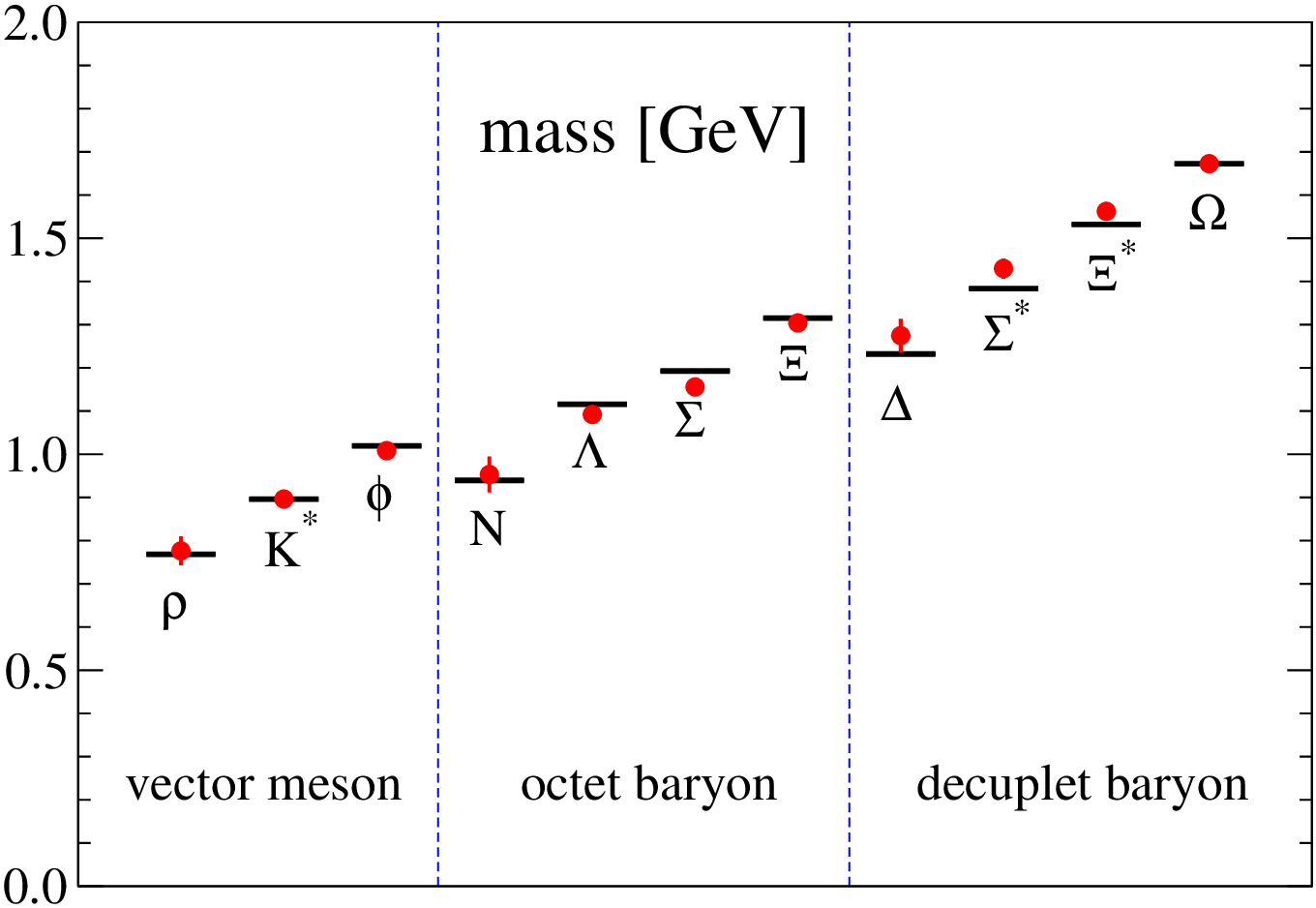} 
\end{tabular}
\end{center}
\vspace{-.5cm}
\caption{Light hadron spectrum extrapolated to the physical point 
using $m_\pi$, $m_K$ and $m_\Omega$ as input. 
Horizontal bars denote the experimental values.}
\label{fig:spectrum}
\end{figure*}

In Fig.~\ref{fig:spectrum} the light hadron spectrum 
extrapolated to the physical point using SU(2) ChPT with the finite
size corrections are compared with the experimental values.
Numerical values with and without the
finite size corrections are listed in Table~\ref{tab:physicalpt_hmass}.
The largest discrepancy between our results and the
experimental values is at most 3\%, albeit errors are still not small 
for the $\rho$ meson, the nucleon and the $\Delta$ baryon. 
The results are clearly encouraging, but  
further work is needed to remove the cutoff errors of $O((a\Lambda_{\rm QCD})^2)$.

\section{Static quark potential}
\label{sec:potential}

\begin{figure*}[h]
\vspace{13mm}
\begin{center}
\begin{tabular}{cc}
\includegraphics[width=90mm,angle=0]{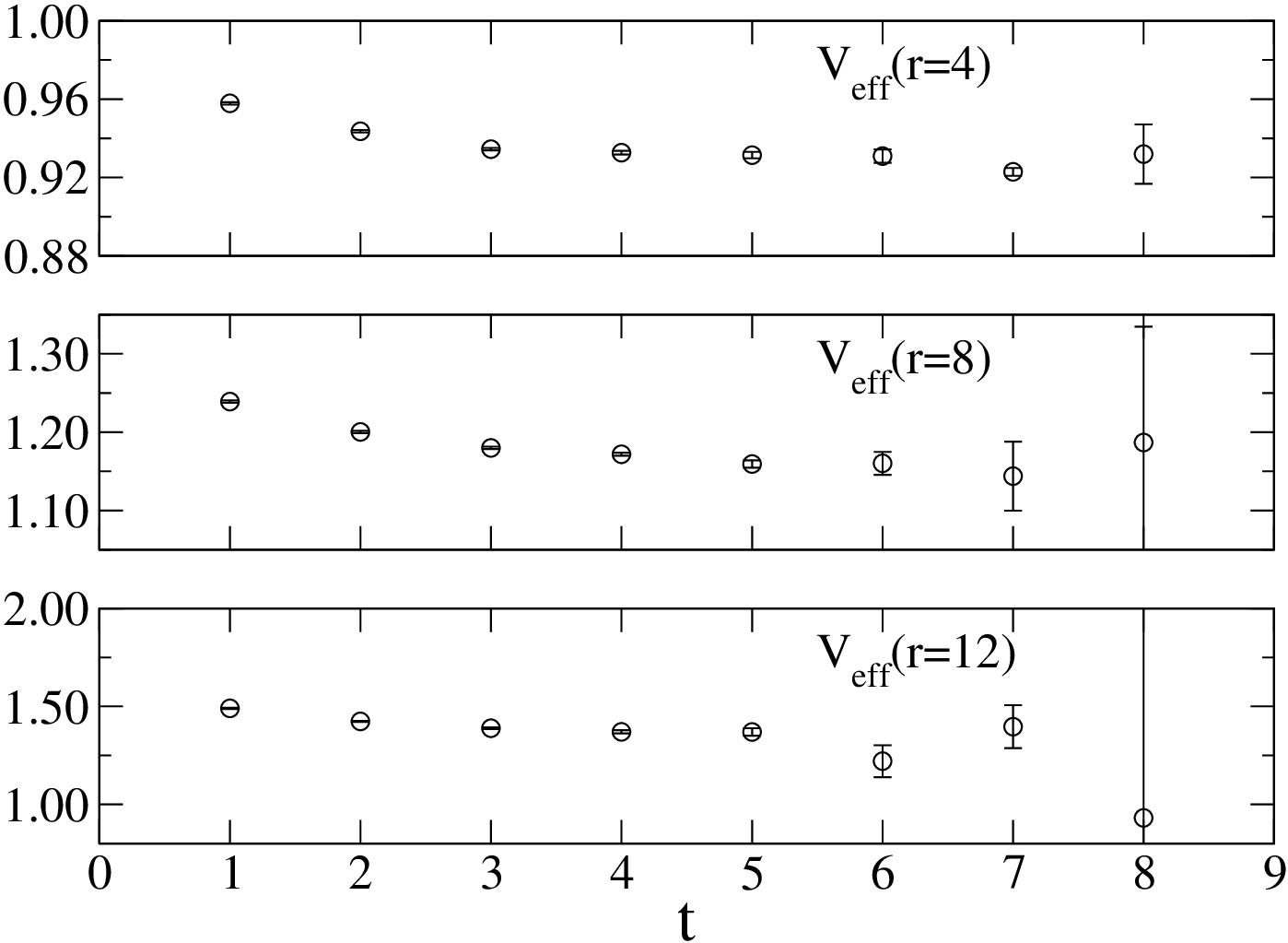}
\end{tabular}
\end{center}
\vspace{-.5cm}
\caption{Effective potential $V_{\rm eff}(r,t)$ with $r=4,8,12$ at
 $\kappa_{\rm ud}=0.13770$ as a representative case.}
\label{fig:v_eff}
\end{figure*}

\begin{figure*}[h]
\vspace{13mm}
\begin{center}
\begin{tabular}{cc}
\includegraphics[width=90mm,angle=0]{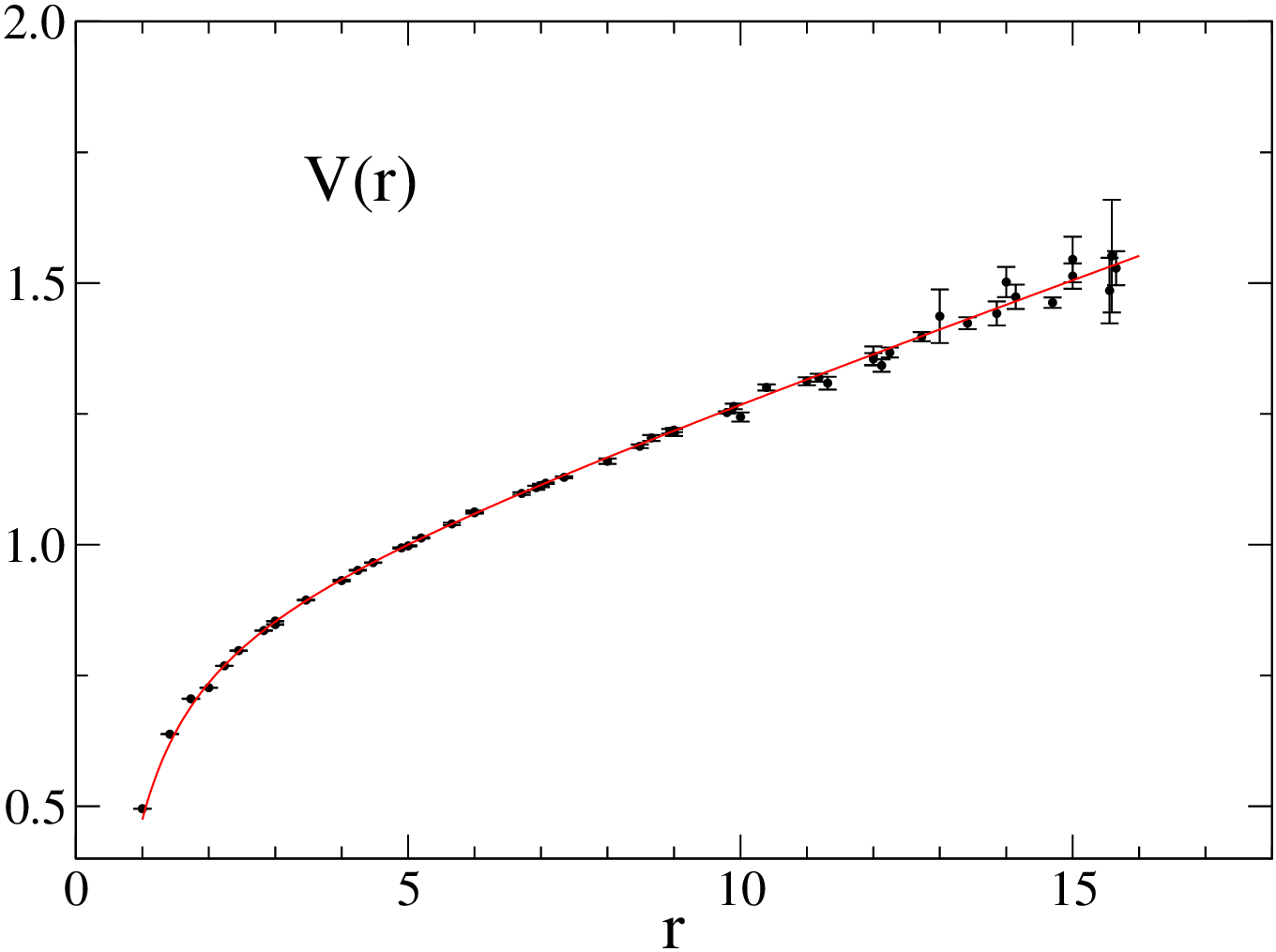}
\end{tabular}
\end{center}
\vspace{-.5cm}
\caption{Static quark potential $V(r)$ at
 $\kappa_{\rm ud}=0.13770$ as a representative case. Solid line
denote the fit result with Eq.~(\protect{\ref{eq:potential}}).}
\label{fig:potential}
\end{figure*}

In addition to the hadronic observables presented so far,
we also calculate the Sommer scale which is
a popular gluonic observable.
In order to calculate the static quark potential we measure 
the temporal and the spatial Wilson loops with the use of the
smearing procedure of Ref.~\cite{smear}.
The number of smearing steps is determined to be 20 after
examining the sufficient overlap of the Wilson loops onto the ground state. 
The potential $V(r)$ is extracted from the Wilson loops 
applying a correlated fit of the form 
\ben
W(r,t)=C(r)\exp(-V(r)t),
\een
where the same fitting range $[t_{\rm min},t_{\rm max}]=[5,8]$ is 
chosen for all the simulations after investigating the 
effective potential
\ben
V_{\rm eff}(r,t)=\ln\left[\frac{W(r,t)}{W(r,t+1)}\right].
\een
Figure~\ref{fig:v_eff} shows a typical case of $V_{\rm eff}(r,t)$ 
with $r=4,8,12$ at $\kappa_{\rm ud}=0.13770$.
We find that plateau starts at $t=4$ and signals are lost beyond $t=7$.
A result of $V(r)$ at $\kappa_{\rm ud}=0.13770$ is plotted
in Fig.~\ref{fig:potential} as a representative case. 
Since good rotational symmetry 
and no sign of the string breaking are observed, 
we employ the following fitting form for the potential:
\ben
V(r)=V_0-\frac{\alpha}{r}+\sigma r,
\label{eq:potential}
\een
where $V_0$, $\alpha$, $\sigma$ are unknown parameters.
The fitting range is $[r_{\min},r_{\max}] = [3,16]$.

The Sommer scale $r_0$ is a phenomenological quantity 
defined by
\ben
r_0^2=\left.\frac{dV(r)}{dr}\right\vert_{r=r_0}=1.65.
\een
Given Eq.~(\ref{eq:potential}) we obtain
\ben
r_0=\sqrt{\frac{1.65-\alpha}{\sigma}}.
\een
In Table~\ref{tab:potential} we list the results for $r_0$  
including the systematic errors due to the choices of $t_{\rm min}$ and 
$r_{\rm min}$. 

At $(\kappa_{\rm ud}, \kappa_{\rm s})=(0.13700, 0.13640)$, our 
result is compared with those of CP-PACS/JLQCD \cite{cp-pacs/jlqcd3} in 
Table~\ref{tab:r0comp}.  The two results are in reasonable agreement
given the sizable magnitude of systematic errors caused by 
the shortness of plateau of effective masses for potentials.

\begin{figure*}[h]
\vspace{13mm}
\begin{center}
\begin{tabular}{cc}
\includegraphics[width=90mm,angle=0]{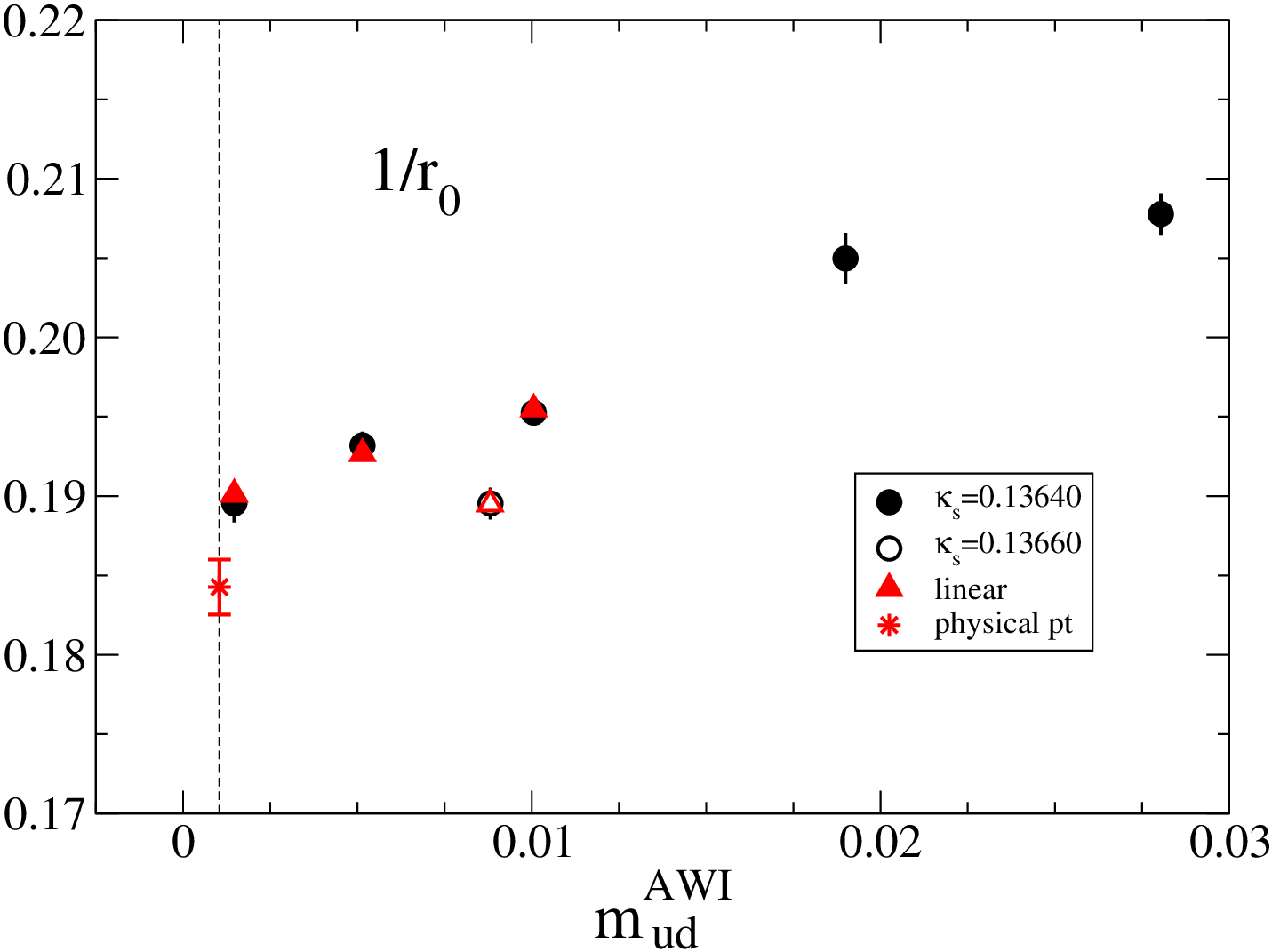}
\end{tabular}
\end{center}
\vspace{-.5cm}
\caption{Linear chiral extrapolation for $1/r_0$ at the physical point.
Red triangles denote the fit results
at the measured values of $m_{\rm ud}^{\rm AWI}$. }
\label{fig:chexp_r0}
\end{figure*}

In order to extrapolate $r_0$ to the physical point
we employ a linear form
$1/r_0=\alpha_r+\beta_r\cdot m_{\rm ud}^{\rm AWI}+\gamma_r\cdot m_{\rm s}^{\rm AWI}$
for the data set at $\kappa_{\rm ud}\ge 0.13754$.
We illustrate the chiral extrapolation in Fig.~\ref{fig:chexp_r0},
where the fit results are plotted by red triangles at the measured 
values of $m_{\rm ud}^{\rm AWI}$. 
The extrapolated result of $r_0$ at the physical point 
is 5.427(51)(+81)($-2$), which is
0.4921(64)(+74)($-2$) fm in physical units 
with the aid of $a^{-1}=2.176(31)$ GeV. The first error is statistical 
and the second and the third ones are the 
 systematic uncertainties originating from 
the choice of $t_{\rm min}$ and $r_{\rm min}$, respectively.

\section{Conclusion}
\label{sec:conclusion}

We have presented the first results of the PACS-CS project which aims at 
a 2+1 flavor lattice QCD simulation at the physical point
using the $O(a)$-improved Wilson quark action. 
The DDHMC algorithm, coupled with several algorithmic improvements, 
have enabled us to reach $m_{\pi}=156$~MeV,  which corresponds to 
$m_{\rm ud}^{\overline{\rm MS}}(\mu=2{\rm ~GeV})= 3.6$~MeV. 
We are almost on the physical point, except that the strange quark mass
is about 20\% larger than the physical value.

We clearly observe the characteristic features of the chiral logarithm 
in the ratios $m_\pi^2/m_{\rm ud}^{\rm AWI}$ and $f_K/f_\pi$. 
We find that our data are not well described by the NLO SU(3) ChPT, 
due to bad convergence of the strange quark contributions. 
We instead employ the NLO SU(2) ChPT for $m_\pi$ and
$f_\pi$, and an analytic expansion around the physical strange quark mass
for $m_K$ and $f_K$ in order to estimate the physical point. 
The low energy constants obtained in this way are compatible
with phenomenological estimates and other recent lattice calculations. 

Thanks to the enlarged physical volume 
compared to the previous CP-PACS/JLQCD work,
we obtain good signals not only for the meson masses 
but also for the baryon masses. 
After linear chiral extrapolations of the vector and baryon masses
the hadron spectrum at the physical point 
shows a good agreement with the experimental values,
albeit some of the hadrons have rather large errors and scaling violations 
remain to be examined.
We find smaller values for the physical quark masses 
compared to the recent estimates in the literature.
This may be due to the one-loop estimate of the renormalization factor.

At present the simulation at the physical point is under way, and
the statistics of the run at $\kappa_{\rm ud}=0.13781$ is being accumulated.
We are evaluating the nonperturbative renormalization factors
for the quark masses and the pseudoscalar meson decay constants in 
order to remove perturbative uncertainties.

Once these calculations are accomplished,
the next step is to investigate the finite size effects 
at the physical point,  and then to reduce the discretization errors by 
repeating the calculations at finer lattice spacings.

\begin{acknowledgments}
Numerical calculations for the present work have been carried out
on the PACS-CS computer 
under the ``Interdisciplinary Computational Science Program'' of 
Center for Computational Sciences, University of Tsukuba. 
We thank T. Sakurai and H. Tadano for a series of informative discussions 
on single precision acceleration of the solver.  
One of the authors (Y.K.) thank
A.~Kennedy for valuable discussions on the algorithmic improvements.
A part of the code development has been carried out on Hitachi SR11000 
at Information Media Center of Hiroshima University. 
This work is supported in part by Grants-in-Aid for Scientific Research
from the Ministry of Education, Culture, Sports, Science and Technology
(Nos.
16740147,   
17340066,   
18104005,   
18540250,   
18740130,   
19740134,   
20340047,   
20540248,   
20740123,   
20740139    
).
\end{acknowledgments}

\appendix
\newcommand{\PQP}{\mathrm{PQP}}
\newcommand{\QPQ}{\mathrm{QPQ}}
\newcommand{\SSOR}{\mathrm{SSOR}}
\newcommand{\SAP}{\mathrm{SAP}}
\newcommand{\KV}{\mathrm{KV}}
\newcommand{\spinpj}{\mathit{spin}}

\section{DDHMC algorithm}
\label{app:DDHMC}

  In Appendix~\ref{app:DDHMC}, we describe our implementation details of 
the L\"{u}scher's DDHMC algorithm\cite{luscher} employed for our 
$\kappa_{\rm ud} \le 0.13770$ runs.

\subsection{Domain decomposed HMC effective action}
In this work we employ the $O(a)$-improved Wilson fermions.
Before applying the domain-decomposition preconditioning for the quark determinant, 
we first apply Jacobi preconditioning to split the local clover term.
The $O(a)$-improved Wilson-Dirac operator $D$ is expressed as
\begin{equation}
  D = 1 + T + M,
\end{equation}
where $T$ is the local clover term, $M$ is the hopping term.
Jacobi preconditioning transforms the up-down quark determinant $|\det[D]|^2$ to
\begin{equation}
  |\det[D]|^2 = |\det[1+T]|^2 |\det[\tilde{D}]|^2
\end{equation}
where $\tilde{D}\equiv 1 + (1+T)^{-1} M = 1 + \tilde{M}$.
By splitting lattice sites into even and odd domains, $\tilde{D}$ has the following $2\times 2$
blocked matrix form,
\begin{equation}
  \tilde{D} = \left(
    \begin{array}{cc}
      \tilde{D}_{EE} & \tilde{D}_{EO} \\
      \tilde{D}_{OE} & \tilde{D}_{OO}
    \end{array}
  \right),
\end{equation}
where the suffix $E$ ($O$) means the even (odd) domain.
Applying the domain decomposition preconditioning for this form, we obtain
\begin{eqnarray}
  |\det[D]|^2 &=& |\det[1+T]|^2 |\det[\tilde{D}_{EE}]|^2 |\det[\tilde{D}_{OO}]|^2\nonumber\\
   && \times |\det[\hat{D}_{EE}]|^2,
\end{eqnarray}
where $\hat{D}_{EE}$ is the Schur complement of $\tilde{D}$ and expressed as
\begin{equation}
\hat{D}_{EE}=1 - (\tilde{D}_{EE})^{-1} \tilde{D}_{EO}(\tilde{D}_{OO})^{-1} \tilde{D}_{OE}.
\end{equation}
Our domain decomposition is based on the four dimensional checkerboard coloring.

The operator $\hat{D}_{EE}$ can be further preconditioned by the spin and hopping structure 
because $\tilde{D}_{EO}$ ($\tilde{D}_{EO}$) only connects domain surface sites. 
Let $P^{\spinpj}_{E}$ ($P^{\spinpj}_{O}$) be the spin and site projection operator to the even (odd)
domain sites,
\begin{widetext}
\begin{equation}
P^{\spinpj}_{E} \psi_n = \left\{
\begin{array}{cc}
\displaystyle
0 & \mbox{if $n$ is located on the bulk site of the even domain},\\
\frac{1}{2}(1+\gamma_{\mu}) \psi_n & 
\mbox{if $n$ is located in the even domain and $n+\hat{\mu}$ is in the odd domain with one value of $\mu$ only},\\
\frac{1}{2}(1-\gamma_{\mu}) \psi_n & 
\mbox{if $n$ is located in the even domain and $n-\hat{\mu}$ is in the odd domain with one value of $\mu$ only},\\
\displaystyle
\psi_n & 
\mbox{otherwise}.
\end{array}\right.
\end{equation}
\end{widetext}
This projection operator satisfies the following relations.
\begin{eqnarray}
(P^{\spinpj}_{E})^2 &=& P^{\spinpj}_{E},\\
\tilde{D}_{OE} &=& \tilde{D}_{OE} P^{\spinpj}_{E},
\end{eqnarray}
and the same relation holds for the odd domain case.
With these properties, $\hat{D}_{EE}$ satisfies
\begin{equation}
\hat{D}_{EE} = 1-P^{\spinpj}_{E} + \hat{D}_{EE}P^{\spinpj}_{E}.
\end{equation}
This means that $\hat{D}_{EE}$ is a triangular matrix in view of 
the projection space.  Thus we have
\begin{equation}
\det[\hat{D}_{EE}] = \det[P^{\spinpj}_{E}\hat{D}_{EE}P^{\spinpj}_{E}],
\end{equation}
where the matrix dimension of the operator 
$P^{\spinpj}_{E}\hat{D}_{EE}P^{\spinpj}_{E}$ is effectively reduced. 
We define 
\begin{equation}
\hat{D}^{\spinpj}_{EE} \equiv P^{\spinpj}_{E}\hat{D}_{EE}P^{\spinpj}_{E},
\label{eq:IRspprojop}
\end{equation}
for the reduced operator.

Since the domain block lattice extent we use is 8 and is an even number,
the domain restricted operator $\tilde{D}_{EE}$ ($\tilde{D}_{OO}$) 
can be further preconditioned by the even-odd site preconditioning 
which is widely used for full lattice case in the literature.
\begin{subequations}
\begin{eqnarray}
\det[\tilde{D}_{EE}] &=& \det[(\hat{D}_{EE})_{ee}],\\
(\hat{D}_{EE})_{ee} &=&
 1 - (\tilde{M}_{EE})_{eo} (\tilde{M}_{EE})_{oe},
\end{eqnarray}
\end{subequations}
where the suffices $eo$ and $oe$ mean hopping from an odd-site to 
an even-site and {\it vice versa}.
For the odd domain operator $\tilde{D}_{OO}$ the same relation exists.
Our even-odd site preconditioning is based on the four dimensional checkerboard
coloring again.

After applying all these preconditioning we obtain the following 
lattice QCD partition function for degenerate up-down quarks.
\begin{widetext}
\begin{subequations}
\begin{eqnarray}
  {\cal Z} &=& \int 
  {\cal D} P
  {\cal D} U
  {\cal D}\phi^{\dag}_{Ee} {\cal D}\phi_{Ee}
  {\cal D}\phi^{\dag}_{Oe} {\cal D}\phi_{Oe}
  {\cal D}\chi^{\dag}_{E}  {\cal D}\chi_{E}
 e^{-H[P,U,\phi_{Ee},\phi_{Oe},\chi_{E}]},\\
 H[P,U,\phi_{Ee},\phi_{Oe},\chi_{E}]&=&
   \frac{1}{2}\Tr[P^2]+S_g[U]+S_{\mathrm{clv}}[U]
   +\sum_{X=E,O}S_{q\mathrm{UV},X}[U,\phi_{Xe}]
   +S_{q\mathrm{IR}}[U,\chi_{E}],
\end{eqnarray}
\end{subequations}
\end{widetext}
where $P$ is the canonical momenta for $U$, $S_g[U]$ the gauge action, and
\begin{subequations}
\begin{eqnarray}
S_{\mathrm{clv}}[U] &=& -2 \log[\det[(1+T)]],\\
S_{q\mathrm{UV},X}[U,\phi_{Xe}] &=& \left|((\hat{D}_{XX})_{ee})^{-1}\phi_{Xe}\right|^2,
\label{eq:UVaction}\\
S_{q\mathrm{IR}}[U,\chi_{E}] &=& \left|(\hat{D}^{\spinpj}_{EE})^{-1}\chi_{E}\right|^2,
\label{eq:IRaction}
\end{eqnarray}
\end{subequations}
where $\chi_{E}$ is projected so as to satisfy $P^{\spinpj}_{E}\chi_{E} = \chi_{E}$.
Our DDHMC algorithm is based on this partition function and the
UVPHMC algorithm for strange quark is simply added to this form.

\subsection{Multi time scale molecular dynamics integrator}
 We employ the Sexton-Weingarten multiple time scale molecular dynamics (MD) 
integrator\cite{sexton}.
The ordering to evolve link variables and momenta is arbitrary 
in the simple leap-frog integrator,
and it is known that the so-called QPQ-ordering has better performance than that of 
the PQP-ordering\cite{UPUvsPUP,forcrandtakaishi,phmc}.
While an actual performance comparison is not made systematically,
this leads us to implement the QPQ-ordered multi time step integrator
expecting better performance.

Suppose that there is a Hamiltonian $H$ expressed as a sum of $N$ potentials:
\begin{equation}
  H = T(p) + \sum^{N-1}_{i=0} V_i(q),
\end{equation}
where $q$ represents dynamical variables, $T(p)$ the kinetic term
$p^2/2$, and $p$ is the canonical momenta.
This leads to the following equation of motion:
\begin{subequations}
\begin{eqnarray}
  \dot{q} &=& p = -\{H,q\}_P, \\
  \dot{p} &=& F = -\{H,p\}_P = \sum^{N-1}_{i=0} F_i, \\
  F_i & = & - \frac{\partial V_i}{\partial q} = -\{V_i,p\}_P,\\
  \{X,Y\}_P &=& \frac{\partial X}{\partial q}\frac{\partial Y}{\partial p}
              -\frac{\partial X}{\partial p}\frac{\partial Y}{\partial q},
\end{eqnarray}
\end{subequations}
where $\{X,Y\}_P$ is Poisson bracket, and the dot $\dot{\ }$ 
is the abbreviation for the time derivative $d/d\tau$.
The formal solution is written as
\begin{equation}
  \left(
   \begin{array}{c}
     q \\
     p \\
   \end{array}
  \right)(\tau) = \exp\{\tau \hat{L}_{H}\}
  \left(
   \begin{array}{c}
     q \\
     p \\
   \end{array}
  \right)(0),
\end{equation}
where $\exp\{\tau \hat{L}_{H}\}$ is the exponentiation of 
the Liouvillean $\hat{L}_H X=-\{H,X \}_{P}$.
In our case $\hat{L}_H X = \hat{L}_T X + \sum_{i=0}^{N-1} \hat{L}_{V_i} X$, 
where $\hat{L}_T X= - \{T,X\}_P$ and $\hat{L}_{V_i} X= - \{V_i,X\}_P$.
We assume that the numbering of the potential $V_i$ is ordered so as to 
satisfy $|F_i|<|F_{i-1}|$.
Any molecular dynamics integrator is an approximation/decomposition of
the operator exponential $\exp\{\tau \hat{L}_{H}\}$ using 
$\exp\{\tau \hat{L}_{T}\}$ and $\exp\{\tau \hat{L}_{V_i}\}$.

To explain symplectic molecular dynamics integrators we introduce the following mapping:
\begin{subequations}
\begin{eqnarray}
  Q(\delta\tau) \equiv \exp\{\tau \hat{L}_{T}\}
\!\! &:& \!\!
 (q,p) \rightarrow (q+ \delta \tau p,  p),\\
P_i(\delta\tau) \equiv \exp\{\tau \hat{L}_{V_i}\}
\!\! &:& \!\!
 (q,p) \rightarrow (q, p+\delta \tau F_i).
\end{eqnarray}
\end{subequations}
Using these operators we can derive the following multi time scale integrators.
\begin{widetext}
\paragraph{PQP-ordering}
The PQP-ordered multi time step integration operator $S^{\PQP}(\tau)$ is defined as
\begin{equation}
S^{\PQP}(\tau,\left(N_0,N_1,\ldots,N_{N-1}\right))  = 
S^{\PQP}_{N-1}(\tau,\left(N_0,N_1,\ldots,N_{N-1}\right)),
\end{equation}
where $S^{\PQP}_{N-1}$ is recursively defined as
\begin{eqnarray}
 S^{\PQP}_i(\tau,\left(N_0,N_1,\ldots,N_{i}\right)) &\equiv& 
\left[       P_i\left(\frac{\tau}{2N_i}\right)
   S^{\PQP}_{i-1}\left(\frac{\tau}{ N_i},\left(N_0,N_1,\ldots,N_{i-1}\right)\right)
             P_i\left(\frac{\tau}{2N_i}\right)\right]^{N_i},\nonumber\\
 S^{\PQP}_0(\tau,N_0) &\equiv& 
\left[ P_0\left(\frac{\tau}{2N_0}\right)
         Q\left(\frac{\tau}{ N_0}\right)
       P_0\left(\frac{\tau}{2N_0}\right)\right]^{N_0},
\end{eqnarray}
where $N_i$ is the step number for each time scale.
The momentum is updated by $\delta \tau_i F_i$ with $\delta \tau_i=\tau/(\prod_{j=0,i} N_j)$ at
depth $i$.

\paragraph{QPQ-ordering}
The QPQ-ordered multi time step integrator is used for our productive runs.
The QPQ-ordered integrator $S^{\QPQ}$ is defined as
\begin{equation}
S^{\QPQ}(\tau,\left(N_0,N_1,\ldots,N_{N-1}\right))  = 
S^{\QPQ}_{N-1}(\tau,\left(N_0,N_1,\ldots,N_{N-1}\right)),
\end{equation}
where $S^{\QPQ}_{N-1}$ is recursively defined as
\begin{eqnarray}
 S^{\QPQ}_i(\tau,\left(N_0,\ldots,N_{i}\right)) &\equiv& 
\left[    
 S^{\QPQ}_{i-1}\left(\frac{\tau}{ N_i},\left(N_0,\ldots,N_{i-1}\right)\right)
           P_i\left(\frac{\tau}{ N_i}\right)
 S^{\QPQ}_{i-1}\left(\frac{\tau}{ N_i},\left(N_0,\ldots,N_{i-1}\right)\right)
 \right]^{\frac{N_i}{2}(1+\delta_{i,N-1})},
 \nonumber\\
 S^{\QPQ}_0(\tau,N_0) &\equiv& 
\left[   Q\left(\frac{\tau}{2N_0}\right) 
       P_0\left(\frac{\tau}{ N_0}\right)
         Q\left(\frac{\tau}{2N_0}\right)\right]^{\frac{N_0}{2}}.
\end{eqnarray}
In this case the division numbers, $N_i$,
should be chosen from even numbers except for the outermost division number $N_{N-1}$.

\end{widetext}

The integrator described above is based on the nesting of 
the simple leap-frog integrator.
We also note that we have not yet tried the so-called Omelyan 
integrator\cite{Omelyan,forcrandtakaishi} for the recurrence kernel,
albeit it is generally known to be a better scheme and may be used for our case. 
The multi time step integrator with the Omelyan kernel has been used
in Refs.~\cite{Jung:2007dj,rbcukqcd08}.

\subsection{UV part solver}
The UV part of the HMC algorithm is governed by the action Eq.~(\ref{eq:UVaction}).
This contains the inversion of $(\hat{D}_{EE})_{ee}$ and  $(\hat{D}_{OO})_{ee}$.
In our parallel implementation of the algorithm, each block lattice is
completely contained in a single node.
This means that there is no ghost site exchange for multiplying $\tilde{D}_{EE}$.
In this case SSOR preconditioning with natural site ordering is more efficient than 
the even-odd site preconditioning\cite{SSORprec}.

  We solve the linear equation 
\begin{equation}
  (\hat{D}_{EE})_{ee} x_e = b_e,
  \label{eq:UVlineqeo}
\end{equation}
using SSOR preconditioned GCR solver where 
$x_e$ and $b_e$ carry the even-site data in the even-domain.
We implemented the SSOR preconditioner with single precision arithmetic.

To solve Eq.~(\ref{eq:UVlineqeo}) with an SSOR preconditioner,
we transform Eq.~(\ref{eq:UVlineqeo}) back to the unpreconditioned form,
\begin{subequations}
\begin{eqnarray}
\tilde{D}_{EE} y &=& c, 
\label{eq:UVlineqorg}
  \\
  c &=& \left(
  \begin{array}{c}
    b_e\\
    0
  \end{array}
  \right), \\
  x_e &=& y_e,
\end{eqnarray}
\end{subequations}
where $y_e$ is the even-site components of the full even-domain vector $y$.
The right hand vector $c$ has zero for the odd-site components and has
$b_e$ for the even-site components.

We make use of the block/domain independence among the computational nodes
and matrix structure of the domain operator $\tilde{D}_{EE}$ to solve 
Eq.~(\ref{eq:UVlineqorg}).
With the natural site-ordering in each block, $\tilde{D}_{EE}$ can be decomposed as
\begin{equation}
\tilde{D}_{EE} = 1 - L - U,
\end{equation}
where $L$ is the forward hopping term and $U$ is the backward hopping term.
The $L$ and $U$ are strictly triangular for the natural site 
ordering because of the Dirichlet boundary condition for each block in a domain.
Eq.~(\ref{eq:UVlineqorg}) is solved by
\begin{subequations}
\begin{eqnarray}
&&d =  (1-\omega L)^{-1} c, \\
&&(\hat{D}_{EE})_{\SSOR} M_{\SSOR} z = d, 
\label{eq:UVSSORlineq}
\\
&&y =  (1-\omega U)^{-1}  M_{\SSOR} z,
\end{eqnarray}
\end{subequations}
where $\omega$ is an over-relaxation parameter to be tuned, 
$(\hat{D}_{EE})_{\SSOR}$ and $M_{\SSOR}$ are defined by
\begin{eqnarray}
\label{eq:UVSSORop}
(\hat{D}_{EE})_{\SSOR} &=& 
\frac{1}{\omega}\left[(1-\omega L)^{-1}
                     +(1-\omega U)^{-1}\right. \nonumber\\
 && \left.
           +(\omega-2)(1-\omega L)^{-1}(1-\omega U)^{-1}\right],\nonumber\\
&&\\
M_{\SSOR} &=& \sum_{j=0}^{N_{\SSOR}}\left[\left(1- (\hat{D}_{EE})_{\SSOR}\right)^{j}\right]_{\mathrm{32bit}}.
\label{eq:UVSSORprec}
\end{eqnarray}

The preconditioner $M_{\SSOR}$ is computed in single precision, and 
Eq.~(\ref{eq:UVSSORlineq}) is solved using GCR solver to double precision.
The inverses $(1-\omega L)^{-1}$ and $(1-\omega U)^{-1}$ are easily solved by forward 
and backward substitutions, and Eq.~(\ref{eq:UVSSORop}) is computed using the Eisenstat trick.
The parameter $\omega$ is tuned to $\sim 1.2$ and $N_{\SSOR}$ to $5\sim 10$ 
to achieve optimal performance. 
The maximal Krylov subspace dimension $N_{\KV}$ for GCR solver is chosen
to avoid frequent restarting and residual stagnation, and
our experience tells that $N_{\KV} \sim O(10)$ is sufficient.

\subsection{IR part solver}

The IR part of the DDHMC algorithm contains the linear equation as
\begin{equation}
  \hat{D}^{\spinpj}_{EE} x_E = b_E,
  \label{eq:IRlineq}
\end{equation}
where $\hat{D}^{\spinpj}_{EE}$ is defined by Eq.~(\ref{eq:IRspprojop}).
Eq.~(\ref{eq:IRlineq}) is solved with the restrictions $(1-P^{\spinpj}_E)x_E=0$
and $(1-P^{\spinpj}_E)b_E=0$. 

As described in Ref.~\cite{luscher}, directly solving Eq.~(\ref{eq:IRlineq})
is rather slow because the operator $\hat{D}^{\spinpj}_{EE}$ contains
the domain inversions $(\tilde{D}_{EE})^{-1}$ and $(\tilde{D}_{OO})^{-1}$ with
double precision.
Instead of solving Eq.~(\ref{eq:IRlineq}), the solution $x_E$ can be expressed 
using the unpreconditioned operator $\tilde{D}$ as
\begin{subequations}
\begin{eqnarray}
  x_E &=& P_E^{\spinpj} (b_E - y_E),\\
  \tilde{D} y &=& w,
  \label{eq:IRlinequnp} \\
  w &=& \left(
  \begin{array}{c}
     0 \\
     \tilde{D}_{OE} b_E
  \end{array}
  \right),
\end{eqnarray}
\end{subequations}
where $y_E$ is the even domain component of $y$, and $w$ is 
the full lattice vector for which the even domain components are set to zero.

Eq.~(\ref{eq:IRlinequnp}) is efficiently solved with the GCR-SAP solver\cite{sap+gcr}
via
\begin{subequations}
\begin{eqnarray}
  \tilde{D} M_{\SAP} z &=& w, \\
  y &=& M_{\SAP} z.
\end{eqnarray}
\end{subequations}
The SAP preconditioner $M_{\SAP}$ is computed in single precision as
\begin{equation}
   M_{\SAP} = \left[K \sum_{j=0}^{N_{\SAP}} (1- \tilde{D}K)^{j}\right]_{\mathrm{32bit}},
\end{equation}
where
\begin{subequations}
\begin{eqnarray}
   K &=& \left(
      \begin{array}{cc}
        A_{EE} & 0 \\
       -A_{OO} \tilde{D}_{OE} A_{EE} & A_{OO}
      \end{array}
   \right), \\
   A_{EE} & = & (1-\omega U)^{-1} M_{\SSOR}(1-\omega L)^{-1},
\end{eqnarray}
\end{subequations}
and $A_{OO}$ similar to $A_{EE}$. 
The operator $A_{EE}$ ($A_{OO}$) is the approximation for 
$(\tilde{D}_{EE})^{-1}$ ($(\tilde{D}_{OO})^{-1}$) via the SSOR fixed iteration
$M_{\SSOR}$ defined in Eq.~(\ref{eq:UVSSORprec}).

Thus the solver for Eq.~(\ref{eq:IRlineq}) contains several tunable parameters;
$\omega$, $N_{\SSOR}$, $N_{\SAP}$, and $N_{\KV}$ the maximal Krylov subspace dimension for GCR.
We observed that $\omega \sim 1.2$, $N_{\SSOR}=1$, $N_{\SAP} = 10 \sim 20$ and 
$N_{\KV}=40\sim 100$ show satisfactory performance.

\subsection{Dead/alive link method}

L\"{u}scher's DDHMC algorithm was originally proposed for 
the plaquette Wilson gauge action and the unimproved Wilson fermion\cite{luscher}.
He restricted the link variables evolved by the MD integrator to a subset. 
The link variables which connect the domain interfaces and 
are located parallel to the domain surfaces are kept fixed during the MD evolution 
(dead links), and only the remaining bulk links are evolved (alive links).
The choice of the set of dead links are dictated by the condition that the alive links 
are decoupled.
The method has the benefit that if the layout of the domain decomposition 
is properly matched to the compute node location there is no need to  
exchange link data during the MD evolution.
Thus the algorithm becomes a semi-local update algorithm. 
To ensure the ergodicity a random parallel translation of the lattice 
coordinate origin is required after each HMC evolution.  

In our case we employed the Iwasaki-gauge action and the $O(a)$-improved 
Wilson fermion. 
These actions have a larger lattice extent compared to the unimproved 
action (the rectangular part of the gauge action and the clover term of the fermion action),
and one may worry about the semi-locality of the MD evolution.
Since the extension is still within two sites, we can conclude that
the dead links are still domain connecting ones and those on the thin surface 
of the domains.
Thus we can apply the same dead/alive link method as that for the unimproved case.
For more extended gauge or fermion action the number of dead links
should be enlarged to decouple the active links.

The efficiency of the dead/alive link method depends on the ratio of the number 
of active links and all links, which is estimated as
\begin{equation}
  (N_{B}-1)(N_{B}-2)^{3}/N_{B}^4
\end{equation}
where $N_{B}$ is a domain block size assuming $N_B^4$ blocking.
In this paper we employed $N_B=8$, which results in $\sim$37\% for the ratio.
We employed the same algorithm for the random parallel translation as in Ref.~\cite{luscher}.

\section{Mass preconditioned DDHMC (MPDDHMC) algorithm}
\label{app:MPDDHMC}

As the up-down quark mass is reduced 
toward the physical point, we observed strong MD instability with 
the DDHMC algorithm.
The origin of the instability is the appearance of near zero or 
negative eigenvalues in the $\tilde{D}$ spectra\cite{Joo:2000dh,Namekawa:2004bi}.
The corresponding eigenmodes yield a strong MD force
and large fluctuations for the IR action (\ref{eq:IRaction}) 
as described in the main text.
We could handle the instability by reducing $\delta \tau_{IR}$. 
However this results in very high values of the HMC acceptance, {\it e.g.,}
$\agt 90\%$,  which is unnecessarily large compared with the optimal acceptance ratio,  
{\it e.g.,} $\sim$ 60--70\% for 2nd order MD integrator\cite{takaishiMD}.

We introduce Hasenbusch's heavy mass preconditioner\cite{massprec1,massprec2}
to stabilize the IR part (\ref{eq:IRaction}), and call the resulting algorithm 
MPDDHMC algorithm.
We also implement several improvements in the algorithm.
Our simulation with the lightest up-down quark mass corresponding to $\kappa_{ud}=0.13781$ 
is finally carried out by the MPDDHMC algorithm.
Here we describe the implementation details of the MPDDHMC algorithm.

\subsection{Hasenbusch's heavy mass preconditioning for DDHMC algorithm}
The mass preconditioner is introduced for the IR part action Eq.~(\ref{eq:IRaction}).
The action is transformed and split into two pieces as
\begin{eqnarray}
 \left|\det[\hat{D}^{\spinpj}_{EE}]\right|^2 &=& 
\left|\det\left[\frac{\hat{D}^{\spinpj}_{EE}}{\hat{D'}^{\spinpj}_{EE}}\right]\right|^2 
\left|\det[\hat{D'}^{\spinpj}_{EE}]\right|^2 \nonumber \\
&=&\left|\det[R_{EE}]\right|^2
   \left|\det[\hat{D'}^{\spinpj}_{EE}]\right|^2,
\end{eqnarray}
where the primed operator $\hat{D'}^{\spinpj}_{EE}$ is defined with the
modified hopping parameter $\kappa' \equiv \rho \kappa$ keeping 
the clover term unchanged.
Introducing the pseudo-fermion fields for each determinant we obtain
\begin{subequations}
\begin{eqnarray}
\left|\det[\hat{D}^{\spinpj}_{EE}]\right|^2 &=& \int
 {\cal D}\chi^{\dag}_E {\cal D}\chi_E
 {\cal D}\zeta^{\dag}_E {\cal D}\zeta_E \nonumber\\
&& \times
  e^{-S_{q\tilde{\mathrm{IR}}}[U,\zeta_E]-S_{q\mathrm{IR'}}[U,\chi_E]},\\
S_{q\tilde{\mathrm{IR}}} &=& \left|(R_{EE})^{-1} \zeta_E   \right|^2,
\label{eq:IRIRaction}
\\
S_{q\mathrm{IR'}} &=& \left|(\hat{D'}^{\spinpj}_{EE})^{-1} \chi_E \right|^2.
\label{eq:IRprimeaction}
\end{eqnarray}
\end{subequations}
The action Eq.~(\ref{eq:IRaction}) is replaced by Eqs.~(\ref{eq:IRIRaction}) 
and (\ref{eq:IRprimeaction}).
Our MPDDHMC algorithm is based on this action.

The parameter $\rho$ is a tunable parameter and should be chosen close to but less than 
$\rho =1$ while keeping $R\sim 1$ so as to achieve optimal performance.
For example, since the DDHMC simulation at $\kappa=0.13770$ ran successfully,
we use $\rho=0.9995$ at $\kappa_{\rm ud}=0.13781$ since we expect 
$\kappa' \sim 0.13770$ would lead to a stabilized behavior 
for Eq.~(\ref{eq:IRprimeaction}).

\subsection{Solver improvements}

As the quark masses are taken small, we encountered a solver stagnation or
failure due to the presence of the near zero modes or the negative (real part) eigenmodes.
In this case the GCR-SAP solver sometimes does not converge.
Although this difficulty could be cured by changing the solver algorithm
or finely tuning the solver parameters, {\it i.e.} $\omega$, $N_{\SSOR}$, {\it etc.},
applying such remedies causes violation of 
the reversibility of the MD evolution when 
a loose stopping condition is adopted for the solver.

To avoid this situation we decided to change the solver algorithm.
Our strategy is to combine (1) a strict stopping condition, 
(2) applying the method of chronological guess, and 
(3) adopting a solver algorithm robust against near zero and negative eigenvalues.
The use of strict stopping condition (1) gives us room to flexibly 
change the solver algorithm without the reversibility violation, 
although this adds an extra computational cost.
A part of the extra cost can be reduced by optimizing the choice of the initial vector 
(the chronological guess method\cite{chronological}). 
It is also required to adopt a solver algorithm which is 
robust and fast against the ill conditioned case.
We employ the inner-outer solver strategy and the deflation technique\cite{deflWilcox,deflMorgan2,deflluscher}
aiming for speed up and taming the difficult eigenmodes.

\subsubsection{Inner-outer strategy}

  The gap between the rapidly increasing floating point capability of processors 
and the memory bandwidth is spreading because 
of the rather slow development of memory speed.
To fill the gap the mixed precision or the inner-outer nested solver strategy 
has been proposed\cite{mixedprec}. 
The outer solver must have the property that the preconditioner can be changed 
from iteration to iteration. 
Since the preconditioner can be replaced by another iterative solver
to make an approximation for the outer problem, 
the preconditioner can be called as the inner-solver for the outer solver.
The inner-outer solver enables us the use of single precision which effectively
doubles the memory bandwidth, data cache size, and processor registers\cite{Durr:2008rw}.
The GCR-SAP solver proposed by L\"{u}scher\cite{sap+gcr} is also along this strategy.
If the solver parameters can be chosen such that most of the computational time is spent 
in the inner-solver, we receive a maximal benefit from the use of single precision 
arithmetic\cite{Durr:2008rw}.

In this work we developed a version of the BiCGStab algorithm which enables us 
to follow the inner-outer strategy\cite{SakuraiTadano}. 
The benefit of BiCGStab compared to GCR (or GMRES) type algorithms
is that BiCGStab has a shorter recurrence iteration, small memory requirement, 
and no restarting.
To make the BiCGStab solver flexible against substitutions of the preconditioner, 
we slightly modify the algorithm. The point of modification is the following.

Any solver algorithm which has the following update point for the solution 
and residual vector can be modified to take the inner-outer solver form.
To solve $ A x = b$, suppose that an algorithm has the lines
\begin{subequations}
\begin{eqnarray}
\mbox{[compute parameter} &\alpha & \mbox{and pre-search vector $p$.]}\nonumber\\
   q & = & A p, \\
   r & = & r - \alpha q, \\
   x & = & x + \alpha p,
\end{eqnarray}
\end{subequations}
where the method to obtain $\alpha$ and $p$ depends on the outer solver algorithm.
To enable a flexible preconditioner replace these lines as
\begin{subequations}
\begin{eqnarray}
\mbox{[compute parameter} &\alpha & \mbox{and pre-search vector $p$.]}\nonumber\\
   v & = & M p, \\
   q & = & A v, \\
   r & = & r - \alpha q, \\
   x & = & x + \alpha v,
\end{eqnarray}
\end{subequations}
where $M$ is a preconditioner and must be an approximation for $A^{-1}$.
The extra vector $v$ is required to hold an intermediate vector.
In this modification the search vector $q$ is produced for the equation $AM y = b$,
while the solution still keeps the solution-residual relation $r = b - A x$ 
of the unpreconditioned equation. 
The preconditioner $M$ can be changed from iteration to iteration in the outer solver,
as far as the the solution-residual relation is kept intact.
In this way, a flexible preconditioner can be introduced for the outer algorithm.

This modification is applicable to many solvers which have similar
local update points (CG, MR, CGS, {\it etc.}).
The iterative refinement or Richardson iteration\cite{Durr:2008rw} is the simplest 
example. Solvers of GMRES type is also modified along this strategy.  
Longer recurrence relations of those algorithms require a series of extra vectors such as 
the above vector $v$ (for ex. GCR, FGMRES), however.

We implement this modification to the BiCGStab solver and replace two update points.
The preconditioner $M$ is replaced by the single precision solver for $Ax=b$ 
with the appropriate precision conversion interface (single to double and {\it vice versa}).
The tolerance of the inner solver can be relaxed as the outer residual approaches
the desired tolerance, and this also reduces the cost of the inner-solver.
We use the following tolerance control method for the inner-solver.
\begin{equation}
  \mathit{tol}_{inner} = 
\min\left(\max\left(\frac{\mathit{err}_{outer}}{\mathit{tol}_{outer}},
10^{-6}\right),10^{-3}\right),
\end{equation}
where $\mathit{err}_{outer}$ is the relative residual norm $|b - Ax|/|b|$
for the outer solver. 
When the residual gap to the desired tolerance is larger than $10^{-6}$,
the inner solver is called with $10^{-6}$ tolerance which is the limit 
of single precision arithmetic. As the outer residual decreases the inner-solver
tolerance is relaxed.

The flexible BiCGStab algorithm is applied to both the IR and the UV problems.
We solve
\begin{equation}
  \tilde{D} (M z) = w,
  \label{eq:IRflex}
\end{equation}
with a flexible preconditioner $M$ for
the IR problem Eq.~(\ref{eq:IRlinequnp})
(heavy mass $\kappa'$ version is also modified),
and
\begin{equation}
  (\hat{D}_{EE})_{\SSOR} (M z) = d,
  \label{eq:UVflex}
\end{equation}
with a flexible preconditioner $M$ 
for the UV problem Eq.~(\ref{eq:UVSSORlineq}).
With these setup the flexible BiCGStab calls the inner solver 3 to 5 times 
to obtain the double precision solution.

\subsubsection{Inner solver and deflation technique}

Because the outer solver is well preconditioned by an inner-solver,
the residual stagnation or convergence failure should not take place.
However the problem of the near zero modes still remains and is left 
to the inner-solver to handle. 

As explained in the main text we use the combination of BiCGStab and
GCRO-DR\cite{deflation} solvers.
The inner solver usually uses BiCGStab. If residual stagnation or breakdown is
detected the solver restarts with the GCRO-DR algorithm.
The use of GCRO-DR is the key point to handle the ill conditioned problem 
in our algorithm. The GCRO-DR incorporates the so-called deflation technique which
removes or deflates the ill conditioned eigenmodes from the matrix 
spectrum as has been described in the literature\cite{deflWilcox,deflMorgan2,deflluscher}.

GCRO-DR solver has the following properties;
(1) solves a linear equation and its eigensubspace simultaneously,
(2) deflates the eigenmodes from the coefficient matrix and 
    reduces the condition number,
(3) can recycle the eigenmodes for another linear equation
with the same or perturbed coefficient matrix 
but different right-hand vectors.

Since the inner-solver is to be called several times by the outer-solver 
and the outer-solver is to be called many times during the MD evolution,
the property (2) might largely help to solve the ill conditioned problem.
The property (3) opens the possibility of reusing the deflation subspace
among the MD evolution steps for possible further speedup.

Unfortunately the performance of GCRO-DR algorithm highly 
depends on the problem to be solved, and we observed that 
the overhead is large compared to normal BiCGStab for well conditioned cases.
One may consider reusing the deflation subspace generated by GCRO-DR 
for the so-called deflated BiCGStab (D-BiCGStab) algorithm.
However, for well conditioned cases
the overhead is still rather large and no improvement is observed.
Moreover we observed that the rate of the occurence of ill conditioned cases  is low.
We, therefore, use the normal BiCGStab algorithm for a first attack, 
and switch the solver to GCRO-DR only when the stagnation or breakdown
is detected as described above, otherwise continue to use the un-deflated BiCGStab.
Once the inner solver is switched to the GCRO-DR solver, GCRO-DR is kept being used
until the outer iteration converges. 
If there is another linear equation with
the same ill-conditioned coefficient matrix in the MD force calculation,
GCRO-DR continues with recycling the deflation subspace.

The actual equation to be solved by the inner solver is as follows.
For the IR problem Eq.~(\ref{eq:IRflex}) to 
obtain $t = Mz \sim (\tilde{D})^{-1} z$, we use
\begin{subequations}
\begin{eqnarray}
  \tilde{D}K s &=& z, \label{eq:IRinner} \\
             t &=& K s,
\end{eqnarray}
\end{subequations}
where Eq.~(\ref{eq:IRinner}) is solved by BiCGStab or GCRO-DR algorithms, and
computation are entirely done with single precision. 
The deflation subspace is spanned for $\tilde{D}K$ when switching occurs.
Similarly we solve
\begin{equation}
  (\hat{D}_{EE})_{\SSOR} t = z, \label{eq:UVinner} \\
\end{equation}
for the UV problem Eq.~(\ref{eq:UVflex})
to obtain $t = Mz \sim (\hat{D}_{EE})_{\SSOR}^{-1} z$.

The parameters for the GCRO-DR algorithm is 
the maximal dimeinsion of Krylov subspace $N_{\KV}$ and
the dimension of deflation/recycling subspace $N_{\mathrm{REC}}$.
The initial value is set to ($N_{\KV},N_{\mathrm{REC}}$)=(40,20),
and it is automatically enlarged when slow convergences are observed.

\bibliography{apssamp}

\clearpage

\begin{table*}[h] 
\setlength{\tabcolsep}{10pt}
\renewcommand{\arraystretch}{1.2}
\centering
\caption{Simulation parameters.  
MD time is the number of
trajectories multiplied by the trajectory length $\tau$.
$\tau_{\rm int}[P]$ denotes the integrated autocorrelation time for 
the plaquette. CPU time for unit $\tau$ using 256 nodes of PACS-CS 
is also listed.}
\label{tab:param}
\begin{ruledtabular}
\begin{tabular}{lllllll} 
$\kappa_{\rm ud}$ & 0.13700 & 0.13727 & 0.13754 & 0.13754 
 & 0.13770 & 0.13781 \\ 
$\kappa_{\rm s}$  & 0.13640 & 0.13640 & 0.13640 & 0.13660 
 & 0.13640 & 0.13640 \\  \hline 
\#run & 1 & 1 & 2 & 4 & 2 & 5 \\
$\tau$   & 0.5 & 0.5 & 0.5 & 0.5 & 0.25 & 0.25 \\
$(N_0,N_1,N_2,N_3)$ &  (4,4,10) &  (4,4,14) &  (4,4,20) & (4,4,28) & (4,4,16) & (4,4,4,6) \\
                    &           &           &           &          &  & (4,4,6,6) \\
$\rho$      & $-$ & $-$ & $-$ & $-$ & $-$ & 0.9995 \\
$N_{\rm poly}$ & 180 & 180 & 180 & 220 & 180 & 200 \\
Replay      & on  & on & on & on & on & off \\
MD time & 2000 & 2000 & 2250 & 2000 & 2000 & 990 \\
$\la P\ra$ & 0.569105(18) & 0.569727(14) & 0.570284(16) & 0.570554(17) &
 0.570573(20) & 0.570868(9) \\
$\la {\rm e}^{-dH}\ra$  &  0.9922(85) & 1.0016(50) & 1.0013(56) &
 0.9993(36)  & 0.9944(53) & 0.970(12) \\
$P_{\rm acc}$(HMC) & 0.8020(63) & 0.8672(47) & 0.8573(52) & 0.9140(44) &
 0.8397(41) & 0.8033(63) \\
$P_{\rm acc}$(GMP) & 0.9529(37)  &  0.9439(34)  & 0.9331(40)  &
 0.9330(41) & 0.9537(26) & 0.9670(32) \\
$\tau_{\rm int}[P]$ & 8.6(3.1)  & 20.9(10.2)  & 9.8(2.8) & 6.3(1.4) &
 25.2(15.2)  & 2.9(1.9) \\
\hline
CPU hour/unit $\tau$ & 0.29  & 0.44 & 1.3 & 1.1 & 2.7 & 7.1 \\
\end{tabular} 
\end{ruledtabular}
\end{table*}

\begin{table*}[h] 
\setlength{\tabcolsep}{10pt}
\renewcommand{\arraystretch}{1.2}
\centering
\caption{Smearing parameters $A$ and $B$ for the ud and s quark
propagators.}
\label{tab:param_smear}
\begin{ruledtabular}
\begin{tabular}{cccccccccccccc} 
$\kappa_{\rm ud}$ & $\kappa_{\rm s}$ & \#source & 
$A_{\rm ud}$ & $B_{\rm ud}$ & $A_{\rm s}$ & $B_{\rm s}$ \\ \hline 
0.13700 & 0.13640 & 1 & 1.2 & 0.21 & 1.2 & 0.28 \\
0.13727 & 0.13640 & 1 & 1.2 & 0.19 & 1.2 & 0.25 \\
0.13754 & 0.13640 & 4 & 1.2 & 0.17 & 1.2 & 0.25 \\
0.13754 & 0.13660 & 4 & 1.2 & 0.17 & 1.2 & 0.25 \\ 
0.13770 & 0.13640 & 4 & 1.2 & 0.09 & 1.2 & 0.21 \\
0.13781 & 0.13640 & 4 & 1.2 & 0.07 & 1.2 & 0.20 \\ 
\end{tabular}
\end{ruledtabular}
\end{table*}

\begin{table*}[h] 
\setlength{\tabcolsep}{10pt}
\renewcommand{\arraystretch}{1.2}
\centering
\caption{Meson and baryon masses in lattice units 
at each combination  of $\kappa_{\rm ud}$ and $\kappa_{\rm s}$. $\chi^2$/dof 
for the fit is also presented in the second row of each channel. The fit range 
is $[13,30]$ for pseudoscalar mesons, $[10,20]$ for vector mesons, $[10,20]$ for 
octet baryons and $[8,20]$ for decuplet baryons.}
\label{tab:hmass}
\begin{ruledtabular}
\begin{tabular}{lllllll} 
$\kappa_{\rm ud}$ & 0.13700 & 0.13727 & 0.13754 & 0.13754 
 & 0.13770 & 0.13781 \\ 
$\kappa_{\rm s}$  & 0.13640 & 0.13640 & 0.13640 & 0.13660 
 & 0.13640 & 0.13640 \\  \hline 
$\pi$           & 0.32242(65) & 0.26191(73) & 0.18903(79) & 0.17671(129)
 & 0.13593(140) & 0.07162(299) \\
                & 0.015       & 0.008       & 0.002       & 0.001   
 & 0.001       & 0.004 \\
$K$             & 0.36269(61) & 0.32785(74) & 0.29190(67) & 0.26729(110)
 & 0.27282(103) & 0.25454(97) \\
                & 0.016       & 0.015       & 0.002       & 0.001 
 & 0.001       & 0.025 \\
$\eta_{\rm ss}$ & 0.39947(58) & 0.38380(74) & 0.36870(71) & 0.33490(93) 
 & 0.36289(103) & 0.35306(82) \\
                & 0.017       & 0.015       & 0.000       & 0.002 
 & 0.001       & 0.016 \\
$\rho$          & 0.5060(30)  & 0.4566(36)  & 0.4108(31)  & 0.3963(53) 
 & 0.3895(94)  & 0.3503(315) \\ 
                & 0.043       & 0.229       & 0.017       & 0.090 
 & 0.005       & 0.418 \\
$K^*$           & 0.5314(23)  & 0.4954(32)  & 0.4665(23)  & 0.4428(37) 
 & 0.4525(35)  & 0.4316(47) \\
                & 0.088       & 0.068       & 0.007       & 0.014 
 & 0.003       & 0.092 \\
$\phi$          & 0.5560(17)  & 0.5325(28)  & 0.5156(21)  & 0.4849(26) 
 & 0.5105(26)  & 0.4949(15) \\ 
                & 0.124       & 0.015       & 0.002       & 0.007 
 & 0.001       & 0.026 \\
$N$             & 0.7277(22)  & 0.6487(56)  & 0.5584(53)  & 0.5331(71) 
 & 0.5025(87)  & 0.4285(360) \\ 
                & 0.077       & 0.027       & 0.358       & 0.014 
 & 0.171       & 1.138 \\
$\Lambda$       & 0.7557(23)  & 0.6913(45)  & 0.6208(36)  & 0.5857(42) 
 & 0.5764(65)  & 0.5240(95) \\ 
                & 0.115       & 0.029       & 0.089       & 0.015 
 & 0.018       & 0.154 \\
$\Sigma$        & 0.7606(20)  & 0.7039(51)  & 0.6437(39)  & 0.6052(48) 
 & 0.6044(71)  & 0.5601(99) \\ 
                & 0.072       & 0.030       & 0.041       & 0.091 
 & 0.043       & 1.377 \\
$\Xi$           & 0.7859(25)  & 0.7399(43)  & 0.6910(30)  & 0.6474(32) 
 & 0.6655(46)  & 0.6405(31) \\ 
                & 0.139       & 0.035       & 0.028       & 0.020 
 & 0.010       & 0.008 \\
$\Delta$        & 0.8290(42)  & 0.7694(84)  & 0.6956(66)  & 0.6731(86) 
 & 0.6438(90)  & 0.5798(378) \\ 
                & 0.046       & 0.022       & 0.102       & 0.038 
 & 0.860       & 0.421 \\
$\Sigma^*$      & 0.8537(35)  & 0.8039(74)  & 0.7464(43)  & 0.7149(74) 
 & 0.7097(67)  & 0.6885(140) \\ 
                & 0.037       & 0.010       & 0.022       & 0.036 
 & 0.179       & 0.031 \\
$\Xi^*$         & 0.8788(30)  & 0.8395(67)  & 0.7964(41)  & 0.7579(60) 
 & 0.7740(58)  & 0.7549(67) \\ 
                & 0.036       & 0.005       & 0.005       & 0.035     
 & 0.022       & 0.255 \\
$\Omega$        & 0.9038(29)  & 0.8754(61)  & 0.8456(37)  & 0.8001(49) 
 & 0.8342(52)  & 0.8142(34) \\ 
                & 0.050       & 0.003       & 0.009       & 0.032  
 & 0.015       & 0.206 \\
\end{tabular}
\end{ruledtabular}
\end{table*}

\begin{table*}[h] 
\setlength{\tabcolsep}{10pt}
\renewcommand{\arraystretch}{1.2}
\centering
\caption{Quark masses in the ${\overline {\rm MS}}$ scheme
 at the scale of $1/a$ and pseudoscalar decay constants  
at each combination  of $\kappa_{\rm ud}$ and $\kappa_{\rm s}$. 
Both are renormalized at one-loop level.
The values for $m_\pi^2/m_{\rm ud}^{\rm AWI}$ and $f_K/f_\pi$ are also
listed.}
\label{tab:qmass}
\begin{ruledtabular}
\begin{tabular}{lllllll} 
$\kappa_{\rm ud}$ & 0.13700 & 0.13727 & 0.13754 & 0.13754 
 & 0.13770 & 0.13781 \\ 
$\kappa_{\rm s}$  & 0.13640 & 0.13640 & 0.13640 & 0.13660 
 & 0.13640 & 0.13640 \\  \hline 
$am_{\rm ud}^{\overline {\rm MS}}$           
& 0.030753(110) & 0.020834(66) & 0.011028(80) & 0.009666(105) 
& 0.005644(120) & 0.001609(118) \\
$am_{\rm s}^{\overline {\rm MS}}$           
& 0.047142(110) & 0.044674(72) & 0.042355(79) & 0.035571(98) 
& 0.041285(94)  & 0.039913(62) \\
$m_{\rm s}/m_{\rm ud}$
& 1.5329(20)  &   2.1443(39)   &    3.841(21)    & 3.680(30)
& 7.32(14)    &   24.8(1.8)    \\
$af_\pi$           
& 0.0898(12) & 0.0853(18) & 0.07481(51) & 0.07262(60) 
& 0.06973(78) & 0.0656(35) \\
$af_K$           
& 0.0942(13) & 0.0916(15) & 0.08432(56) & 0.08058(40) 
& 0.08089(57) & 0.0777(13) \\
$am_\pi^2/m_{\rm ud}^{\rm AWI}$           
& 3.708(13) & 3.610(14) & 3.558(16) & 3.542(25) 
& 3.585(29) & 3.732(73) \\ 
$f_K/f_\pi$           
& 1.0485(13) & 1.0739(57) & 1.1271(16) & 1.1095(64)
& 1.1601(73) & 1.186(48) \\
\end{tabular}
\end{ruledtabular}
\end{table*}

\begin{table*}[h!]
\centering
\caption{PACS-CS and CP-PACS/JLQCD results for 
the hadron masses at 
$(\kappa_{\rm ud}, \kappa_{\rm s})=(0.13700, 0.13640)$.
[$t_{\rm min}$,$t_{\rm max}$] denotes the fitting range.}
\label{tab:comp}
\begin{ruledtabular}
\begin{tabular}{ccccc}  
   & lattice size      &  $m_{\pi}$ & $m_{\rho}$ & $m_{\rm N}$  \\ \hline
PACS-CS     &$32^3\times 64$& 0.32242(65) & 0.5060(30) & 0.7277(22) \\
$[t_{\rm min},t_{\rm max}]$ & & [13,30] & [10,20] & [10,20]  \\ 
CP-PACS/JLQCD&$20^3\times 40$ &0.32247(74)& 0.5157(21) & 0.7337(28) \\ 
$[t_{\rm min},t_{\rm max}]$ & & [9,17] & [9,15] & [9,15] \\ 
\end{tabular}
\end{ruledtabular}
\end{table*}

\begin{table*}[h!]
\centering
\caption{Results for the low energy constants in the SU(3) ChPT 
together with the
phenomenological estimates and the RBC/UKQCD and MILC results.
$f_0$ is perturbatively renormalized at one-loop level.
$L_{4,5,6,8}$ are in units of $10^{-3}$ at the scale of 770MeV.
$\la {\bar u}u \ra$ and $\la {\bar u}u \ra_0$  are
 renormalized in the ${\overline {\rm MS}}$ scheme at 2 GeV.}
\label{tab:fit_su3chpt}
\begin{ruledtabular}
\begin{tabular}{cccccc}
& \multicolumn{2}{c}{PACS-CS} & phenomenology\cite{colangelo05,amoros01} & 
 RBC/UKQCD\cite{rbcukqcd08} & MILC\cite{milc07} \\
& w/o FSE  & w/ FSE   &  & &   \\
\hline
$aB_0$      & 1.789(34)  & 1.778(34)  & $-$   & 2.35(16)   & $-$ \\
$af_0$      & 0.0534(38) & 0.0546(39) & $-$   & 0.0541(40) & $-$ \\
$f_0$[GeV]  & 0.1160(88) & 0.1185(90) & 0.115 & 0.0935(73) & $-$ \\
$f_\pi/f_0$  & 1.159(57) & 1.145(56) & 1.139 &  1.33(7)   &  $1.21(5)\left(^{+13}_{-3}\right)$ \\
$m_{\rm ud}^{\rm ph}B_0$[GeV$^2$] & 0.00859(10) & 0.00859(11) & 0.0181 & $-$ & $-$ \\ 
$m_{\rm s}^{\rm ph}B_0$[GeV$^2$]  & 0.2550(36) & 0.2534(36) & 0.434  & $-$ & $-$ \\  
$a^3\langle {\bar u}u\rangle_0$ & $-(0.132(6))^3$ & $-(0.133(6))^3$ & $-$ & $-$ & $-$ \\
$\langle {\bar u}u\rangle_0$[GeV$^3$]  & $-(0.286(15))^3$ &
 $-(0.290(15))^3$    & $-$ & $-$ &
$-\left(0.242(9)\left(^{+5}_{-17}\right)(4)\right)^3$\\
$L_4$        & $-0.04(10)$  & $-0.06(10)$       & 0.00(80)  
& 0.139(80) & $0.1(3)\left(^{+3}_{-1}\right)$           \\
$L_5$        & 1.43(7)  & 1.45(7)       & 1.46(10)  
& 0.872(99) & $1.4(2)\left(^{+2}_{-1}\right)$           \\
$2L_6-L_4$   &  0.10(2)  &  0.10(2)         & 0.0(1.0)     
& $-$0.001(42) & $0.3(1)\left(^{+2}_{-3}\right)$ \\ 
$2L_8-L_5$    &  $-0.21(3)$  & $-0.21(3)$      & 0.54(43)  
& 0.243(45) & 0.3(1)(1)    \\
\hline
$\chi^2$/dof & 4.2(2.7) & 4.4(2.8)  & $-$ & 0.7 & $-$  \\
\end{tabular}
\end{ruledtabular}
\end{table*}

\begin{table*}[h!]
\centering
\caption{Results for the low energy constants 
in the SU(2) ChPT obtained by the conversion from those
in the SU(3) ChPT. The RBC/UKQCD and MILC results are also given 
for comparison. $f$ and $f_0$ are perturbatively renormalized at one-loop level.
$\la {\bar u}u \ra$ and $\la {\bar u}u \ra_0$  are
 renormalized in the ${\overline {\rm MS}}$ scheme at 2 GeV.}
\label{tab:lec_su2_conv}
\begin{ruledtabular}
\begin{tabular}{ccccc}
& \multicolumn{2}{c}{PACS-CS} & 
 RBC/UKQCD\cite{rbcukqcd08} & MILC\cite{milc07} \\
& w/o FSE  & w/ FSE   &   &   \\
\hline
$aB$ & 1.950(31) & 1.935(30) &  2.457(78) & $-$ \\
$af$ & 0.0582(19) & 0.0588(19) & 0.0661(18) & $-$ \\
$f$[GeV] & 0.1263(51) & 0.1277(51) & 0.1143(36) & $-$ \\
$f_\pi/f$  & 1.065(8) & 1.062(8) & $-$    &  $1.052(2)\left(^{+6}_{-3}\right)$ \\
$m_{\rm ud}^{\rm ph}B$[GeV$^2$] & 0.009364(36) & 0.009352(34) & $-$ & $-$ \\ 
$m_{\rm s}^{\rm ph}B$[GeV$^2$]  & 0.2780(52) & 0.2758(49) & $-$ & $-$ \\  
$a^3\langle {\bar u}u\rangle$ &  $-(0.143(3))^3$ &  $-(0.144(3))^3$ & $-$ & $-$ \\
$\langle {\bar u}u\rangle$[GeV$^3$] & $-(0.310(9))^3$ &
 $-(0.312(10))^3$ & $-$ & $-\left(0.278(1)\left(^{+2}_{-3}\right)(5)\right)^3$\\
${\bar l}_3$ & 3.50(11) & 3.47(11) & 2.87(28) & $1.2(6)\left(^{+1.0}_{-1.5}\right)$ \\
${\bar l}_4$ & 4.22(10) & 4.21(11) & 4.10(5) & $4.4(4)\left(^{+4}_{-1}\right)$ \\
\hline
$B/B_0$ & 1.090(15) & 1.089(15) & $-$ & $-$ \\
$f/f_0$  & 1.089(45) & 1.078(44)  & $-$ & $1.15(5)\left(^{+13}_{-3}\right)$ \\
$\langle {\bar u}u\rangle$/$\langle {\bar u}u\rangle_0$ & 1.268(10) &
 1.245(10) & $-$ & $1.52(17)\left(^{+38}_{-15}\right)$\\
\end{tabular}
\end{ruledtabular}
\end{table*}

\begin{table*}[h!]
\centering
\caption{Comparison of ${\bar l}_{3,4}$.}
\label{tab:l_34}
\begin{ruledtabular}
\begin{tabular}{llllll}
group  & \#flavor & quark action & ChPT &${\bar l}_3$ & ${\bar l}_4$ \\
\hline
this work & 2+1 & NP clover & SU(2) w/o FSE & 3.23(21) & 4.10(20) \\
          &     &  & SU(2) w/  FSE & 3.14(23) & 4.04(19) \\
          &     &  & SU(3) w/o FSE & 3.50(11) & 4.22(10) \\
          &     &  & SU(3) w/ FSE  & 3.47(11) & 4.21(11) \\
RBC/UKQCD\protect{\cite{rbcukqcd08}} & 2+1 & DWF       & SU(3) &
 2.87(28) & 4.10(5) \\
                           &     &           & SU(2) &
 3.13(33)(24) & 4.43(14)(77) \\
MILC\protect{\cite{milc07}} & 2+1 & KS       & SU(3) &
 $1.1(6)\left(^{+1.0}_{-1.5}\right)$ & $4.4(4)\left(^{+4}_{-1}\right)$ \\
JLQCD\protect{\cite{jlqcd08}} & 2 & Overlap  & SU(2) & 
$3.44(57)\left(^{+0}_{-68}\right)\left(^{+32}_{-0}\right)$ & 
$4.14(26)\left(^{+49}_{-0}\right)\left(^{+32}_{-0}\right)$ \\
ETM\protect{\cite{urbach07}} & 2 & TM        & SU(2) &
 3.44(8)(35) & 4.61(4)(11) \\
CERN\protect{\cite{del07}} & 2 & Wilson + NP clover   & SU(2) & 3.0(5)(1) &
 $-$ \\
\hline
CGL\protect{\cite{colangelo01}} & $-$ & $-$ $-$  & SU(2) & $-$ & 4.4(2) \\
GL\protect{\cite{chpt_nf2}} & $-$ & $-$ $-$  & SU(2) &
 2.9(2.4) & 4.3(9) \\
\end{tabular}
\end{ruledtabular}
\end{table*}

\begin{sidewaystable*}[h]
\centering
\caption{Results for the low energy constants in the SU(2) ChPT fit
together with the
phenomenological estimates and the RBC/UKQCD results.
$B=B_{\rm s}^{(0)}+m_{\rm s}B_{\rm s}^{(1)}$ and 
$f=f_{\rm s}^{(0)}+m_{\rm s}f_{\rm s}^{(1)}$
are given at the physical strange quark mass. 
$f$ and $f_0$ are perturbatively renormalized at one-loop level.
Range I, II, III denote the selection of data sets corresponding to 
$\kappa_{\rm ud}\ge 0.13754$,
$\kappa_{\rm ud}\ge 0.13727$, $0.13770\ge \kappa_{\rm ud}\ge 0.13727$,
 respectively. $\la {\bar u}u \ra$ and $\la {\bar u}u \ra_0$  are
 renormalized in the ${\overline {\rm MS}}$ scheme at 2 GeV.}
\label{tab:fit_su2chpt}
\begin{ruledtabular}
\begin{tabular}{ccccccccc}
& \multicolumn{6}{c}{PACS-CS} & phenomenology & 
 RBC/UKQCD\cite{rbcukqcd08} \\
& \multicolumn{2}{c}{Range I} & \multicolumn{2}{c}{Range II} &
 \multicolumn{2}{c}{Range III} & & \\ 
& w/o FSE  & w/ FSE   & w/o FSE  & w/ FSE   &  w/o FSE  & w/ FSE   & &    \\
\hline
$aB$      & 1.907(36) & 1.891(35) & 1.941(20) & 1.931(21) & 1.947(20) &
 1.942(20) & $-$   & 2.414(61)(115)  \\
$af$      & 0.0573(23) & 0.0581(21) & 0.0547(13) & 0.0553(14) &
 0.0541(13) & 0.0544(13) & $-$   & 0.0665(21)(47) \\
$f$[GeV]  & 0.1248(51) & 0.1264(47) & 0.1181(30) & 0.1194(31) &
0.1158(28) & 0.1165(28) & 0.1219(7)\cite{colangelo04} &  0.1148(41)(81)\\
$f_\pi/f$  & 1.063(8) & 1.060(7) & 1.074(5) & 1.072(5) & 1.078(5) &
 1.077(5) & 1.072(7)\cite{colangelo04} &  1.080(8)  \\
$m_{\rm ud}^{\rm ph}B$[GeV$^2$] & 0.009345(27) & 0.009332(26) &
0.009381(16) & 0.009372(17) & 0.009391(17) & 0.009387(16) & $-$ & 0.00937(57)(64) \\ 
$m_{\rm s}^{\rm ph}B$[GeV$^2$]  & 0.2709(43) & 0.2686(43) &
0.2782(25) & 0.2768(26) & 0.2794(26) & 0.2787(26) & $-$ & 0.270(16)(18) \\  
$a^3\langle {\bar u}u\rangle$  & $-(0.141(3))^3$ & $-(0.142(3))^3$ & 
$-(0.138(2))^3$ & $-(0.138(2))^3$ & 
$-(0.137(2))^3$ & $-(0.137(2))^3$ & $-$ & $-$ \\
$\langle {\bar u}u\rangle$[GeV$^3$] & $-(0.307(8))^3$ & $-(0.309(7))^3$ & 
$-(0.297(5))^3$ & $-(0.299(5))^3$ &
$-(0.293(5))^3$ & $-(0.294(4))^3$& $-$ & $-\left(0.255(8)(8)(13)\right)^3$ \\
${\bar l}_3$    & 3.23(21) & 3.14(23) & 3.32(10) & 3.28(11) & 
3.31(10) & 3.30(10) & 2.9(2.4)\cite{chpt_nf2} & 3.13(33)(24) \\
${\bar l}_4$    & 4.10(20) & 4.04(19) & 4.32(9) & 4.28(10) & 
4.36(9) & 4.34(9) & 4.4(2)\cite{colangelo01} & 4.43(14)(77) \\
\hline
$B/B_0$ & 1.066(15) & 1.064(15) & $-$ & $-$ & $-$ & $-$ & $-$ & 1.03(5) \\
$f/f_0$ & 1.073(55) & 1.065(58) & $-$ & $-$ & $-$ & $-$ & $-$ & 1.229(59) \\
$\langle {\bar u}u\rangle$/$\langle {\bar u}u\rangle_0$ & 
1.228(13) & 1.205(14) & $-$ & $-$ & $-$ & $-$ & $-$ & 1.55(21) \\
\hline
$\chi^2$/dof & 0.33(68) & 0.43(77) & 2.0(1.0) & 2.3(1.1) & 2.8(1.8) & 3.0(1.8) & $-$ & 0.3  \\
\end{tabular}
\end{ruledtabular}
\end{sidewaystable*}

\begin{table*}[h!]
\centering
\caption{Cutoff, renormalized 
quark masses, pseudoscalar meson decay constants determined with 
$m_\pi$, $m_K$, $m_\Omega$ inputs. Quark masses are renormalized at 2 GeV.}
\label{tab:physicalpt_qmass}
\begin{ruledtabular}
\begin{tabular}{cccc}  
 &  \multicolumn{2}{c}{physical point} & experiment\protect{\cite{pdg}} \\ 
 &   w/o FSE &  w/ FSE  &  \\ 
\hline 

$a^{-1}$~[GeV]  & 2.176(31) & 2.176(31) & $-$  \\ 
$m^{\overline{\rm MS}}_{\rm ud}$~[MeV] & 2.509(46) & 2.527(47) & $-$  \\ 
$m^{\overline{\rm MS}}_{\rm s}$~[MeV]  & 72.74(78) & 72.72(78) & $-$  \\ 
$m_{\rm s}/m_{\rm ud}$ & 29.0(4) & 28.8(4) & $-$  \\
$f_\pi$~[MeV] & 132.6(4.5) & 134.0(4.2) & $130.7\pm 0.1\pm 0.36$  \\ 
$f_K$~[MeV]   & 159.2(3.2) & 159.4(3.1) & $159.8\pm 1.4\pm 0.44$  \\ 
$f_K/f_\pi$   & 1.201(22)  & 1.189(20)  & 1.223(12)  \\ 
\end{tabular}
\end{ruledtabular}
\end{table*}

\begin{table*}[h] 
\setlength{\tabcolsep}{10pt}
\renewcommand{\arraystretch}{1.2}
\centering
\caption{Meson and baryon masses at the physical point in physical
 units. $m_\pi$, $m_K$, $m_\Omega$ are inputs.}
\label{tab:physicalpt_hmass}
\begin{ruledtabular}
\begin{tabular}{llll} 
channel & experiment~[GeV]\protect{\cite{pdg}} & \multicolumn{2}{c}{physical point~[GeV]} \\ 
 &  &  w/o FSE &  w/ FSE \\ 
\hline 
$\pi$           & 0.1350 & $-$ & $-$ \\ 
$K$             & 0.4976 & $-$ & $-$ \\
$\rho$          & 0.7755 & 0.776(34) & 0.776(34) \\   
$K^*$           & 0.8960 & 0.896(9) & 0.896(9) \\   
$\phi$          & 1.0195 & 1.0084(40) & 1.0084(40) \\   
$N$             & 0.9396 & 0.953(41) & 0.953(41) \\   
$\Lambda$       & 1.1157 & 1.092(20) & 1.092(20) \\   
$\Sigma$        & 1.1926 & 1.156(17) & 1.156(17) \\   
$\Xi$           & 1.3148 & 1.304(10) & 1.304(10) \\   
$\Delta$        & 1.232  & 1.274(39) & 1.275(39) \\   
$\Sigma^*$      & 1.3837 & 1.430(23) & 1.430(23) \\   
$\Xi^*$         & 1.5318 & 1.562(9)  & 1.562(9)  \\   
$\Omega$        & 1.6725 & $-$ & $-$ \\   
\end{tabular}
\end{ruledtabular}
\end{table*}

\begin{table*}[h!]
\centering
\caption{$r_0$ at each hopping parameter and the physical point.
The first error at the physical point is statistical 
and the second and the third ones are the 
 systematic uncertainties due to the choice 
of $t_{\rm min}$ and $r_{\rm min}$, respectively.}
\label{tab:potential}
\begin{ruledtabular}
\begin{tabular}{lll} 
$\kappa_{\rm ud}$ & $\kappa_{\rm s}$ & $r_0$ \\ 
\hline 
0.13700 & 0.13640 & 4.813(30)(+40)(+13) \\
0.13727 & 0.13640 & 4.879(38)(+35)(+74) \\
0.13754 & 0.13640 & 5.121(21)(+82)(+9)  \\
0.13754 & 0.13660 & 5.276(28)(+85)(+8) \\ 
0.13770 & 0.13640 & 5.176(23)(+54)(+8)  \\
0.13781 & 0.13640 & 5.276(33)(+112)($-3$) \\
\hline 
\multicolumn{2}{l}{physical point} &  5.427(51)(+81)($-2$) \\
\end{tabular}
\end{ruledtabular}
\end{table*}

\begin{table*}[h!]
\centering
\caption{PACS-CS and CP-PACS/JLQCD results for $r_0$ in lattice units at 
$(\kappa_{\rm ud}, \kappa_{\rm s})=(0.13700, 0.13640)$.  Meaning of errors 
are the same as in Table~\ref{tab:potential}.}
\label{tab:r0comp}
\begin{ruledtabular}
\begin{tabular}{ccccc}  
   & lattice size      &  $r_0$ \\ \hline
PACS-CS     &$32^3\times 64$& 4.813(30)(+40)(+13) \\
CP-PACS/JLQCD&$20^3\times 40$ &4.741(33)(+323)(+30) \\ 
\end{tabular}
\end{ruledtabular}
\end{table*}

\clearpage

\clearpage

\end{document}